\documentclass[a4paper,11pt]{article}
\pdfoutput=1 
\usepackage{jheppub}
\usepackage{amsmath,amssymb,amsthm,amscd}

\usepackage{tikz} 
\usepackage{dynkin-diagrams}

\usepackage{array}
\usepackage{amsfonts}
\usepackage{color}
\usepackage{braket}
\usepackage[utf8]{inputenc}
\usepackage{indentfirst}
\usepackage{hyperref} 
\usepackage{comment}

\input
\usepackage{subcaption}
\usepackage{graphicx}
\usepackage[export]{adjustbox}
\usepackage{colortbl}
\usepackage{booktabs}                       
\usepackage{multirow,bigdelim}              
\usepackage{longtable}                      
\usepackage{units}                          
\usepackage{slashed}                        
\usepackage{float}
\usepackage[export]{adjustbox}              
\usepackage{url}                            
\usepackage{fancyhdr}

\def \be {\begin{equation}}
\def \ee {\end{equation}}
\def \ba {\begin{aligned}}
\def \ea {\end{aligned}}

\newcommand{\re}{{\rm e}}
\newcommand{\ri}{{\rm i}}
\newcommand{\rd}{{\rm d}}

\def\({\left(}
\def\){\right)}
\def\[{\left[}
\def\]{\right]}

\newcommand{\RN}[1]{\textup{\uppercase\expandafter{\romannumeral#1}}}

\title{\huge TBA equations and quantum periods for D-type Argyres-Douglas theories}
\author{Katsushi Ito, Jingjing Yang}

\affiliation{Department of Physics, Tokyo Institute of Technology,\\Tokyo, 152-8551, Japan}

\emailAdd{ito@th.phys.titech.ac.jp, j.yang@th.phys.titech.ac.jp}

\abstract{
We construct TBA equations for D-type Argyres-Douglas theories with an SU(2) flavor symmetry based on their spectral networks. We show that the solutions of these TBA equations agree with the quantum periods of the corresponding quantum Seiberg-Witten curves defined in the Nekrasov-Shatashvili limit of the Omega background, including a centrifugal correction. We study the variety of TBA systems across the Coulomb branch moduli space and find that they correspond to the Dynkin diagrams of $D_n$ Lie algebras in the minimal chamber, and reproduce the TBA equations for reflectionless D scattering theories at the maximally symmetric point. Numerical computations demonstrate that the quantum periods obtained from the Borel-Pad{\'e} resummation and their WKB expansions are in agreement with the solutions of the TBA equations.}

\begin{document}

\maketitle

\flushbottom

\section{Introduction}

Four-dimensional $\mathcal{N}=2$ supersymmetric gauge theory provides a rich framework for studying the non-perturbative structures of quantum field theory. The low-energy effective action of these theories is described by the prepotential, which is completely determined by the period integrals on the Riemann surface, known as the Seiberg-Witten (SW) curve \cite{SW941, SW942}. At the critical points on the Coulomb branch moduli space, the mutually non-local BPS states become massless, and their dynamics are captured by a superconformal field theory (SCFT) known as the  Argyres-Douglas (AD) theory \cite{AD95, ADSW95}. AD theories can be realized near the superconformal points of $\mathcal{N}=2$ gauge theories \cite{EHIY96}; from the compactification of the six-dimensional (2,0) theory on a Riemann surface \cite{Gaiotto09, GMN09, Bonelli:2011aa, Xie12, WX15}; and by compactifying type $\RN{2}\mathrm{B}$ string theory on a singular Calabi-Yau threefold, a process known as geometric engineering \cite{Alim11, KLPV96, KKV96}. We focus on AD theories compactified from the $A_1$ (2,0) theory, whose SW curve is a Riemann surface with an irregular puncture for type A theories, and both an irregular and a regular puncture for type D theories, denoted as $(A_1, A_r)$ or $(A_1, D_r)$ theories, respectively.

We introduce the $\Omega$ background, which provides rich structures for studying supersymmetric gauge theories under various deformations. The Nekrasov partition function from instanton counting reproduces the prepotential of the low-energy effective theory in the limit of the $\Omega$-background parameters $\epsilon_1, \epsilon_2\to0$ \cite{Nekrasov02}. When both parameters $(\epsilon_1, \epsilon_2)$ are turned on with $\epsilon_1=-\epsilon_2$, the Nekrasov partition function gives the free energy of topological string theory. We focus on the Seiberg-Witten (SW) theory embedded in the Nekrasov-Shatashvili (NS) limit, $\epsilon_1=\hbar$ and $\epsilon_2=0$, which provides a deformation of the SW theory \cite{NS09}. The SW curve is quantized into the quantum SW curve, a second-order differential equation of the Schr{\"o}dinger type. SW periods are quantized into quantum periods through quantum corrections, obtained via the Borel resummation of all-orders WKB (Wentzel–Kramers–Brillouin) expansions. The WKB method was initiated as a semiclassical approximation for solving problems in quantum mechanics \cite{WKB}. It has been developed into an exact approach in mathematics and physics due to a series of works \cite{BW69, BPV79, Voros81, Voros83, DDP93, DDP97, DP99, KT05}. Recent developments include its connection to cluster algebras \cite{IN14, IN15}, resurgent properties \cite{SKMU20, SKMU21}, and application in supersymmetric gauge theories and topological string theory \cite{GGM19, GHN21, Yan20, ML22}. In this paper, we obtain quantum periods of the $(A_1, D_r)$ AD theories via the Borel resummation of the WKB expansions.

Interestingly, it has been discovered that these quantum periods are governed by solutions to a set of integral equations known as Thermodynamic Bethe Ansatz (TBA) equations. These equations were inspired by the work of Gaoitto, Moore, and Neitzke \cite{GMN08, GMN09}, and were specifically proposed in \cite{Gaiotto14}.  In \cite{GGM19}, the authors established the Gaoitto-Moore-Neitzke (GMN) TBA equations for pure SU(2) SYM in their standard form (See also \cite{Fioravanti:2019vxi,Imaizumi:2020fxf}), while detailed studies on TBA for SQCD with one flavor were conducted in \cite{GHN21}. Further applications have been explored in works such as \cite{ML22, OS22}. The GMN approach provides a systematic framework for investigating the BPS spectrum and quantum periods of quantum SW curves. A key tool in this approach is the spectral network \cite{GMN09, GMN12, LP16}, which elucidates the composition of the BPS spectrum and analytic properties of quantum periods, along with their dependence on moduli parameters and wall-crossing phenomena \cite{IK20, IK21}. In this paper, we establish the TBA equations for $(A_1, D_r)$ theories with SU(2) flavor symmetry across different chambers of the moduli space and investigate their wall-crossing. The validity of these TBA equations is confirmed by comparing their solutions with quantum periods derived from the WKB analysis. The BPS spectra of $(A_1, D_r)$ theories with SU(2) symmetry are always finite and were studied in detail in \cite{MPY13}. 

The relationship between the WKB method and the Bethe Ansatz Equation (BAE) was explored through the ODE/IM correspondence, which relates the WKB analysis of an ordinary differential equation (ODE) to the TBA equations of an integrable model (IM) \cite{DT96, DT98, DT07}. Initially proposed for monomial potentials in Schr{\"o}dinger-type equations, this correspondence was later generalized to include polynomial cases. In particular, \cite{IMS18} derived the TBA equations from the Y-system and discussed their wall-crossing in detail, with additional examples provided in \cite{Emery20}. These equations pertain to AD theories of type A. The same approach was applied to derive the TBA equations for $(A_1, D_r)$ theories in the minimal chamber \cite{IS19}. However, wall-crossing for those TBA is not straightforward there. This issue is naturally addressed in the GMN framework, revealing a new type of wall-crossing distinct from those in polynomial cases. In this paper, we present the TBA equations for $(A_1, D_r)$ theories across various chambers and provide specific examples. We also discuss their potential extension to any $r$ and any point in the moduli space, thus extending the family of D-type TBA equations. We also compare the GMN TBA equations derived from the ODE with those obtained from $(1+1)$-dimensional integrable field theories. Notably, at the maximally symmetric point in the moduli space, GMN TBA equations take a simple form, with kernel functions given by the reflectionless ADE scattering S-matrices \cite{Klassen:1989ui}. The universal property of the TBA equations for these scattering theories was well studied by Al. B. Zamolodchikov in \cite{Zam91}.

This paper is organized as follows. Section \ref{sc: TBA} outlines the general considerations, including a summary of the spectral network and the construction of the GMN TBA equations. We also provide a concise review of exact WKB analysis and Borel resummation, and give an overview of the TBA equations for the diagonal scattering theories. We apply this approach to construct the TBA equations for $(A_1, D_r)$ with $r=3,4$ and $r>4$, as detailed in Sections \ref{sc:d3}, \ref{sc:d4}, and \ref{sc:dn}, respectively. We discuss the BPS spectrum and the corresponding TBA equations case by case, following the classification in \cite{MPY13}, with a particular focus on the minimal and maximal chambers. We conclude our discussion in section \ref{sc:concl}, and an appendix on the differential operator method to compute the quantum corrections for the quantum periods is given in \ref{sc: pf}. The numerical results are summarized in Table \ref{tab:d3tbawkb}, \ref{tab:d4tbawkb}, and \ref{tab:d5tbawkb}.

\vspace{0.5cm}

\begin{center}
    Summary of numerical results
\end{center}

\setlength{\arrayrulewidth}{1.0pt} 
\begin{table}[H]
\centering
\begin{tabular}{c c c c c c}
\hline
\multicolumn{6}{c}{Minimal chamber: $u_1=3, u_2=2$}\\
\hline
$\ell$ & $n$ & $\Pi_{\gamma_1}^{(n)}$ & $Z_{\gamma_1}^{(n)}$ & $\Pi_{\gamma_2}^{(n)}$ & $ Z_{\gamma_2}^{(n)}$ \\
\hline
\multirow{3}{*}{$-\frac{1}{2}$} &  $1$ &   $-0.1779896649446$ &  $-0.1779896649456$ &$0.4775247236285\ri$ &  $0.4775247236312\ri$  \\
         
           & $2$ &  $0.0223026467159$  &$0.0223026467160$  &$2.3234312816024\ri$& $ 2.3234312816144\ri$\\
            
            &$3$ & $-0.0154730597043$  &  $-0.0154730597044$ &$67.615114532114\ri$&$67.615114531972\ri$\\
            \hline 
\multirow{3}{*}{$-\frac{1}{5}$} &  $1$ &   $-0.0060807644383$ &  $-0.0060807644386$ &$0.41344844424842\ri$ &  $0.4134484442496\ri$  \\
         
            &$2$ &  $-0.0051315560227$  &$-0.0051315560227$  &$2.2178269880265\ri$& $2.2178269880318\ri$\\
            
            &$3$ & $0.0043851532260$  &  $ 0.0043851532260$ &$65.839049568380\ri$&$ 65.839049568068\ri$\\
            \hline 
\multirow{3}{*}{$\frac{1}{5}$} &  $1$ &   $0.7579587933672$ &  $0.7579587933671$ &$0.1286649803371\ri$ &  $0.1286649803372\ri$  \\
         
            &$2$ &  $0.0292128542235$  &$ 0.0292128542235$  &$1.797681422195\ri$& $1.7976814221949\ri$\\
            
            &$3$ & $0.0064020255292$  &  $0.0064020255291$ &$58.52102248036\ri$&$ 58.5210224799055\ri$\\
            
            \hline 
            \hline
\multicolumn{6}{c}{Maximal chamber: $u_1=0, u_2=1$}\\
\hline
$\ell$ & $n$ & $\Pi_{\gamma_1}^{(n)}$ & $Z_{\gamma_1}^{(n)}$ & $\lvert\Pi_{\gamma_2}^{(n)}\rvert$ & $\lvert Z_{\gamma_2}^{(n)}\rvert$ \\
\hline
\multirow{3}{*}{$-\frac{1}{2}$} &  $1$ &   $0.2118032711985$ &  $0.2118032711979$ &$0.1497675293419$ &  $0.1497675293415$  \\
         
           & $2$ &  $0.0265557570968$  &$0.0265557570967$  &$0.0187777559227$& $ 0.0187777559226$\\
            
            &$3$ & $-0.0323841134381 $  &  $-0.0323841134378$ &$0.0228990262148$&$0.0228990262146$\\
            \hline 
\multirow{3}{*}{$-\frac{1}{5}$} &  $1$ &   $0.0593049159356$ &  $0.0593049159354$ &$0.0419349082157$ &  $0.0419349082156$  \\
         
            &$2$ &  $-0.0193325911664$  &$-0.0193325911664$  &$0.0136702063117$& $0.0136702063117$\\
            
            &$3$ & $0.0166017656574$  &  $ 0.0166017656572$ &$0.0117392210760$&$ 0.0117392210759$\\

            \hline 
\end{tabular}
\caption{Expansions of the quantum periods for the $(A_1, D_3)$ theory. $\Pi_\gamma^{(n)}$ and $Z_\gamma^{(n)}$, defined in \eqref{eq:piexp} and \eqref{eq:tbaexp} respectively, denote the $n$-th WKB expansions of quantum periods and the large $\theta$ expansions of the pseudo-energies for BPS state $\gamma$. We compute these expansions up to the third order for different values of the parameter $\ell$. TBA equations are numerically solved using Fourier discretization with $2^{20}$ points and iterated $300$ times. The integration regions are chosen to be $[-50,50]$ for all examples. This convention is also applied in Tables \ref{tab:d4tbawkb} and \ref{tab:d5tbawkb}.}
\label{tab:d3tbawkb}
\end{table}

\setlength{\arrayrulewidth}{1.2pt} 
\begin{table}[htbp]
\centering
\begin{tabular}{c c c c c c}
\hline
\multicolumn{6}{c}{Intermediate chamber \RN{1}: $u_1=5+2\ri, u_2=8+7\ri, u_3=3+3\ri$}\\
\hline
$\ell$ & $n$ & \multicolumn{2}{c}{$\Pi_{\gamma_1}^{(n)}$} & \multicolumn{2}{c}{$Z_{\gamma_1}^{(n)}$} \\
\hline
\multirow{3}{*}{$-\frac{1}{2}$}
           & $1$ &   
\multicolumn{2}{c}{$-0.1595922521573 + 0.0036274096046\ri$} &\multicolumn{2}{c}{$-0.1595922521549 + 0.0036274096070\ri$}  \\
         
           & $2$ &  
\multicolumn{2}{c}{$0.0021730362142 - 0.0139242943357\ri$} &\multicolumn{2}{c}{$0.0021730362140 - 0.0139242943358\ri$}  \\

          & $3$ &  
\multicolumn{2}{c}{$0.0024942395470 + 0.0100571677303\ri$} &\multicolumn{2}{c}{$0.0024942395471 + 0.0100571677302\ri$}  \\
            \hline 
            \hline
\multicolumn{6}{c}{Intermediate chamber \RN{4}: $u_1=3+\ri/2, u_2=9/4+3\ri/8, u_3=9/16-\ri/8$}\\
\hline
& $n$ & \multicolumn{2}{c}{$\Pi_{\gamma_1}^{(n)}$} & \multicolumn{2}{c}{$Z_{\gamma_1}^{(n)}$} \\
\hline
\multirow{3}{*}{$-\frac{1}{2}$}
           & $1$ &   
\multicolumn{2}{c}{$-0.8469717532939 + 0.0393667667059\ri$} &\multicolumn{2}{c}{$-0.8469717532860 + 0.0393667667062\ri$}  \\
         
           & $2$ &  
\multicolumn{2}{c}{$0.2850546262003 + 0.0627058254705\ri$} &\multicolumn{2}{c}{$0.2850546261889 + 0.0627058254692\ri$}  \\

          & $3$ &  
\multicolumn{2}{c}{$3.9939327211427 - 1.7741540108092\ri$} &\multicolumn{2}{c}{$3.9939327211669 - 1.7741540107961\ri$}  \\
            \hline 
            \hline
\multicolumn{6}{c}{Maximal chamber: $u_1=u_2=0, u_3=1$}\\
\hline
$\ell$ & $n$ & $\Pi_{\gamma_1}^{(n)}$ & $Z_{\gamma_1}^{(n)}$ & $\lvert\Pi_{\gamma_2}^{(n)}\rvert$ & $\lvert Z_{\gamma_2}^{(n)}\rvert$ \\
\hline
\multirow{3}{*}{$-\frac{1}{2}$} &  $1$ &   $0.4311849265383$ &  $0.4311849265367$ &$0.2489447334074$ &  $0.2489447334065$  \\
         
           & $2$ &  $0$  &$0$  &$0$& $ 0$\\
            
            &$3$ & $0.0842463322541$  &  $0.0842463322531$ &$0.0486396426052$&$0.0486396426045$\\
            
            \hline 
\multirow{3}{*}{$-\frac{1}{5}$} &  $1$ &   $0.1983450662076$ &  $0.1983450662069$ &$0.1145145773674$ &  $0.1145145773670$  \\
         
            &$2$ &  $0$  &$0$  &$0$& $0$\\
            
            &$3$ & $0.0096685660970$  &  $ 0.0096685660970$ &$0.0055821492388$&$  0.0055821492388$\\
            
            \hline 
\end{tabular}
\caption{Expansions of the quantum periods for the $(A_1, D_4)$ theory up to the third order.}
\label{tab:d4tbawkb}
\end{table}

\setlength{\arrayrulewidth}{1.2pt} 
\begin{table}[htbp]
\centering
\begin{tabular}{c c c c c c}
\hline
\multicolumn{6}{c}{Maximal chamber: $u_1=u_2=u_3=0, u_4=1$}\\
\hline
$\ell$ & $n$ & $\Pi_{\gamma_1}^{(n)}$ & $Z_{\gamma_1}^{(n)}$ & $\Pi_{\gamma_2}^{(n)}$ & $Z_{\gamma_2}^{(n)}$ \\
\hline
\multirow{3}{*}{$-\frac{1}{2}$} &  $1$ &   $0.2598322449433$ &  $0.2598322449728$ &$0.6272905296839$ &  $0.6272905297552$  \\
         
           & $2$ &  $0.0179987960634$  &$0.0179987961361$  &$-0.007455345436$& $-0.0074553454660$\\
            
            &$3$ & $-0.0586777298768$  &  $-0.0586777298761$ &$0.0243051115243$&$0.0243051115239$\\
            
            \hline 
\end{tabular}
\caption{Expansions of the quantum periods for the $(A_1, D_5)$ theory up to the third order.}
\label{tab:d5tbawkb}
\end{table}

\section{GMN TBA and exact WKB}
\label{sc: TBA}

In this section, we briefly review the exact WKB method and the construction of GMN TBA equations. We also discuss the relation between the TBA for D-type AD theories and those for integrable models. The general theory for exact WKB analysis can be found in several references, particularly \cite{IN14, IN15}. The construction of GMN TBA is developed in the works \cite{GMN08, GMN09, GMN12, Gaiotto14, GGM19, GHN21}. 

\subsection{BPS spectrum and spectral network}

Let us consider the Seiberg-Witten curve for the $(A_1, G)$ theories:
\begin{equation}
\label{eq:sw}
 \Sigma=\{  x^2-\phi_2(z)=0\}.
\end{equation}
The SW differential $\lambda$ is determined by the quadratic differential:
\begin{equation}
\lambda^2-\phi_2(z) \mathrm{d} z^2=0.
\end{equation}
The SW curve for the Argyres-Douglas theory of the $(A_1, A_{N-1})$ class defines a Riemann surface with an irregular singularity, where $\phi_2(z)$ is a polynomial of degree $N$:
\begin{equation}
\lambda^2-\left(z^{N}+u_2 z^{N-2}+\cdots+u_N\right) \rd z^2=0.
\end{equation}
Quantizing this curve results in a second-order differential equation by promoting $x\to \hbar \partial_z$ in \eqref{eq:sw}:
\begin{equation}
\left(-\hbar^2 \frac{\rd^2}{\rd z^2}+\phi_2(z)\right) \psi(z)=0.
\label{eq:ode2-4}
\end{equation}
In this paper, we focus on the Argyres-Douglas theory of $(A_1, D_{N+2})$ type, whose SW curve defines a Riemann surface with one irregular singularity and one regular singularity, determining the SW differential as
\begin{equation}
\label{eq:dsw}
\lambda^2-\left(z^{N}+u_1 z^{N-1}+\cdots+u_N+\frac{u_{N+1}}{z}+\frac{m^2}{z^2}\right) \rd z^2=0.
\end{equation}
We constrain our consideration to the limit $m\to 0$ throughout this paper, where the residue of the regular singularity vanishes and the BPS spectrum is finite. Quantizing this curve leads to a Schr{\"odinger} equation:
\begin{equation}
\label{eq:dqsw}
\left(-\hbar^2 \frac{\rd^2}{\rd z^2}+\left(\phi_2(z) -\frac{\hbar^2}{4z^2}\right)\right) \psi(z)=0.
\end{equation}

There is an extra term $-\frac{\hbar^2}{4 z^2}$, well-known as the Langer modification in the WKB method, which appears in \cite{Gaiotto14} as $t_0(z)$. This term is also necessary for the consistency of the quantization scheme of the D-type quantum SW curve in the AD limit \cite{IKO19}. One can consider a deformation of this centrifugal term parametrized by $\ell$:
\begin{equation}
\label{eq:leq}
\left(-\hbar^2 \frac{\rd^2}{\rd z^2}+\left(\phi_2(z) +\frac{\ell(\ell+1)\hbar^2}{z^2}\right)\right) \psi(z)=0.
\end{equation}
$\ell=-\frac{1}{2}$ corresponds to the original quantum SW curve \eqref{eq:dqsw}. Since the BPS spectrum is entirely determined by the Seiberg-Witten curve, or $\phi_2(z)$, this deformation \eqref{eq:leq} does not alter the BPS states. We associate an electromagnetic charge lattice as a one-dimensional homology space on the SW curve $\Sigma_{\boldsymbol{u}}$ with fixed moduli parameters $\boldsymbol{u}=\{u_1,\dots, u_{N+1}\}$:
\begin{equation}
    \Gamma=H_1\left(\Sigma_{\boldsymbol{u}}, \mathbb{Z}\right).
\end{equation}
We define the one-cycles $\alpha_i$ and $\beta_i$ with $i$ from 1 to $g$, the genus of the SW curve \eqref{eq:sw}. These form a canonical basis of the homology class for the charge lattice. The contour integrals along these one-cycles are defined as
\begin{equation}
    a^i=\oint_{\alpha_i} \lambda, \quad a_{D, i}=\oint_{\beta_i} \lambda, \quad i=1,2,\cdots, g.
\end{equation}
The non-trivial intersection numbers for these canonical cycles are given by  
$\left\langle\alpha_i, \beta_j\right\rangle=\delta_{ij}$. A BPS state corresponds to a vector in this charge lattice $ \gamma\in \Gamma$, which can be represented by a linear combination of the canonical basis $\alpha_i$ and $\beta_i$: 
\begin{equation}
    \gamma=\sum_{i=1}^{g}n_i \alpha_i+m_i \beta_i.
\end{equation}
The central charge of the BPS state $\gamma$ is given by 
\begin{equation}
    Z_{\gamma}(\boldsymbol{u})=\oint_\gamma\lambda=\sum_{i=1}^{g}n_i a^i+m_i a_{D,i}.
\end{equation}
The BPS spectrum consists of all BPS states, which depends on the moduli parameters $\boldsymbol{u}$ of the theory. Within the moduli space, the spectrum remains constant in specific regions, known as chambers. These chambers are separated by walls of marginal stability, across which the BPS spectrum can change discontinuously—a phenomenon known as wall-crossing. This can be described by a BPS index $\Omega(\gamma, \boldsymbol{u})$, which accounts for the degeneracy of a BPS state (or the number of saddle connections discussed later) corresponding to the charge $\gamma$. This index remains invariant within a given chamber but changes discontinuously when crossing into another chamber. BPS states can be identified using the Stokes graph or the spectral network\footnote{The term Stokes graph is used in WKB analysis and describes the Stokes phenomenon of the Borel-resummed WKB solutions of differential equations. The spectral network, introduced in \cite{GMN09, GMN12}, refers to the same concept. These terms are used interchangeably throughout this paper.}. In particular, the complete BPS spectrum for Argyres-Douglas theories of type A and D was established in \cite{MPY13}, and our analysis primarily relies on their classification.

A Stokes line is a trajectory starting at a turning point $z_0$ in the $z$ plane along which the 1-form $\lambda/\hbar$ is real, that is,
\begin{equation}
    {\rm Im} \frac{1}{\hbar}\int_{z_0}^{z}\lambda=0.
\end{equation}
We define the turning points as the set of zeros of $\phi_{2}(z)$ and the origin point 0, which are also the branch points of the SW differential; we will use both terms interchangeably in this paper. The orientation of a Stokes line can be specified by choosing a particular sheet of the Riemann surface. Changing to another sheet implies changing $\lambda \to -\lambda$, a transformation that does not affect our discussion, as will be elaborated on later. Hence, the orientation of the Stokes line will not be a primary concern in this paper. The Stokes lines emanate from a turning point to infinity or end at another turning point\footnote{A Stokes line can terminate at its originating turning point, although this does not occur in this paper.}. Along a Stokes line, the WKB solutions either increase or decrease depending on their orientation. The collection of all the Stokes lines starting from each turning point constitutes the Stokes graph $\mathcal{W}_\vartheta$, whose configuration depends on $\vartheta=\arg \hbar$. The Stokes graphs for $\vartheta$ and $\vartheta+\pi$ are identical except for their orientations, which result in the exchange of BPS states and their anti-states. The ambiguity in sign selection is counterbalanced by the signs arising from the choice of the Riemann sheet and the orientations of the integral contour. Thus, this does not impact the final TBA equations. Therefore, it is sufficient to consider the Stokes graphs for $\vartheta \in [0, \pi)$. A notable feature is the saddle connection, where a Stokes line begins at one turning point and terminates at another, with two Stokes lines merging to form a one-cycle on the punctured Riemann surface. This signifies the presence of a BPS state at phase $\vartheta$. As $\vartheta$ varies from 0 to $\pi$, like a radar scan, all BPS states are revealed. For further details, see \cite{GMN09, GMN12, IK20, IK21}.

\begin{figure}[htbp]
\centering
    \begin{subfigure}{0.3\linewidth}
        \centering
        \includegraphics[width=0.9\linewidth]{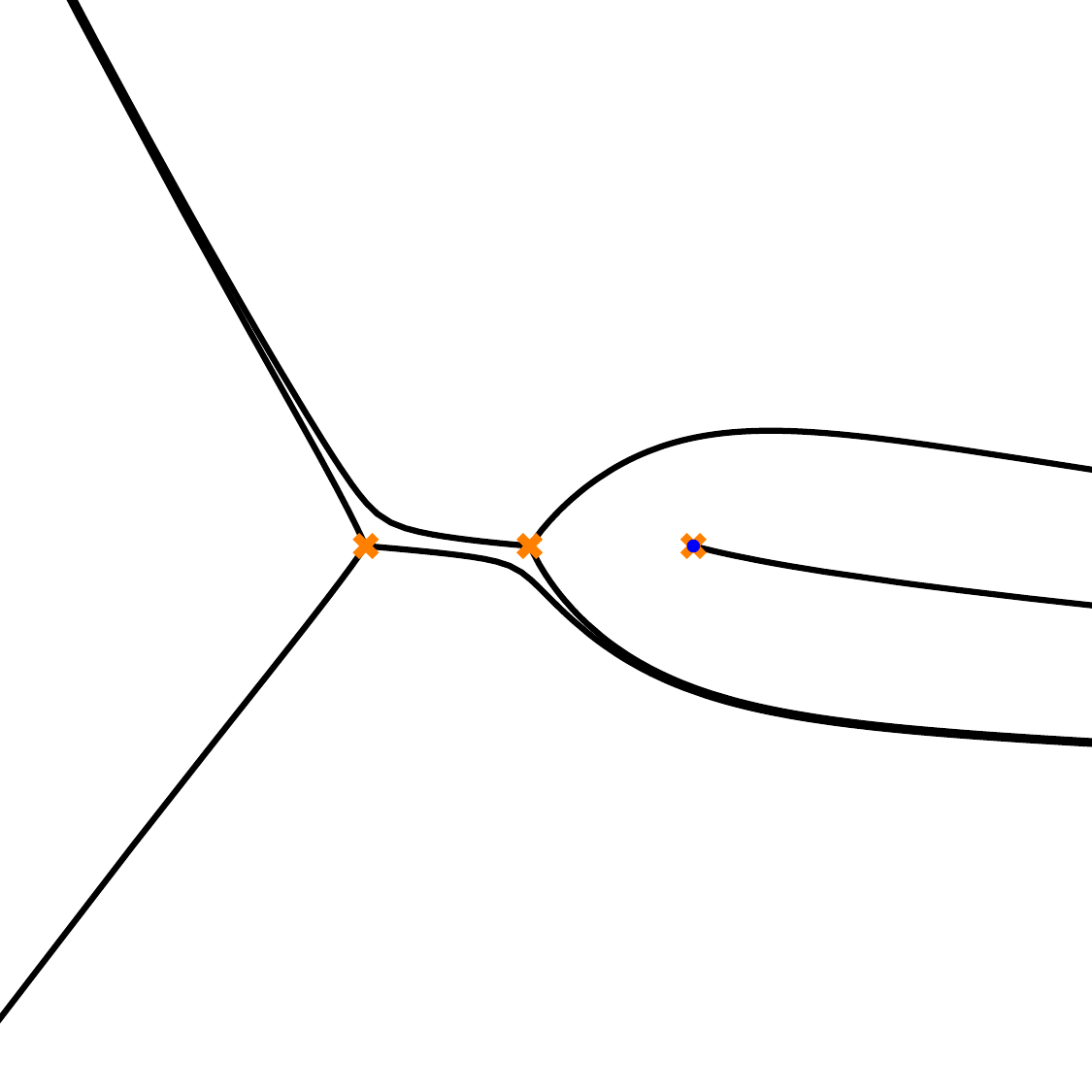}
        \caption{$\vartheta=\arg Z_{\gamma_1}-\delta$}
        \label{fig:d3minbps1}
    \end{subfigure}
    \begin{subfigure}{0.3\linewidth}
        \centering
        \includegraphics[width=0.9\linewidth]{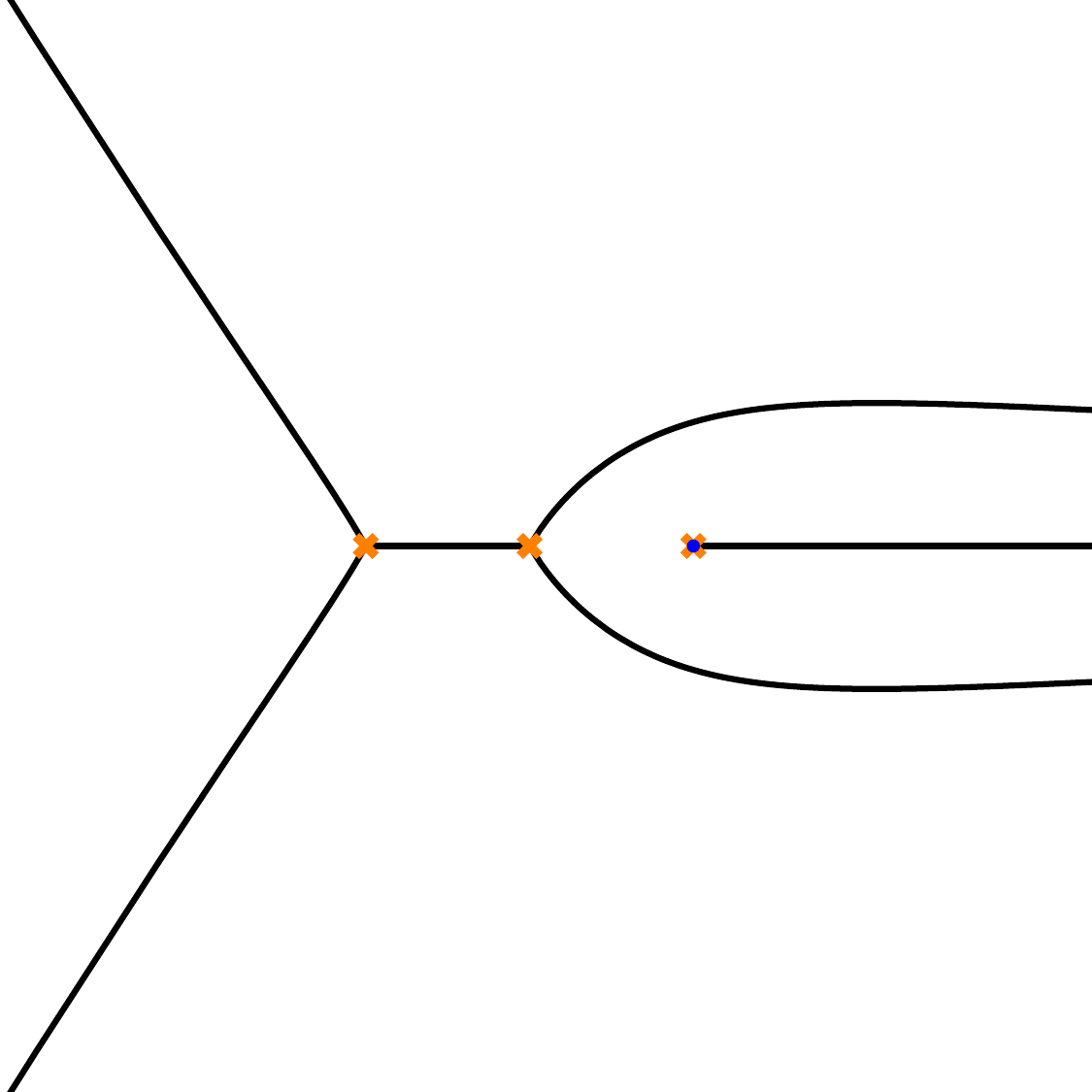}
        \caption{$\vartheta=\arg Z_{\gamma_1}$}
        \label{fig:d3minbps2}
    \end{subfigure}
     \begin{subfigure}{0.3\linewidth}
        \centering
        \includegraphics[width=0.9\linewidth]{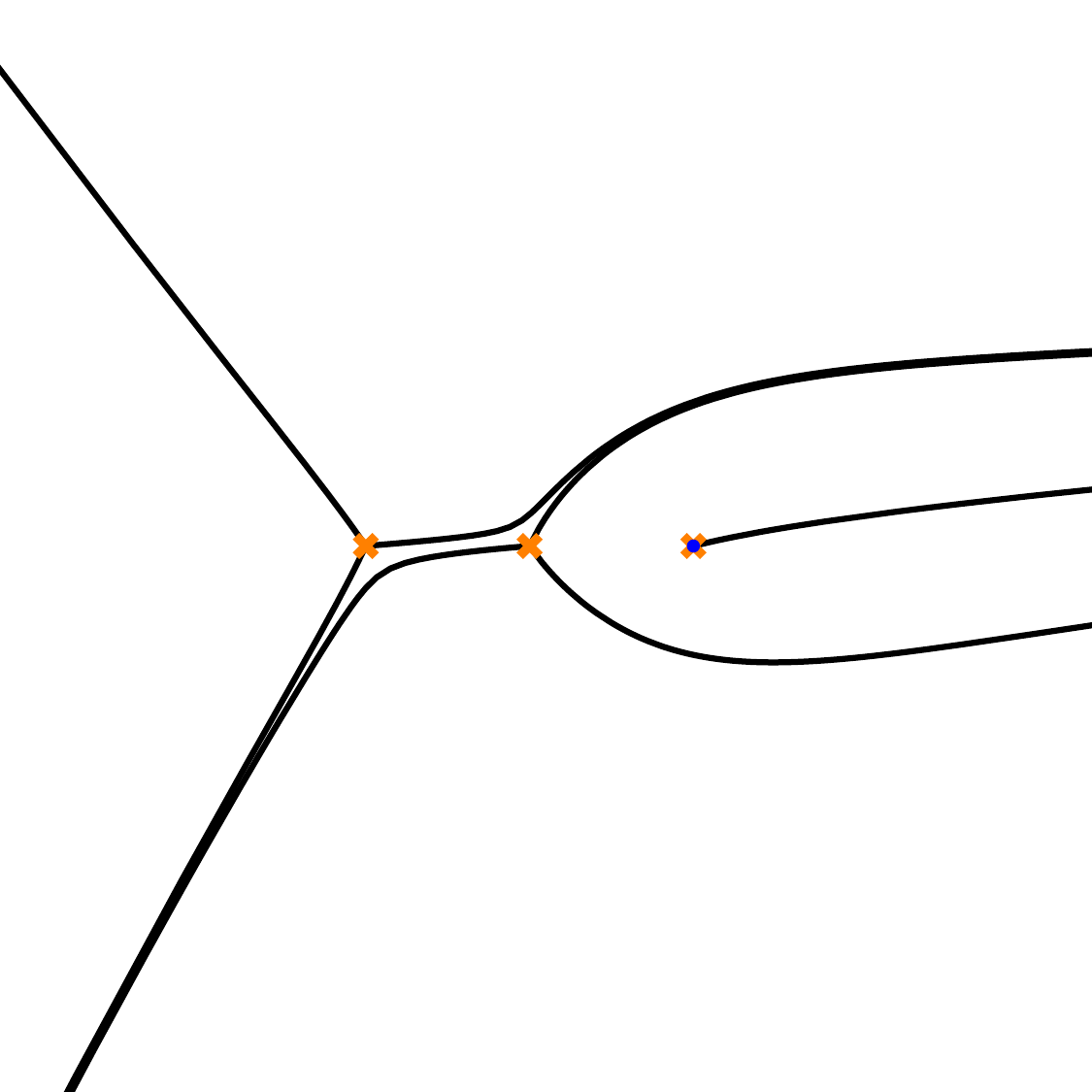}
        \caption{$\vartheta=\arg Z_{\gamma_1}+\delta$}
        \label{fig:d3minbps3}
    \end{subfigure}
    
    \vspace{1em} 

     \begin{subfigure}{0.3\linewidth}
        \centering
        \includegraphics[width=0.9\linewidth]{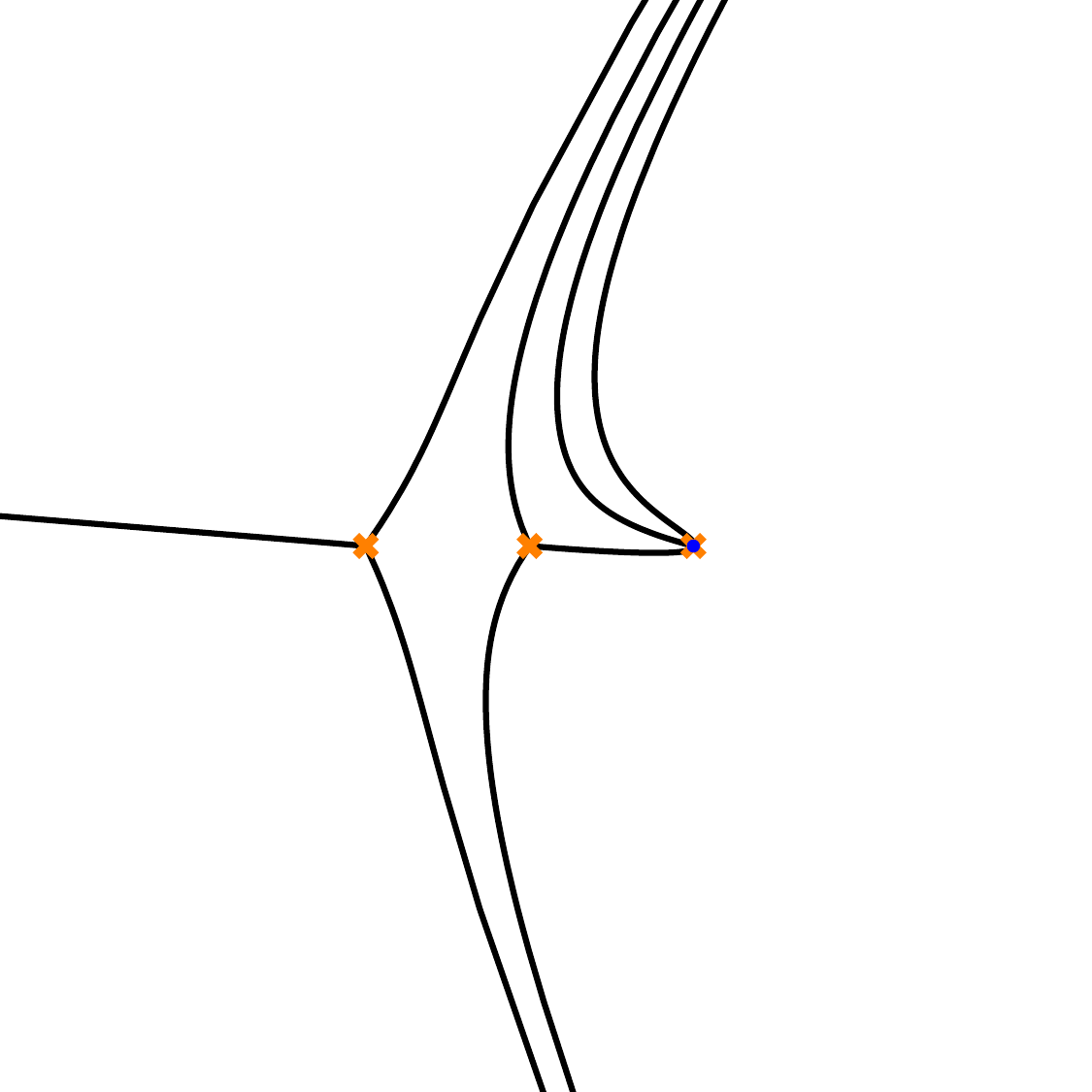}
        \caption{$\vartheta=\arg Z_{\gamma_2}-\delta$}
        \label{fig:d3minbps4}
    \end{subfigure}
    \begin{subfigure}{0.3\linewidth}
        \centering
        \includegraphics[width=0.9\linewidth]{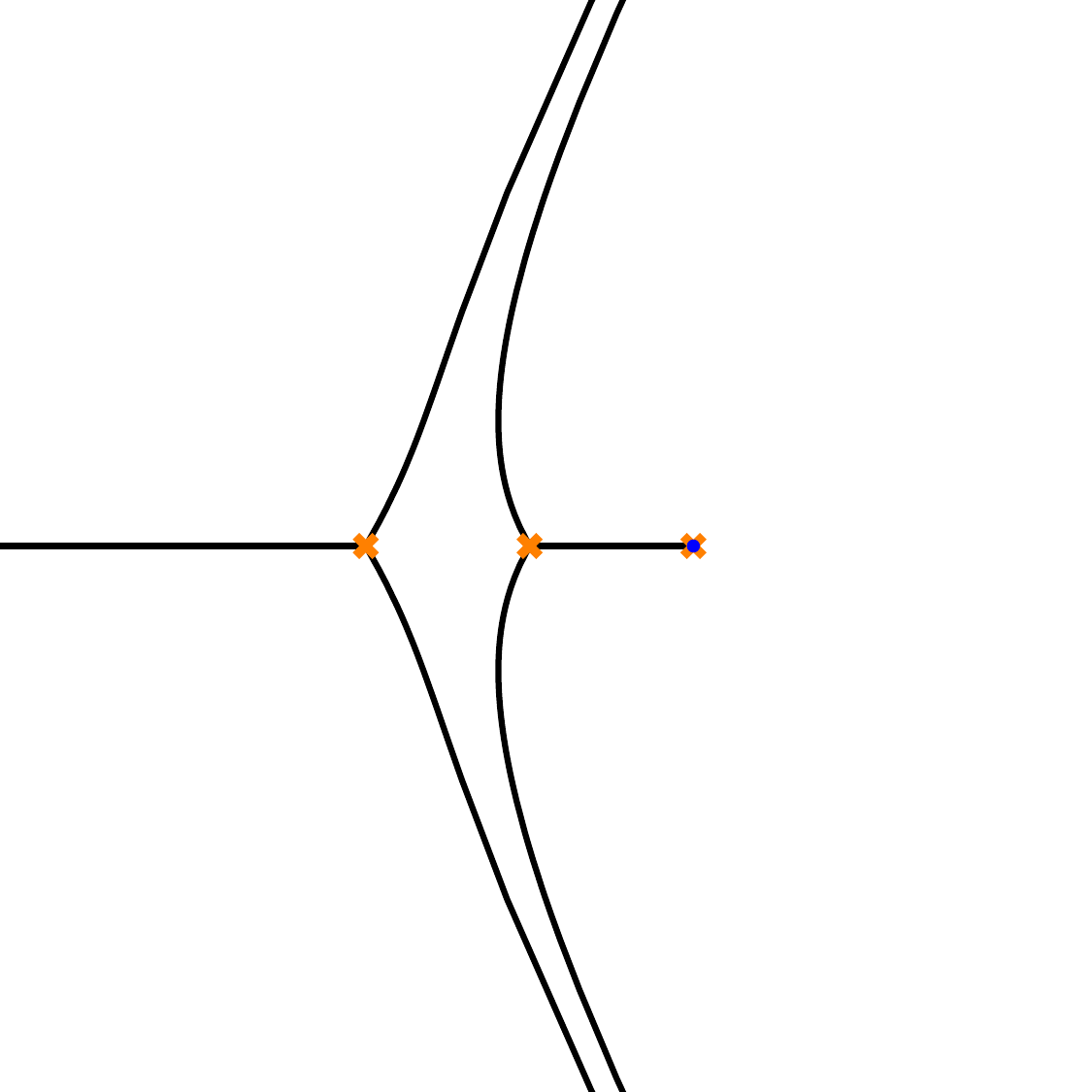}
        \caption{$\vartheta=\arg Z_{\gamma_2}$}
        \label{fig:d3minbps5}
    \end{subfigure}
     \begin{subfigure}{0.3\linewidth}
        \centering
        \includegraphics[width=0.9\linewidth]{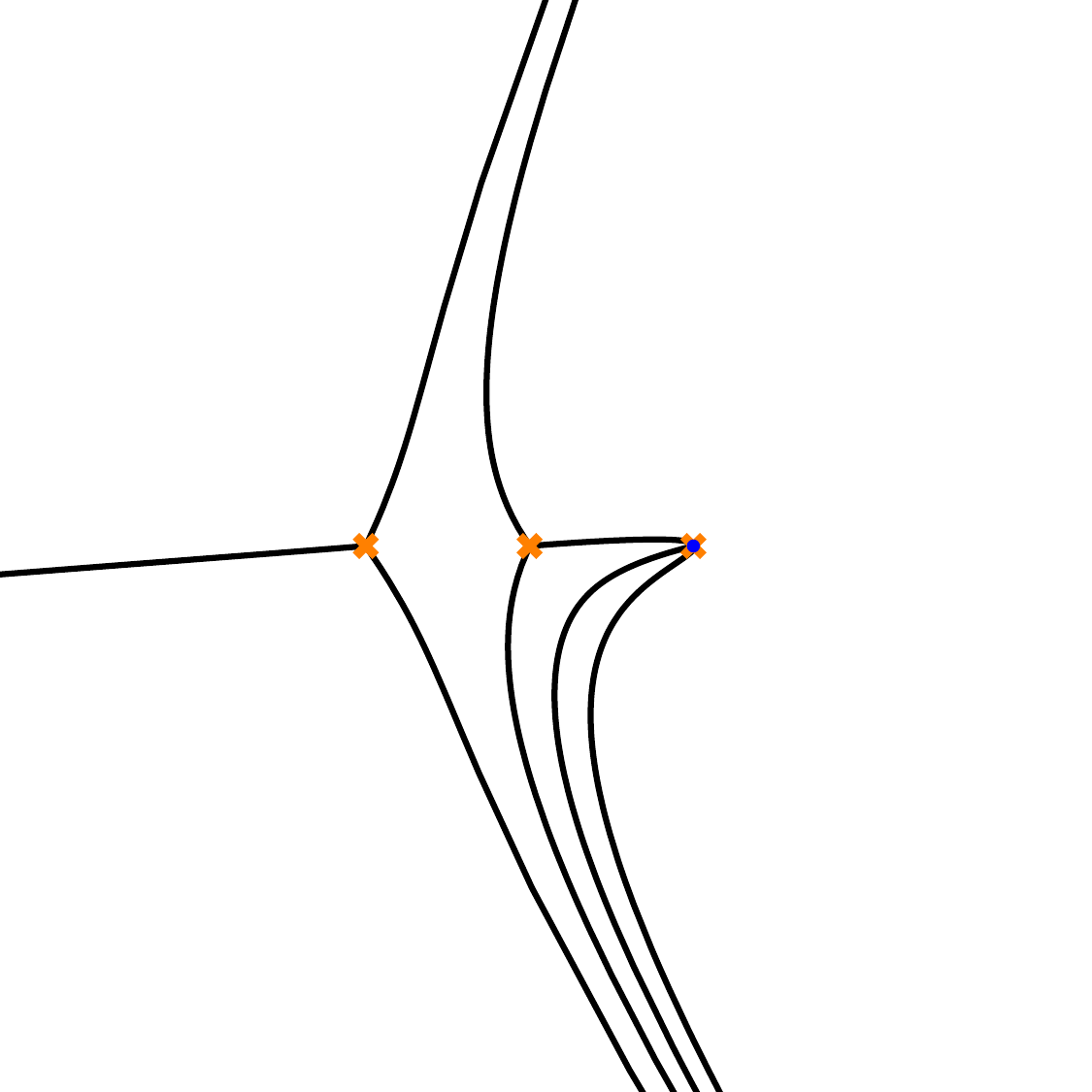}
        \caption{$\vartheta=\arg Z_{\gamma_2}+\delta$}
        \label{fig:d3minbps6}
    \end{subfigure}

      \vspace{1em} 
    
    \begin{subfigure}{0.4\linewidth}
        \centering
       \includegraphics[width=0.9\linewidth]{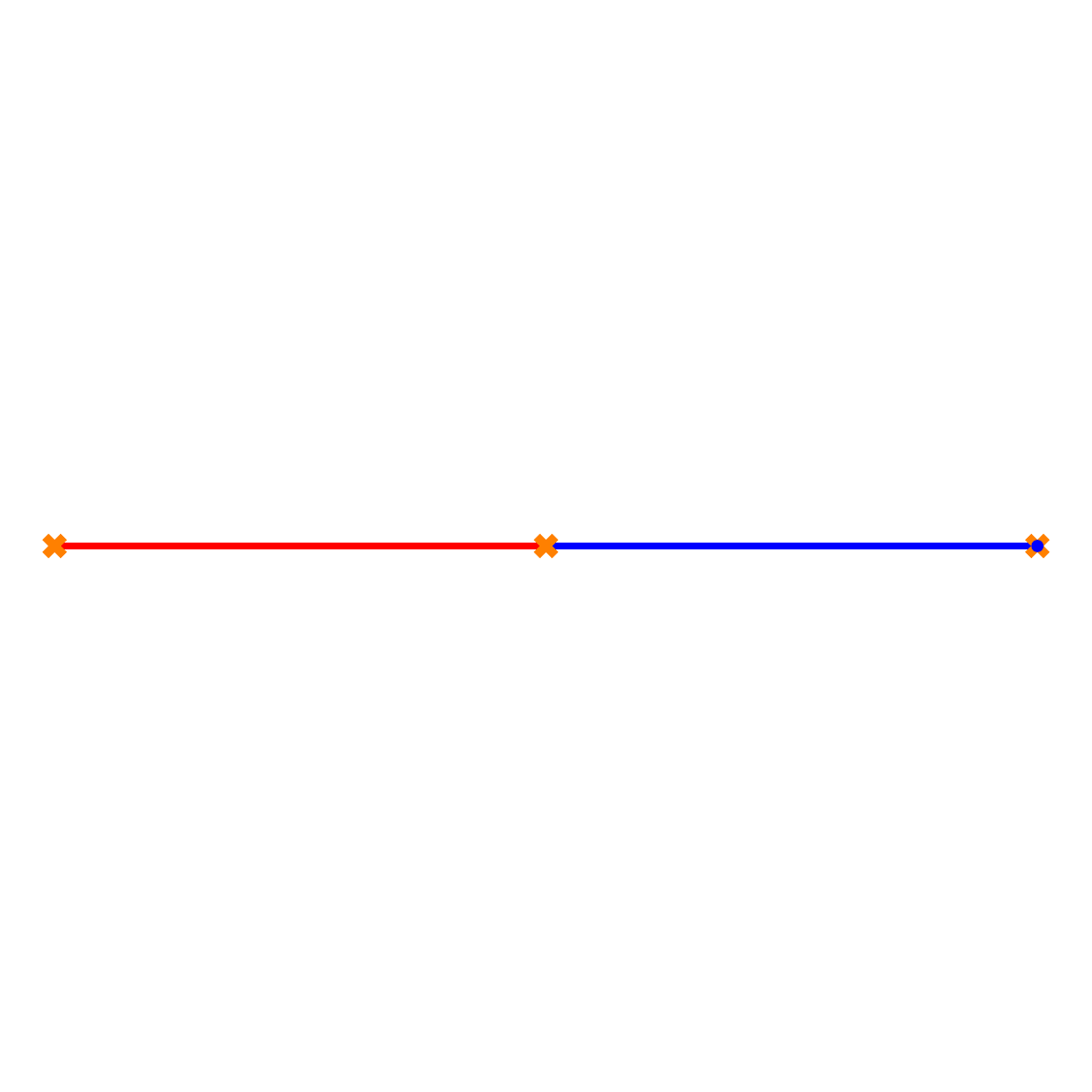}
        \caption{BPS states}
        \label{fig:d3minbpsplot}
    \end{subfigure}
    \begin{subfigure}{0.4\linewidth}
        \centering
       \includegraphics[width=0.9\linewidth]{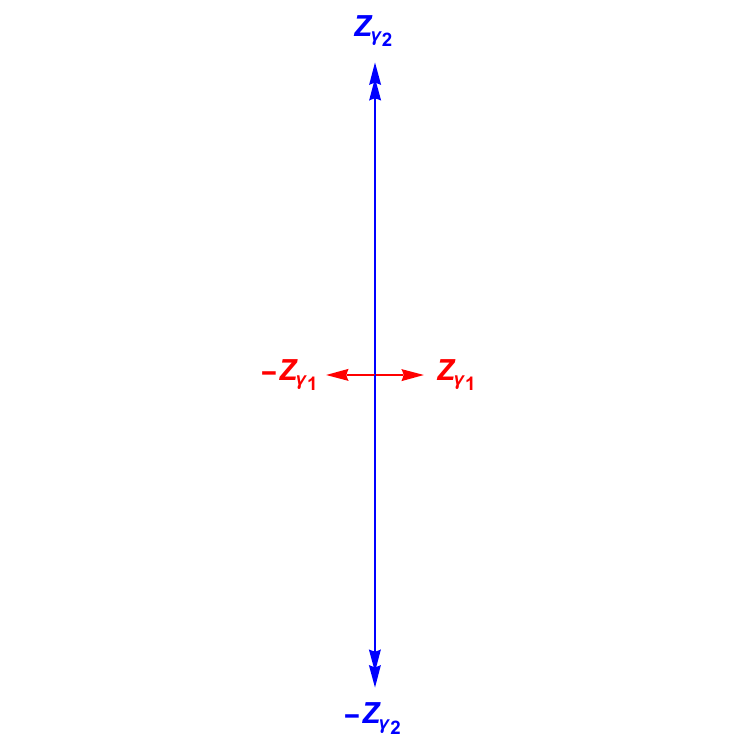}
        \caption{Central charges}
        \label{fig:d3minbpscharge}
    \end{subfigure}
    \caption{BPS spectrum in the minimal chamber for the $(A_1,D_3)$ theory with SU(2) flavor symmetry. The orange crosses represent turning points, and the blue dot denotes the origin, which is also a turning point. The double arrow indicates a doublet. }
    \label{fig:d3minbps}
\end{figure}

Figure \ref{fig:d3minbps} illustrates the determination of the BPS states of the $(A_1, D_3)$ theory within the minimal chamber at the limit $m\to 0$, using Stokes graphs. It features three turning points including $0$, with two BPS states $\gamma_1$ and $\gamma_2$, corresponding to the turning points shown in Figure \ref{fig:d3minbpsplot}. Each BPS state is associated with a one-cycle on the punctured Riemann surface; here, we depict only its trivialization onto the $z$ plane. The choice of branch cut and orientation for one-cycle is made to align the central charge as indicated in Figure \ref{fig:d3minbpscharge}. Figure \ref{fig:d3minbps1} shows the Stokes graph for $\vartheta$ slightly smaller than the central charge of $\gamma_1$; and all Stokes lines initially extend to infinity. As $\vartheta$ increases to $\vartheta_{\mathrm{c}} = \arg Z_{\gamma_1}$, two Stokes lines originating from the left and middle turning points converge and become finite. This saddle connection indicates the presence of a BPS state $\gamma_1$ at $\vartheta_{\mathrm{c}}$. The configuration of the Stokes graph changes discontinuously as $\vartheta$ varies around $\vartheta_{\mathrm{c}}$, which refers to the mutation of Stokes graphs. Similarly, another BPS state, $\gamma_2$, appears at $\vartheta_{\mathrm{c}} = \arg Z_{\gamma_2}$, as shown in Figure \ref{fig:d3minbps5}. This saddle connection involves a turning point with a pole $0$ (it is also a turning point). The mutation of the Stokes graph around $\arg Z_{\gamma_2}$ differs from that around $\arg Z_{\gamma_1}$. Specifically, it corresponds to a BPS state with a degeneracy of 2, referred to as a doublet in \cite{MPY13}, with the corresponding BPS index $\Omega(\gamma_2) = 2$. In contrast, $\gamma_1$ is a singlet with $\Omega(\gamma_1) = 1$. These constitute all the BPS states for $(A_1, D_3)$ in the limit $m \to 0$. This procedure can be straightforwardly extended to generic AD theories of type D, which will be discussed later. The property of Stokes graphs near $m=0$ is thoroughly detailed in Section 3.2.9 of \cite{GMN09}.

\subsection{GMN TBA}

In this part, we follow the derivations in \cite{GGM19, GHN21} to present the GMN TBA equations and fix our notation. We associate a ray 
\begin{equation}
\ell_\gamma=\left\{\zeta: \frac{Z_\gamma(\boldsymbol{u})}{\zeta} \in \mathbb{R}_{-}\right\}
\end{equation}
to the central charge $Z_\gamma$.
The spectral coordinates $\mathcal{X}_\gamma$ satisfy a non-linear, TBA-like integral equation as 
\begin{equation}
\label{eq:gmntba}
\mathcal{X}_\gamma(\zeta)=\exp \left(\frac{Z_\gamma}{\zeta}+\frac{1}{4 \pi \mathrm{i}} \sum_{\gamma^{\prime} \in \Gamma} \Omega\left(\gamma^{\prime}, \boldsymbol{u}\right)\left\langle\gamma, \gamma^{\prime}\right\rangle \mathcal{I}_{\gamma^{\prime}}(\zeta)\right),
\end{equation}
where 
\begin{equation}
\mathcal{I}_{\gamma}=\int_{\ell_{\gamma}} \frac{\mathrm{d} \zeta^{\prime}}{\zeta^{\prime}} \frac{\zeta^{\prime}+\zeta}{\zeta^{\prime}-\zeta} \log \left(1-\sigma(\gamma) \mathcal{X}_\gamma\left(\zeta^{\prime}\right)\right).
\end{equation}
The Dirac pairing $\left\langle\gamma, \gamma^{\prime}\right\rangle $ for two BPS states $\gamma$ and $\gamma^\prime$ is defined by the intersection number. This TBA-like integral equation can be viewed as the conformal limit of the integral equation in \cite{GMN09}. The BPS indices $\Omega\left(\gamma,\boldsymbol{u}\right)$ obey the following the charge conjugation symmetry:
\begin{equation}
\Omega\left(\gamma\right)= \Omega\left(-\gamma\right).
\end{equation}
In other words, they take the same value for a BPS state and its anti-state. The equations and their solutions exhibit an additional $\mathbb{Z}_2$ symmetry
\begin{equation}
    \mathcal{X}_\gamma(\zeta)=\mathcal{X}_{-\gamma}(-\zeta),
\end{equation}
which allows us to sum contributions from a BPS state $\gamma$ and its anti-state $-\gamma$ for \eqref{eq:gmntba}:
\begin{equation}
\mathcal{C}_{\gamma}:=\mathcal{I}_{\gamma}-\mathcal{I}_{-\gamma}=4\zeta\int_{\ell_\gamma}\frac{\mathrm{d} \zeta^{\prime}}{\left(\zeta^{\prime}\right)^2-\zeta^2} \log \left(1-\sigma(\gamma) \mathcal{X}_\gamma\left(\zeta^{\prime}\right)\right).
\end{equation}
Denoting central charge as 
\begin{equation}
Z_\gamma=\left|Z_\gamma\right| \mathrm{e}^{\mathrm{i} \phi_\gamma},
\end{equation}
and changing variables
\begin{equation}
\zeta=-\mathrm{e}^{\mathrm{i}\phi-\theta},\quad \zeta^\prime=-\mathrm{e}^{\mathrm{i} \phi^\prime-\theta^\prime},
\end{equation}
$\mathcal{C}_\gamma$ can be written as
\begin{equation}
    \mathcal{C}_{\gamma}= 2\int_{\mathbb{R}} \frac{\log \left(1-\sigma(\gamma) \mathcal{X}_{\gamma}\left(\theta^\prime\right)\right)}{\sinh \left(\theta-\theta^{\prime}+\mathrm{i} \phi_{\gamma}-\mathrm{i} \phi\right)} \mathrm{d} \theta^{\prime}.
\end{equation}
We define another function $\tilde{\epsilon}(\theta)$ related to the spectral coordinate as
\begin{equation}
    \exp\left(-\tilde{\epsilon}_\gamma(\theta)\right)=\mathcal{X}_{\gamma}\left(-\mathrm{e}^{\mathrm{i} \phi_\gamma-\theta}\right).
\end{equation}
$\tilde{\epsilon}_\gamma(\theta)$ represents pseudo-energy of a particle in TBA equations. The quadratic refinement $\sigma(\gamma)$ is fixed to be $-1$, the TBA equations can be written down uniformly 
\begin{equation}
\label{eq:gmntba2}
\tilde{\epsilon}_\gamma(\theta)=\left|Z_\gamma\right| \mathrm{e}^\theta-\frac{1}{2 \pi \mathrm{i}} \sum_{\gamma^{\prime}>0}\left\langle\gamma, \gamma^{\prime}\right\rangle \Omega\left(\gamma^{\prime}, u\right) \int_{\mathbb{R}} \frac{\log \left(1+\mathrm{e}^{-\tilde{\epsilon}_{\gamma^{\prime}}\left(\theta^{\prime}\right)}\right)}{\sinh \left(\theta-\theta^{\prime}-\mathrm{i} \phi_{\gamma,\gamma^\prime}\right)} \mathrm{d} \theta^{\prime},
\end{equation}
where we introduce a short notation
\begin{equation}
\label{eq:phase}
    \phi_{\gamma,\gamma^\prime}=\phi_\gamma-\phi_{\gamma^\prime}.
\end{equation}
For simplicity, these TBA equations are typically written as
\begin{equation}
\label{eq:kernel}
\begin{aligned}
     &\tilde{\epsilon}_{\gamma}(\theta)=\left|Z_{\gamma}\right| \mathrm{e}^\theta- \sum_{\gamma^{\prime}>0}\left\langle\gamma, \gamma^{\prime}\right\rangle \Omega\left(\gamma^{\prime}, u\right)K_{\gamma,\gamma^\prime}\star \tilde{L}_{\gamma^\prime},\\
       &K_{\gamma,\gamma^\prime}=\frac{1}{2\pi \ri}\frac{1}{\sinh\left(\theta-\ri \phi_{\gamma, \gamma^\prime}\right)},\quad 
     \tilde{L}_{\gamma^\prime}=\log \left(1+\mathrm{e}^{-\tilde{\epsilon}_{\gamma^\prime}\left(\theta\right)}\right),
    \end{aligned}
\end{equation}
where $\star$ represents the convolution: $(f\star g)(\theta)=\int_{\mathbb R}f(\theta-\theta')g(\theta')\rd\theta'$.
One defines $\epsilon_\gamma\left(\theta\right)=\tilde{\epsilon}_\gamma\left(\theta+\ri \phi_\gamma\right)$, which is related to the quantum periods in Section \ref{sc:wkb}. It can be evaluated by direct integration once one knows the solutions of TBA equations:
\begin{equation}
\label{eq:epsilonfun}
\epsilon_\gamma\left(\theta\right)=Z_\gamma\re^\theta-\frac{1}{2 \pi \mathrm{i}} \sum_{\gamma^{\prime}>0}\left\langle\gamma, \gamma^{\prime}\right\rangle \Omega\left(\gamma^{\prime}, u\right) \int_{\mathbb{R}} \frac{\log \left(1+\mathrm{e}^{-\tilde{\epsilon}_{\gamma^{\prime}}\left(\theta^{\prime}\right)}\right)}{\sinh \left(\theta-\theta^{\prime}+\mathrm{i} \phi_{\gamma^{\prime}}\right)} \mathrm{d} \theta^{\prime}.
\end{equation}
The integration may encounter singularities when $\phi_{\gamma^\prime}=0$ or $\pi$,
 but one can avoid them using the Cauchy principal value. The following expansion of the shifted kernel:
\begin{equation}
        \frac{1}{2\pi \ri \sinh(\theta-\theta^\prime+\ri \phi_{\gamma^{\prime}})}
        =\frac{1}{\pi \ri} \sum_{n=1}^{\infty}\re^{(1-2n)\theta+(2n-1)\theta^\prime+(1-2n)\ri\phi_{\gamma^{\prime}}},
\end{equation}
yields the large $\theta$ expansion of $\epsilon(\theta)$ \eqref{eq:epsilonfun}:
\begin{equation}
\label{eq:tbaexp}
    \begin{aligned}
       & \epsilon_{\gamma}(\theta)\sim Z_{\gamma} \re^{\theta}+ \sum_{n=1}^{\infty} Z_{\gamma}^{(n)} \re^{(1-2n)\theta}, \\
       & Z_{\gamma}^{(n)}=-\frac{1}{\pi \ri}\sum_{\gamma^{\prime}>0}\left\langle\gamma, \gamma^{\prime}\right\rangle \Omega\left(\gamma^{\prime}, u\right)  \re^{\ri(1-2n)\phi_{\gamma^{\prime}}} \int_{\mathbb{R}}\re^{(2n-1)\theta^\prime}\tilde{L}_{\gamma^{\prime}}(\theta^\prime)\rd \theta^\prime.
    \end{aligned}
\end{equation}

\subsection{All-orders WKB and Borel resummation}
\label{sc:wkb}

In this part, we review the all-orders WKB expansions and Borel resummation, demonstrating their relationship to the TBA equations. We start with the Schr{\"o}dinger equation including a centrifugal correction:
\begin{equation}
\label{eq:seq}
\left(-\hbar^{2} \frac{\rd^{2}}{\rd z^{2}}+ \phi_2(z)+Q_2(z)\hbar^2 \right) \psi(z)=0, \quad Q_2(z)=\frac{\ell(\ell+1)}{z^2},
\end{equation}
where $\phi_2(z)$ is determined by \eqref{eq:dsw}. We set a WKB ansatz for the solution of \eqref{eq:seq}, determined by a formal power series $P(z,\hbar)$ in $\hbar$:
\begin{equation}
\label{eq:ansatz}
    \psi(z)=\exp\left( \frac{1}{\hbar}\int^{z}P(z,\hbar)\rd z\right), \quad   P(z,\hbar)=\sum_{n=0}^{\infty} p_n(z)\hbar^n.
\end{equation}
Substituting the wave ansatz into \eqref{eq:seq}, we obtain the 
Riccati equation for $P(z)$:
\begin{equation}
\label{eq:riccati}
    \hbar P^{\prime}(z)=\phi_2(z)-P^2(z)+\hbar^2Q_2(z).
\end{equation}
The expansion coefficients $p_n(z)$ are determined by a recursive relation:
\begin{equation}
    p_n(z)=-\frac{1}{2p_0(z)}\left(\sum_{i=1}^{n-1}p_{n-i}(z)p_i(z)+p_{n-1}^{\prime}(z)\right), \quad n\geq3,
\end{equation}
with the initial terms given by
$$
p_0^2(z)=\phi_2(z),\quad
p_1(z)=-\frac{p_0^{\prime}(z)}{2p_0(z)}, \quad p_2(z)=-\frac{1}{2p_0(z)}\left(p_1^{2}(z)+p_1^{\prime}(z)-Q_2(z)\right).
$$
One can split the odd and even power of $P(z,\hbar)$:
\begin{equation}
     P_{\mathrm{even}}(z,\hbar)=\sum_{n=0}^{\infty}p_{2n}(z)\hbar^{2n},\quad
      P_{\mathrm{odd}}(z,\hbar)=\sum_{n=0}^{\infty}p_{2n+1}(z)\hbar^{2n+1},
\end{equation}
which are related by
\begin{equation}
    P_{\mathrm{odd}}(z,\hbar)=-\frac{\hbar}{2}\frac{\rd}{\rd z}\log P_{\mathrm{even}}(z,\hbar).
\end{equation}
We define the quantum WKB periods as
\begin{equation}
\label{eq:qp}
    \Pi_{\gamma}(\hbar)=\oint_{\gamma}P_{\mathrm{even}}(z)\rd z=\sum_{n=0}^{\infty}\Pi_{\gamma}^{(n)} \hbar^{2 n},
\end{equation}
where 
\begin{equation}
\label{eq:piexp}
    \Pi_{\gamma}^{(n)}=\oint_\gamma p_{2n}(z)\rd z.
\end{equation}
The WKB periods defined above exhibit a divergent double-factorial growth:
\begin{equation}
    \Pi_\gamma^{(n)}\sim(2n)!.
\end{equation}
Borel resummation is used to resum the divergent series into a well-defined function.
We define the Borel transform of the quantum WKB periods as
\begin{equation}
\widehat{\Pi}_\gamma(\xi)=\sum_{n=0}^{\infty} \frac{1}{(2 n)!} \Pi_\gamma^{(n)} \xi^{2 n}.
\end{equation}
This series is convergent within a certain domain near the origin and can be analytically continued into the complex $\xi$-plane, known as the Borel plane. If there are no singularities along the ray $\varphi$, the Borel resummation of the WKB periods in this direction is defined by the Laplace transform of its Borel transform
\begin{equation}
s_\varphi\left(\Pi_\gamma\right)(\hbar)=\frac{1}{\lvert\hbar\rvert}\int_{0}^{\infty}\re^{-\xi/\lvert\hbar\rvert}\widehat{\Pi}_\gamma\left(\re^{\ri\varphi}\xi\right)\rd \xi, \quad \varphi=\arg \hbar.
\end{equation}
In this case, the WKB period is Borel-summable along the direction of $\varphi$. However, the integral is not well-defined if there are singularities along the ray $\varphi$. To bypass these singularities, one can rotate the integration line by a small angle $\delta$ and define two lateral resummations as
\begin{equation}
s_{\varphi \pm}\left(\Pi_\gamma\right)\left( \hbar\right)=\lim _{\delta \rightarrow 0} s_{\varphi \pm\delta}\left(\Pi_\gamma\right)\left( \hbar\right).
\end{equation}
Typically, the subscript $\varphi$ is omitted when it is $0$. In practice, we use median summation to represent the Borel resummation of quantum WKB periods when they are non-summable. It can be written as 
\begin{equation}
s_{\mathrm{med}}\left(\Pi_\gamma\right)\left( \hbar\right)=\frac{1}{2}\left(s_{\varphi +}\left(\Pi_\gamma\right)\left( \hbar\right)+s_{\varphi -}\left(\Pi_\gamma\right)\left( \hbar\right)\right),
\end{equation}
its generic definition involves the alien derivatives and the Stokes automorphism \cite{Dorigoni14}. We refer to the formal series in \eqref{eq:qp} as quantum WKB periods, or simply WKB periods, while the term quantum periods emphasize their (median) Borel resummation. If the Borel transform displays singularities along a ray $\arg \hbar=\vartheta_{\mathrm{c}}$, the two lateral resummations will generally differ. This difference indicates a jump in the resummed quantum period as $\vartheta$ crosses $\vartheta_{\mathrm{c}}$. The discontinuity is denoted as 
\begin{equation}
\operatorname{disc}_{\vartheta_{\mathrm{c}}}\left(\Pi_\gamma\right)=s_{\vartheta_{\mathrm{c}}+}\left(\Pi_\gamma\right)-s_{\vartheta_{\mathrm{c}}-}\left(\Pi_\gamma\right),
\end{equation}
and it takes the form
\begin{equation}
\operatorname{disc}_{\vartheta_{\mathrm{c}}}\left(\Pi_\gamma\right)=\sum_{\gamma^{\prime}>0}\left\langle\gamma, \gamma^{\prime}\right\rangle \Omega\left(\gamma^{\prime}, u\right) \log \left(1+\mathcal{X}_{\gamma^\prime}\right).
\end{equation}

This formula is derived in \cite{DP99} for $\Omega(\gamma)=1$ based on exact WKB analysis and dubbed as Delabaere-Pham (DP) formula. It has since been generalized and is recognized as the Kontsevich-Soibelman (KS) wall-crossing formula \cite{KS08, GMN09}. The discontinuity formula, combined with the asymptotic behavior of the spectral coordinate $\mathcal{X}$, forms a Riemann-Hilbert problem, whose solutions are governed by the TBA equations. This is how the TBA equations discussed in the last part are established. 

Quantum periods defined from the all-orders WKB are expected to be related to the solutions of TBA equations:
\begin{equation}
\label{eq:wkbtbarelation}
\frac{s\left(\Pi_\gamma\right)(\hbar)}{\hbar}=\epsilon_\gamma(\theta), \quad \Pi_\gamma^{(n)}=Z_\gamma^{(n)}, \quad \hbar=\mathrm{e}^{-\theta},
\end{equation}
where $\Pi_{\gamma}^{(n)}$ and $Z_{\gamma}^{(n)}$ are defined in \eqref{eq:piexp} and \eqref{eq:tbaexp} respectively. This relationship will be verified through numerical calculations in the subsequent sections. The Borel singularities of quantum WKB periods correspond to the poles in the kernel of the TBA equations. These relationships are preserved through wall-crossing and analytic continuation in $\ell$.
To compute the quantum corrections to the quantum WKB periods, one notes that the WKB expansions for \eqref{eq:seq} satisfy the Picard-Fuchs equations:
\begin{equation}
\label{eq:pfeq}
   p_{2n}(z)= \mathcal{O}_n p_{0}(z)+\rd (*):=\sum_{k=1}^{N+1}c_k^{(n)} \frac{\partial}{\partial u_k}p_0(x)+\rd (*).
\end{equation}
where the $\rd(*)$ represents a total derivative, which vanishes in the contour integral. Consequently, the expansions of quantum periods satisfy
\begin{equation}
     \Pi_\gamma^{(n)}=\mathcal{O}_n \Pi_\gamma^{(0)}.
\end{equation}

This differential operator method is used to compute quantum WKB periods and provides an efficient way to determine corrections to WKB periods at higher orders. The coefficients $c_k^{(n)}$ up to the third order for $N=1,2,3$ are listed in Appendix \ref{sc: pf}. Although quantum periods are defined by the Borel resummation of the formal series, a closed form of the Borel transform is typically not available. In practice, the diagonal Pad{\'e} approximant of the Borel transform is often employed, followed by its Laplace transform to approximate the exact quantum periods. This approach, known as the Borel-Pad{\'e} resummation, also allows us to infer information about the singularities in the Borel plane from the poles of the Pad{\'e} approximant. Concrete examples are discussed in the subsequent sections.

\subsection{TBA equations for integrable field theories}

In the following sections, we will compare the TBA equations derived from the ODE with those obtained from $(1+1)$-dimensional integrable field theories. In particular, at the maximally symmetric point in the moduli space, TBA equations take a simple form whose kernel functions are given by the reflectionless ADE scattering S-matrices \cite{Klassen:1989ui}. 
For the ODE \eqref{eq:ode2-4} with monomial $\phi_2(z)=z^{2M}+u_N$, the $A_{2M-1}$-type Y-system appears \cite{Dorey:1999uk} and the related TBA equations are obtained in \cite{Zam91}. In this subsection, we review the TBA equations related to the ADE scattering matrices and discuss the D-type TBA equations.

\paragraph{TBA for diagonal scattering theories}
There are 
diagonal scattering theories for a simply-laced Lie algebra $\mathfrak{g}$.
The TBA equations for pseudo-energies $\epsilon_a(\theta)$ with rapidity $\theta$ are written as
\begin{equation}
\label{eq:zamtba1}
    -\epsilon_a+\nu_a-\frac{1}{2\pi}\sum_{b}\varphi_{ab}\star\log\left(1+\re^{-\epsilon_b}\right)=0,\quad a=1,\cdots,N(\mathfrak{g}),
\end{equation}
where $\nu_a(\theta)=R m_a \cosh \theta$ are the source term with a scale $R$ and masses $m_a$ of particles, and $N(\mathfrak{g})$ represents the number of particles in the $\mathfrak{g}$-related scattering theory. The kernel is related to the scattering matrix $S_{ab}(\theta)$ by
\begin{equation}
    \varphi_{ab}(\theta)=-\ri \frac{\rd}{\rd\theta}\log S_{ab}(\theta),
\end{equation}
where explicit form of $S_{ab}(\theta)$ is given in \cite{Klassen:1989ui}.
The Fourier transform  of the kernel $\varphi_{ab}(\theta)$ 
\begin{equation}
    \tilde{\varphi}_{ab}(k)=\int_{-\infty}^{\infty}\varphi_{ab}(\theta)\re^{\ri k\theta}\rd \theta.
\end{equation}
satisfies
\begin{equation}
    \left(\delta_{ab}-\frac{1}{2\pi}\tilde{\varphi}_{ab}(k)\right)^{-1}=\delta_{ab}-\frac{1}{2\cosh(k/h)}l_{ab},
\end{equation}
where $l_{ab}$ is the incidence matrix of the Dynkin diagram of $\mathfrak{g}$ and $h$ is the Coxeter number of $\mathfrak{g}$.
The masses $m_a$ obey
\begin{align}
    \sum_b l_{ab}m_b=2m_a\cos{\frac{\pi}{\hbar}}.
\end{align}

For later convenience, we denote the TBA kernel by
\begin{equation}
    \Phi_{ab}=-\frac{1}{2\pi}\varphi_{ab}.
\end{equation}

This set of TBA systems for integrable field theories coincides with GMN TBA equations for D-type Argyres-Douglas theories in the maximally symmetric point of the moduli space. This is clarified in the subsequent sections for $D_3$, $D_4$, and $D_5$ cases explicitly. We write down here the explicit form of the kernels for $\mathfrak{g}=D_3$ and $D_4$ for comparison.
\begin{equation}
\label{eq:d3kernel}
\Phi^{D_3}(\theta)=\left(\begin{array}{ccc}
\frac{1}{\pi \cosh \theta} & \frac{\sqrt{2}\cosh \theta}{\pi \cosh 2 \theta} & \frac{\sqrt{2}\cosh \theta}{\pi \cosh 2 \theta} \\[0pt]
\frac{\sqrt{2}\cosh \theta}{\pi \cosh 2 \theta} & \frac{1}{2\pi \cosh \theta} & \frac{1}{2\pi \cosh \theta} \\[0pt]
\frac{\sqrt{2}\cosh \theta}{\pi \cosh 2 \theta} & \frac{1}{2\pi \cosh \theta} & \frac{1}{2\pi \cosh \theta}
\end{array}\right).
\end{equation}
One can identify $\epsilon_2=\epsilon_3$ for $\mathfrak{g}=D_3$, and the TBA system \eqref{eq:zamtba1} is simplified into two equations for $\epsilon_1$ and $\epsilon_2$.
\begin{equation}
\label{eq:d4kernel}
\Phi^{D_4}(\theta)=\left(
\begin{array}{cccc}
 \frac{2 \sqrt{3} \cosh \theta}{\pi  \left(2 \cosh 2\theta+1\right)} & \frac{3 \cosh 2\theta}{\pi \cosh 3\theta} & \frac{2 \sqrt{3} \cosh \theta}{\pi  \left(2 \cosh 2\theta+1\right)} & \frac{2 \sqrt{3} \cosh \theta}{\pi  \left(2 \cosh 2\theta+1\right)} \\
 \frac{3 \cosh 2\theta}{\pi \cosh 3\theta} &  \frac{6 \sqrt{3} \cosh \theta}{\pi  \left(2 \cosh 2\theta+1\right)}  & \frac{3 \cosh 2\theta}{\pi \cosh 3\theta} & \frac{3 \cosh 2\theta}{\pi \cosh 3\theta}  \\
  \frac{2 \sqrt{3} \cosh \theta}{\pi  \left(2 \cosh 2\theta+1\right)}  & \frac{3 \cosh 2\theta}{\pi \cosh 3\theta}  &  \frac{2 \sqrt{3} \cosh \theta}{\pi  \left(2 \cosh 2\theta+1\right)}  &  \frac{2 \sqrt{3} \cosh \theta}{\pi  \left(2 \cosh 2\theta+1\right)}  \\
 \frac{2 \sqrt{3} \cosh \theta}{\pi  \left(2 \cosh 2\theta+1\right)} &\frac{3 \cosh 2\theta}{\pi \cosh 3\theta}  &  \frac{2 \sqrt{3} \cosh \theta}{\pi  \left(2 \cosh 2\theta+1\right)} &  \frac{2 \sqrt{3} \cosh \theta}{\pi  \left(2 \cosh 2\theta+1\right)}  \\
\end{array}
\right).
\end{equation}
One can identify $\epsilon_1=\epsilon_3=\epsilon_4$ for $\mathfrak{g}=D_4$, and the TBA system \eqref{eq:zamtba1} is simplified into two equations for $\epsilon_1$ and $\epsilon_2$.

\paragraph{Effective central charge} Associated with the GMN TBA equations and TBA equations \eqref{eq:zamtba1}, one can define the corresponding effective central charge \cite{Zam90, Klassen:1989ui}
\begin{equation}
         c_{\mathrm{eff}}=\frac{6}{\pi^2}\sum_{\gamma_a>0} \Omega\left(\gamma_a, u\right) \lvert Z_{\gamma_a}\rvert \int_{\mathbb{R}} \re^{\theta}\tilde{L}_a(\theta) \rd \theta,
\end{equation}
which can be evaluated exactly. One notes the asymptotic behavior of $\tilde{\epsilon}_\gamma(\theta)$ as $\theta\rightarrow-\infty$
\begin{equation}
\tilde{\epsilon}_{\gamma_a}^{\star}=\lim _{\theta \rightarrow-\infty} \tilde{\epsilon}_{\gamma_a}(\theta).
\end{equation}
These are constant values and can be determined by a set of algebraic equations reduced from the TBA equations. The effective central charge is expressed as
\begin{equation}
     c_{\mathrm{eff}}=\frac{6}{\pi^2}\sum_{\gamma_a>0} \Omega\left(\gamma_a, u\right) \mathcal{L}\left(\frac{1}{1+\re^{\epsilon_{\gamma_a}^\star}}\right),
\end{equation}
where 
\begin{equation}
    \mathcal{L}(x)=-\frac{1}{2}\int_0^x\rd y\left(\frac{\ln y}{1-y}+\frac{\ln (1-y)}{y}\right)
\end{equation}
is Rogers' dilogarithm function. The above-defined effective central charge is independent of the moduli parameters; namely, it remains the same for TBA systems in different chambers from the same SW curve. This provides a way to validate the family of D-type TBA equations. Interestingly, the effective central charge defined above is related to the so-called PNP relation \cite{BD15, CM16, BD17, IMS18}, which reveals the connection between WKB expansions and TBA equations from another perspective.

\section{$(A_1,D_3)$ theory}
\label{sc:d3}

\begin{figure}[htbp]
\centering
    \begin{subfigure}{0.45\linewidth}
        \centering
        \includegraphics[width=1.0\linewidth]{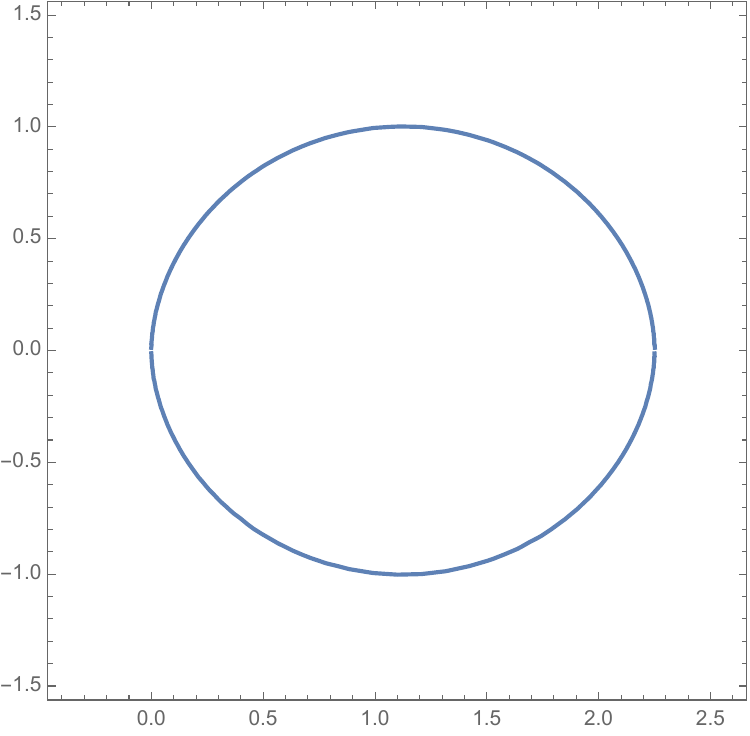}
        \caption{}
        \label{fig:d3wall1}
    \end{subfigure}
    \hspace{0.05\linewidth} 
    \begin{subfigure}{0.45\linewidth}
        \centering
        \includegraphics[width=1.0\linewidth]{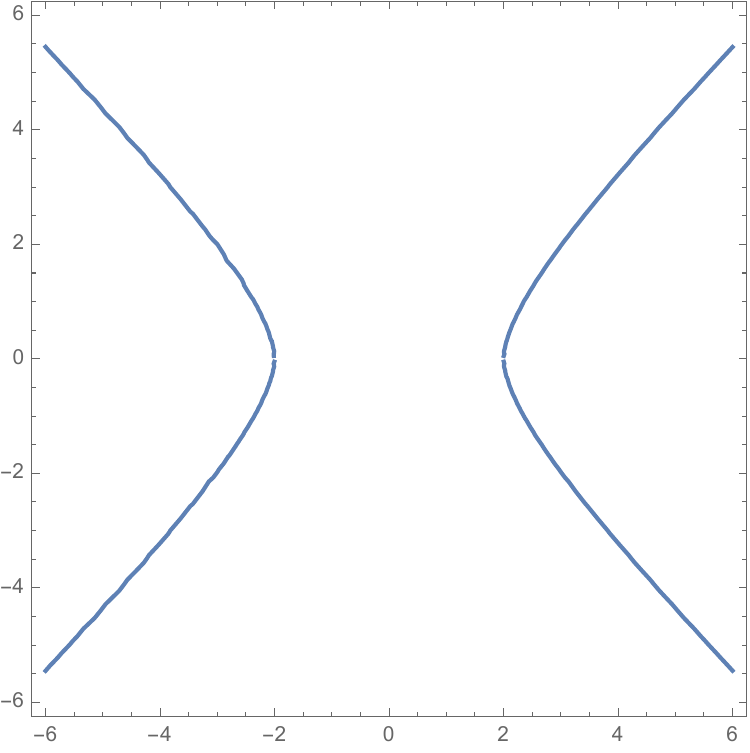}
        \caption{}
        \label{fig:d3wall2}
    \end{subfigure}

    \caption{The walls of marginal stability for $(A_1, D_3)$. The left figure shows the wall in the $u_2$ plane for a fixed value of $u_1 = 3$. The area inside the curve represents the minimal chamber, while the region outside corresponds to the maximal chamber. The right figure illustrates the curve in the $u_1$ plane with a fixed value of $u_2=1$. The area between the two curves constitutes the maximal chamber, whereas the regions to the left and right of the curves represent the minimal chambers.}
    \label{fig:d3wall}
\end{figure}

Let us consider the first non-trivial example of AD theories of type D, the $(A_1, D_3)$ theory. Its Seiberg-Witten differential $\lambda$ is defined by 
\begin{equation}
\label{eq:d3sw}
\lambda^2-\left(z+u_1+\frac{u_2}{z}+\frac{m^2}{z^2}\right) \mathrm{d} z^2=0,
\end{equation}
where $u_2$ is the vacuum expectation value of the relevant operator and $u_1$ is the coupling. The parameter $m$, the residue of the regular puncture at $z=0$, is associated with an SU(2) flavor symmetry. We consider the massless limit $m=0$ when one branch point collides with $0$ and the SU(2) flavor symmetry is restored. In this case, there are only two chambers: the minimal chamber and the maximal chamber. They are divided by the wall of marginal stability. The walls in $u_2$ and $u_1$ planes for fixed $u_1$ and $u_2$, respectively, are shown in Figure \ref{fig:d3wall}.

\subsection{TBA equations in the minimal chamber}

In the minimal chamber, there are two BPS states (and their anti-states) as indicated in Figure \ref{fig:d3minbps}. The intersection number is an essential ingredient in writing down TBA equations. One can assign an electromagnetic charge to each of the BPS states so that their Dirac pairing gives the intersection number. The charges of the two BPS states are
\begin{equation}
    \gamma_1=(0, 1), \quad   \gamma_2=(1, 0),
\end{equation}
where $\gamma_1$ is a singlet, and $\gamma_2$ is a doublet as we explained in Section \ref{sc: TBA}.
Their intersection pairings are
\begin{equation}
    \left\langle\gamma_1, \gamma_2\right\rangle=-\left\langle\gamma_2, \gamma_1\right\rangle=-1.
\end{equation}
The BPS indices are given by
\begin{equation}
\Omega(\gamma)= \begin{cases}
2 & \text { for } \gamma \in\left\{\gamma_2,-\gamma_2\right\}, \\ 

1 & \text { for } \gamma \in\left\{\gamma_1,-\gamma_1\right\}, \\

0 & \text{ otherwise.}
\end{cases}
\end{equation}
From \eqref{eq:gmntba2}, the TBA equations can be written down explicitly as
\begin{equation}
\label{eq:d3mintba1}
\begin{aligned}
    \tilde{\epsilon}_{\gamma_1}(\theta)&=\left|Z_{\gamma_1}\right| \mathrm{e}^\theta+\frac{1}{2 \pi \mathrm{i}} 2 \int_{\mathbb{R}} \frac{\log \left(1+\mathrm{e}^{-\tilde{\epsilon}_{\gamma_2}\left(\theta^{\prime}\right)}\right)}{\sinh \left(\theta-\theta^{\prime}-\mathrm{i} \phi_{12}\right)} \mathrm{d} \theta^{\prime},\\
     \tilde{\epsilon}_{\gamma_2}(\theta)&=\left|Z_{\gamma_2}\right| \mathrm{e}^\theta-\frac{1}{2 \pi \mathrm{i}}  \int_{\mathbb{R}} \frac{\log \left(1+\mathrm{e}^{-\tilde{\epsilon}_{\gamma_1}\left(\theta^{\prime}\right)}\right)}{\sinh \left(\theta-\theta^{\prime}-\mathrm{i} \phi_{21}\right)} \mathrm{d} \theta^{\prime},
\end{aligned}
\end{equation}
where the factor 2 comes from the BPS index of the doublet $\Omega(\gamma_2)=2$. $\phi_{12}$ and $\phi_{21}$ are phase differences defined in \eqref{eq:phase}, with $\gamma$ omitted. We first test these TBA equations numerically by choosing parameters $u_1=3, u_2=2, m=0$ in \eqref{eq:d3sw}, resulting in three branch points located at $-2$, $-1$ and $0$. The two BPS states $\gamma_1$ and $\gamma_2$ correspond to the one-cycles specified in Figure \ref{fig:d3minbpsplot}. Their central charges are
\begin{equation}
    Z_{\gamma_1}=\frac{4\sqrt{2}}{3}\left(3 \mathbb{E}\left(\frac{1}{2}\right)-2 \mathbb{K}\left(\frac{1}{2}\right)\right), \quad  Z_{\gamma_2}=\ri \frac{4\sqrt{2}}{3}\left(3 \mathbb{E}\left(\frac{1}{2}\right)- \mathbb{K}\left(\frac{1}{2}\right)\right),
\end{equation}
where $\mathbb{E}$ and $\mathbb{K}$ represent the complete elliptic integrals of the first and second kinds, respectively. The phases of the two central charges are $0$ and $\frac{\pi}{2}$,
and the TBA equations become
\begin{equation}
\label{eq:d3mintba2}
\begin{aligned}
    \tilde{\epsilon}_{\gamma_1}(\theta)&=\left|Z_{\gamma_1}\right| \mathrm{e}^\theta-\frac{1}{2 \pi } 2 \int_{\mathbb{R}} \frac{\log \left(1+\mathrm{e}^{-\tilde{\epsilon}_{\gamma_2}\left(\theta^{\prime}\right)}\right)}{\cosh \left(\theta-\theta^{\prime}\right)} \mathrm{d} \theta^{\prime},\\
     \tilde{\epsilon}_{\gamma_2}(\theta)&=\left|Z_{\gamma_2}\right| \mathrm{e}^\theta-\frac{1}{2 \pi }  \int_{\mathbb{R}} \frac{\log \left(1+\mathrm{e}^{-\tilde{\epsilon}_{\gamma_1}\left(\theta^{\prime}\right)}\right)}{\cosh \left(\theta-\theta^{\prime}\right)} \mathrm{d} \theta^{\prime}.
\end{aligned}
\end{equation}
It is also recognized as the TBA of the $(A_1, A_3)$ theory in the minimal chamber with SU(2) symmetry, corresponding to a quantum mechanical system with a symmetric double-well potential in its minimal chamber, as detailed in equation (5.5) of \cite{IMS18}. This arises from the equivalence of their Lie algebras, $D_3=A_3$. 

We solve these TBA equations numerically and compare the large $\theta$ expansions $Z_{\gamma}^{(n)}$ of the pseudo-energies to the WKB expansions $\Pi_{\gamma}^{(n)}$ of the quantum periods for \eqref{eq:dqsw}, as shown in Table \ref{tab:d3tbawkb}. The numerical results illustrate the connection between the WKB expansions and the TBA equations. Furthermore, the TBA equations provide insight into Borel summability and Borel singularities through their kernels, which can be numerically detected from the Borel-Pad{\'e} singularities. We plot the poles of the diagonal Pad{\'e} approximant of the truncated Borel transforms in Figure \ref{fig:d3d4pole}. For example, the kernel $K_{1,2}$ exhibits a singularity at $\arg Z_{\gamma_2}=\frac{\pi}{2}$, which is evident in the Pad{\'e} approximant of the Borel transform of the quantum period $\Pi_{\gamma_1}$, as shown in Figure \ref{fig:d3minpole1}. Similarly, the singularities for $\gamma_2$ can be located at $\arg Z_{\gamma_1}=0$. This analysis indicates that the quantum WKB period for $\gamma_1$ is Borel summable along the positive real ray, while that for $\gamma_2$ is not. It is evident that the TBA equations predict Borel singularities that are consistent with the Borel resummation of the quantum WKB periods obtained from the WKB expansions, demonstrating the resurgent structure of the quantum periods. We perform Borel-Pad{\'e} summation for $\gamma_1$ to find an approximation of the exact quantum period:
\begin{equation}
    \Pi_{\gamma_1}^{\mathrm{ex}}\left(\hbar=1\right)=\underline{0.485357754}31\cdots  
\end{equation}
where we use 50 terms of the WKB expansions (up to $\hbar^{100}$) and diagonal [50/50] Pad{\'e} approximant. We  compare this value with 
\begin{equation}
    \epsilon_{\gamma_1}\left(\theta=0\right)= \underline{0.485357754}94\cdots
\end{equation}
to find agreement, validating the first equality of \eqref{eq:wkbtbarelation}.

Moreover, \eqref{eq:d3mintba2} corresponds to the TBA equations in \cite{IS19} when $\ell=-\frac{1}{2}$. In other words, we can deform the first TBA equation in \eqref{eq:d3mintba1} to the following form:
\begin{equation}
\label{eq:d3mintba3}
\tilde{\epsilon}_{\gamma_1}(\theta) =\left|Z_{\gamma_1}\right| \re^\theta+\frac{1}{2\pi}\int_{\mathbb{R}}  \frac{\log \left(\left(1-\re^{2 \pi \ri \ell} \re^{-\tilde{\epsilon}_{\gamma_2}\left(\theta^{\prime}\right)}\right)\left(1-\re^{-2 \pi \ri \ell} \re^{-\tilde{\epsilon}_{\gamma_2}\left(\theta^{\prime}\right)}\right)\right)}{\sinh \left(\theta-\theta^{\prime}-\mathrm{i} \phi_{12}\right)}\rd \theta^{\prime},
\end{equation}
which corresponds to the deformation of the quantum SW curve \eqref{eq:leq} for the $(A_1, D_3)$ theory, parameterized by $\ell$. This relationship persists for $\ell\in (-1,0)$ and is confirmed by numerical computations for $\ell=-\frac{1}{5}$, as shown in Table \ref{tab:d3tbawkb}. However, the physical interpretation of the deformation parameter $\ell$ in the AD theory remains unclear to us.

\paragraph{Analytic continuation of $\ell$}

As $\ell$ increases and moves beyond the strip $(-1,0)$, poles in logarithmic terms of \eqref{eq:d3mintba3} and the second equation of \eqref{eq:d3mintba1} emerge, introducing corrections to the corresponding TBA equations \cite{DT96}. The first pole, $\alpha(\ell)$, is determined by the zero of $1-\re^{2 \pi \ri \ell} \re^{-\tilde{\epsilon}_{\gamma_2}(\theta)}
$, and satisfies the equation
\begin{equation}
    2\pi \ri \ell= \tilde{\epsilon}_{\gamma_2}\left(\alpha(\ell)\right)=\left|Z_{\gamma_2}\right| \mathrm{e}^{\alpha(\ell)}-\frac{1}{2 \pi \mathrm{i}}  \int_{\mathbb{R}} \frac{\log \left(1+\mathrm{e}^{-\tilde{\epsilon}_{\gamma_1}\left(\theta^{\prime}\right)}\right)}{\sinh \left(\alpha(\ell)-\theta^{\prime}-\mathrm{i} \phi_{21}\right)} \mathrm{d} \theta^{\prime}.
\end{equation}
This pole introduces a residue correction to $\tilde{\epsilon}_{\gamma_1}(\theta)$ given by
\begin{equation}
    \Delta \tilde{\epsilon}_{\gamma_1}(\theta)=\coth^{-1} \left( \cosh(\alpha(\ell)-\theta+\ri \phi_{12})\right)=\log\left[\frac{\re^{\theta}+\re^{\alpha(\ell)+\ri \phi_{12}}}{\re^{\theta}-\re^{\alpha(\ell)+\ri \phi_{12}}}\right],
\end{equation}
which comes back to equation (5.11) in \cite{GY21} when $\phi_{12}=-\frac{\pi}{2}$. This term also modifies the large $\theta$ expansion of $\epsilon_{\gamma_1}(\theta)$ as given in \eqref{eq:tbaexp} by 
\begin{equation}
    \Delta Z_{\gamma_1}^{(n)} = \frac{2}{2n-1} \exp\left[\left(2n-1\right)\left(\alpha(\ell) - \ri \phi_2\right)\right].
\end{equation}
A similar discussion is also presented in \cite{Fendley:1997ys, IY23, IS24}. We perform numerical calculations for $\ell=\frac{1}{5}$, as shown in Table \ref{tab:d3tbawkb}. The corresponding pole is located at
\begin{equation}
    \alpha\left(\ell=\frac{1}{5}\right)\approx -1.99159112927+ \ri\frac{\pi}{2}.
\end{equation}
As $\ell$ increases, additional poles emerge, each corresponding to further correction to the TBA equations \cite{GY21}. It is feasible to investigate the analytic continuation of $\ell$ in the complex plane, but it is beyond the scope of this paper.

\paragraph{Effective central charge}
It is straightforward to find the behaviors of $\tilde{\epsilon}_{\gamma}(\theta)$ \eqref{eq:d3mintba1} in the limit $\theta\to-\infty$:
\begin{equation}
    \tilde{\epsilon}_{\gamma_1}^\star=-\log 3, \quad   \tilde{\epsilon}_{\gamma_2}^\star=-\log 2.
\end{equation}
The effective central charge associated with the TBA system \eqref{eq:d3mintba1} can be computed as follows:
\begin{equation}
\label{eq:d3mineff}
\begin{aligned}
    c_{\mathrm{eff}}&=\frac{6}{\pi^2}\int\left|Z_{\gamma_1}\right|\re^\theta \tilde{L}_1(\theta)\rd \theta+\frac{6}{\pi^2}2\left|Z_{\gamma_2}\right|\int\re^\theta \tilde{L}_2(\theta)\rd \theta,\\
    &=\frac{6}{\pi^2}\left(\mathcal{L}_1\left(\frac{1}{1+\re^{\tilde{\epsilon}_{\gamma_1}^\star}}\right)+2\mathcal{L}_1\left(\frac{1}{1+\re^{\tilde{\epsilon}_{\gamma_2}^\star}}\right)\right)=2,
    \end{aligned}
\end{equation}
where the factor 2 in front of $L_2$ accounts for the doublet $\gamma_2$.

\subsection{TBA equations in the maximal chamber}

\begin{figure}[htbp]
\centering
  \begin{subfigure}{0.4\linewidth}
        \centering
        \includegraphics[width=0.9\linewidth]{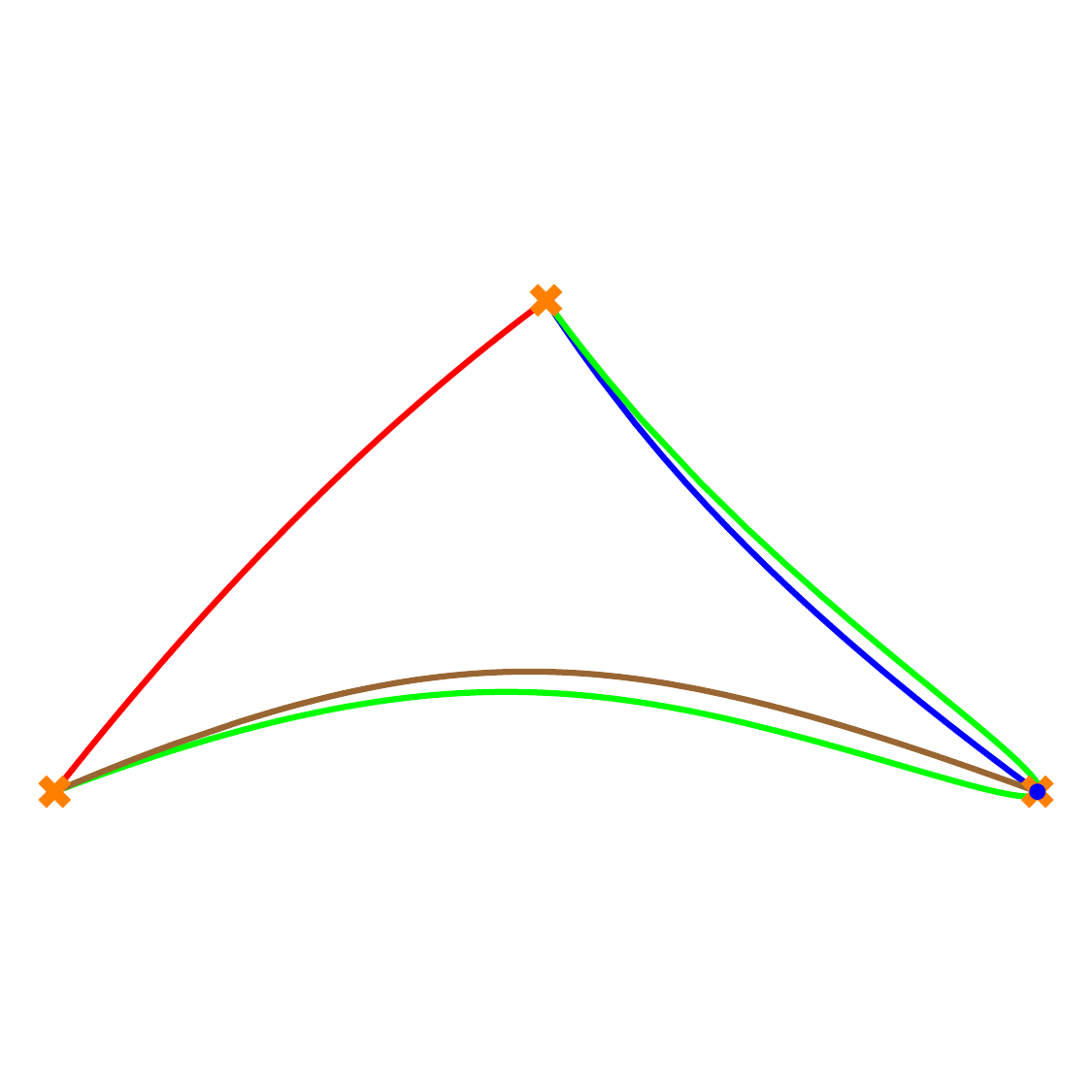}
        \caption{BPS states}
        \label{fig:d3cham1bpsplot}
    \end{subfigure}
    \begin{subfigure}{0.4\linewidth}
        \centering
        \includegraphics[width=0.9\linewidth]{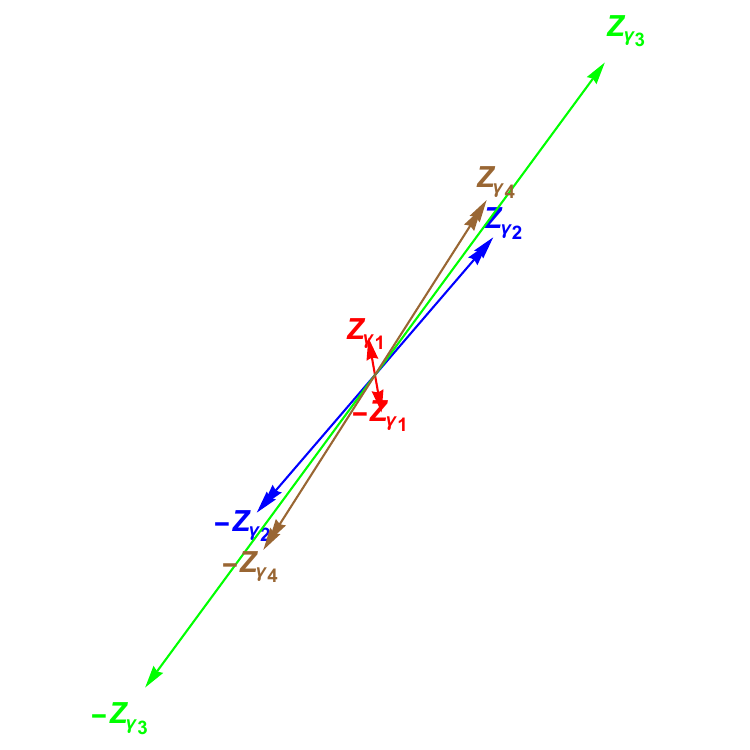}
        \caption{Central charges}
        \label{fig:d3cham1bpscharge}
    \end{subfigure}

\vspace{1em}
    \begin{subfigure}{0.4\linewidth}
        \centering
        \includegraphics[width=0.9\linewidth]{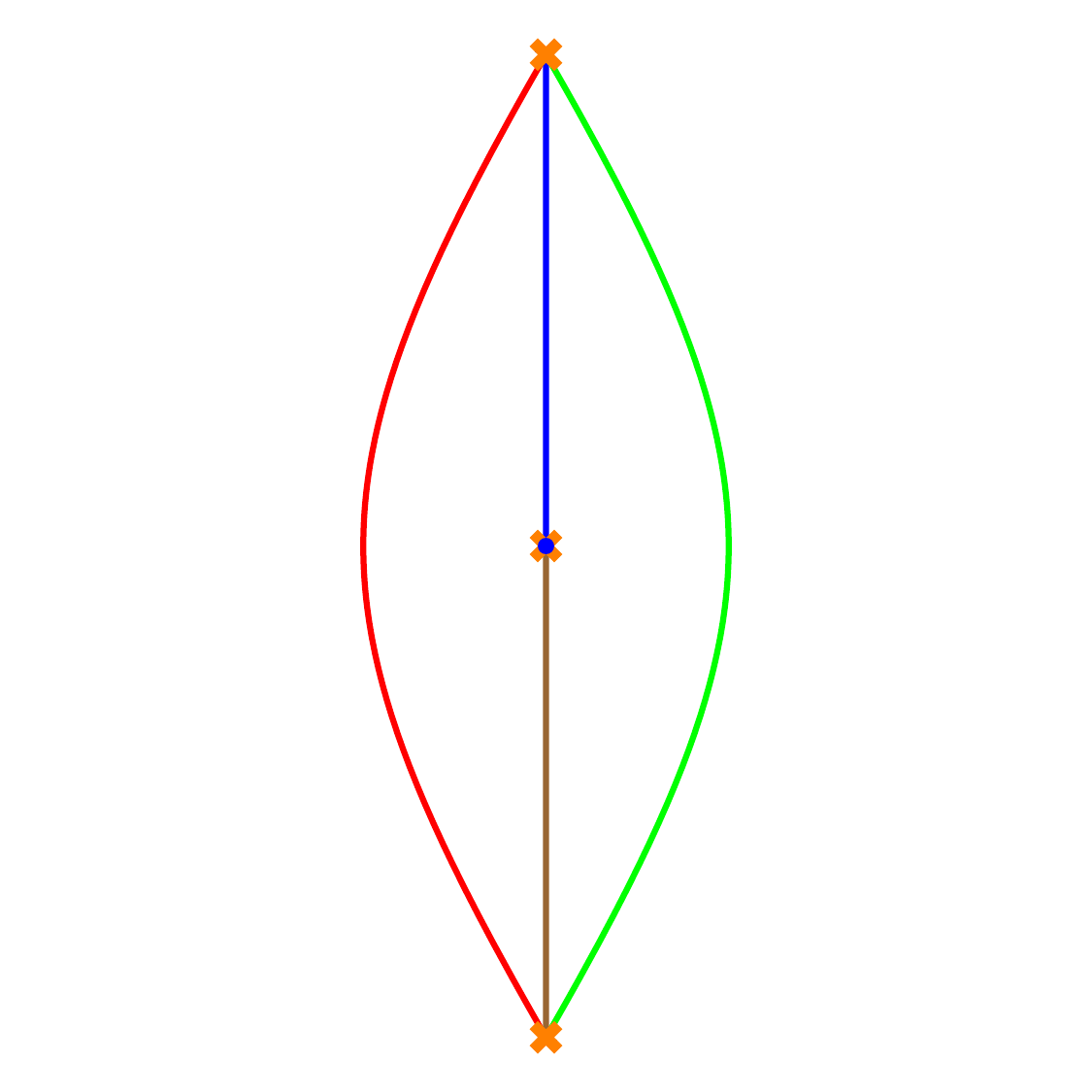}
        \caption{BPS states}
        \label{fig:d3maxbpsplot}
    \end{subfigure}
    \begin{subfigure}{0.4\linewidth}
        \centering
        \includegraphics[width=0.9\linewidth]{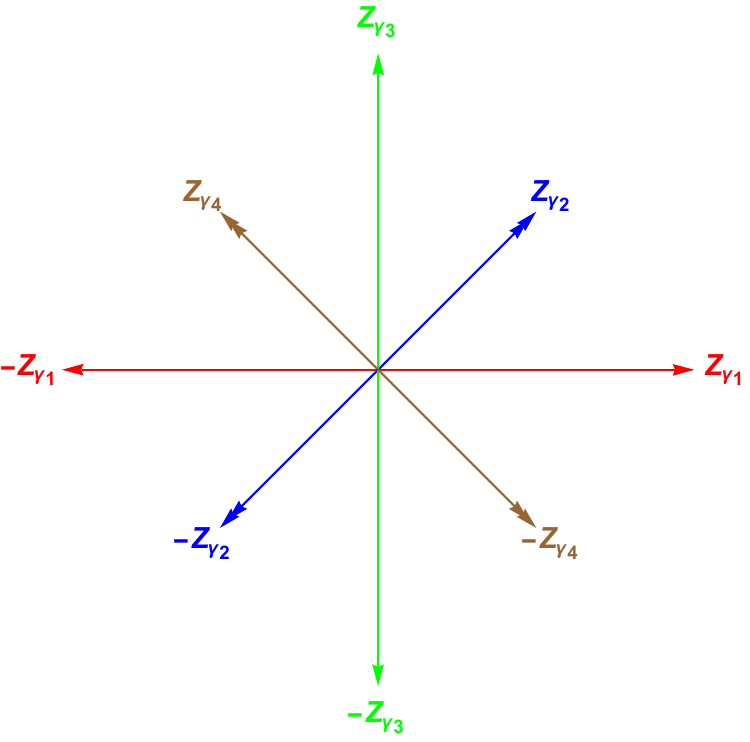}
        \caption{Central charges}
        \label{fig:d3maxbpscharge}
    \end{subfigure}

    \caption{BPS spectrum in the maximal chamber for the $(A_1,D_3)$ theory}
    \label{fig:d3maxbps}
\end{figure}
We begin by considering the BPS spectrum in the minimal chamber, as shown in Figure \ref{fig:d3minbpscharge}. Moving in the moduli space, we can rotate $Z_{\gamma_1}$ anticlockwise and $Z_{\gamma_2}$ clockwise. Upon reaching the wall of the marginal stability, the two central charges align. After crossing this wall, two new BPS states emerge: a singlet $\gamma_3$ and a doublet $\gamma_4$, as depicted in Figures \ref{fig:d3cham1bpsplot} and \ref{fig:d3cham1bpscharge}. We refer to this as a type D wall-crossing, in contrast to the ordinary type A wall-crossing, when only a singlet is created. Examples of type A wall-crossing will be discussed in Section \ref{sc:d4}. 

In the maximal chamber, there exists a maximally symmetric point where the potential takes the form of a monomial plus an inverse term. The configuration of turning points and BPS spectrum at this point are illustrated in Figure \ref{fig:d3maxbpsplot} and \ref{fig:d3maxbpscharge}. We emphasize that, for later convenience, the choice of branch cuts and electromagnetic charges here differs from those in Figures \ref{fig:d3cham1bpsplot} and \ref{fig:d3cham1bpscharge}. Specifically, we order the central charges according to their phase in the interval $[0,\pi)$. This convention will also be used for the maximally symmetric points in the $(A_1, D_4)$ and $(A_1, D_5)$ theories. This choice does not alter the TBA equations except for the exchange between a BPS state and its anti-BPS state: $\gamma_1 \leftrightarrow-\gamma_1$. We label the electromagnetic charge in the maximal chamber as
\begin{equation}
     \gamma_1=(1,0), \quad \gamma_2=(1,1), \quad \gamma_3=(2,1),\quad  \gamma_4=(0,1).
\end{equation}
Using $\gamma_2$, $\gamma_4$ as basis, the charges can be represented as
\begin{equation}
   \gamma_1=\gamma_2-\gamma_4,\quad \gamma_3=\gamma_2+\gamma_4.
\end{equation}
Here $\gamma_1$ and $\gamma_3$ are singlets, while $\gamma_2$ and $\gamma_4$ are doublets. Their BPS indices are
\begin{equation}
\Omega(\gamma)= \begin{cases}
2 & \text { for } \gamma \in\left\{\pm \gamma_2,\pm \gamma_4\right\}, \\ 

1 & \text { for } \gamma \in\left\{\pm \gamma_1,\pm \gamma_3\right\}, \\

0 & \text { otherwise. }
\end{cases}
\end{equation}
The TBA equations are written as
\begin{equation}
\label{eq:d3maxtba1}
\begin{aligned}
    \tilde{\epsilon}_{\gamma_1}(\theta)&=\left|Z_{\gamma_1}\right| \mathrm{e}^\theta- 2 K_{1,2}\star \tilde{L}_2-2 K_{1,3}\star \tilde{L}_3-2 K_{1,4}\star \tilde{L}_4,\\
     \tilde{\epsilon}_{\gamma_2}(\theta)&=\left|Z_{\gamma_2}\right| \mathrm{e}^\theta+ K_{2,1}\star \tilde{L}_1- K_{2,3}\star \tilde{L}_3-2  K_{2,4}\star \tilde{L}_4,\\
      \tilde{\epsilon}_{\gamma_3}(\theta)&=\left|Z_{\gamma_3}\right| \mathrm{e}^\theta+2 K_{3,1}\star \tilde{L}_1+ 2 K_{3,2}\star \tilde{L}_2-2 K_{3,4}\star \tilde{L}_4,\\
       \tilde{\epsilon}_{\gamma_4}(\theta)&=\left|Z_{\gamma_4}\right| \mathrm{e}^\theta+ K_{4,1}\star \tilde{L}_1+2 K_{4,2}\star \tilde{L}_2+ K_{4,3}\star \tilde{L}_3,
\end{aligned}
\end{equation}
where $K_{a,b}$ and $\tilde{L}_{a}$ are defined as in \eqref{eq:kernel}, with $\gamma$ omitted. This TBA system describes the wall-crossing from the minimal TBA \eqref{eq:d3mintba1} to the maximal chamber, extending the discussion in \cite{IS19}. It also represents the TBA for the $(A_1, A_3)$ theory in the maximal chamber with $SU(2)$ symmetry, corresponding to the TBA equations for the symmetric double well in its maximal chamber.

\paragraph{Maximally symmetric point} 
This TBA system applies to any point within the maximal chamber.  Let us consider the maximally symmetric point in the moduli space, specifically at $u_1=0, u_2=1$. At this point, the magnitude of the two singlets and the two doublets are equal, respectively:
\begin{equation}
    \left|Z_{\gamma_1}\right|=    \left|Z_{\gamma_3}\right|=\frac{2\sqrt{2\pi}\Gamma\left(\frac{5}{4}\right)}{\Gamma\left(\frac{7}{4}\right)}, \quad   \left|Z_{\gamma_2}\right|=    \left|Z_{\gamma_4}\right|=\frac{2\sqrt{\pi}\Gamma\left(\frac{5}{4}\right)}{\Gamma\left(\frac{7}{4}\right)}.
\end{equation}
The phases of the central charges are
\begin{equation}
    \phi_{\gamma_1}=0, \quad   \phi_{\gamma_2}=\frac{\pi}{4}, \quad  
    \phi_{\gamma_3}=\frac{\pi}{2}, \quad 
    \phi_{\gamma_4}=\frac{3\pi}{4}.
\end{equation}
One can identify $\tilde{\epsilon}_{\gamma_1}=\tilde{\epsilon}_{\gamma_3}$ and $\tilde{\epsilon}_{\gamma_2}=\tilde{\epsilon}_{\gamma_4}$, which simplifies the TBA equations further, resulting in the TBA system collapsing into two equations as follows:
\begin{equation}
\label{eq:d3maxsym}
\begin{aligned}
    \tilde{\epsilon}_{\gamma_1}(\theta)&=\left|Z_{\gamma_1}\right| \mathrm{e}^\theta+\frac{1}{\pi } \int_{\mathbb{R}} \frac{\tilde{L}_1\left(\theta^\prime\right)}{\cosh \left(\theta-\theta^{\prime}\right)} \mathrm{d} \theta^{\prime}+\frac{2\sqrt{2}}{\pi } \int_{\mathbb{R}} \frac{\cosh \left(\theta-\theta^{\prime}\right)}{\cosh \left(2(\theta-\theta^{\prime})\right)}\tilde{L}_2\left(\theta^\prime\right) \mathrm{d} \theta^{\prime},\\
     \tilde{\epsilon}_{\gamma_2}(\theta)&=\left|Z_{\gamma_2}\right| \mathrm{e}^\theta+\frac{\sqrt{2}}{\pi } \int_{\mathbb{R}} \frac{\cosh \left(\theta-\theta^{\prime}\right)}{\cosh \left(2(\theta-\theta^{\prime})\right)} \tilde{L}_1\left(\theta^\prime\right)\mathrm{d} \theta^{\prime}+\frac{1}{\pi } \int_{\mathbb{R}} \frac{\tilde{L}_2\left(\theta^\prime\right)}{\cosh \left(\theta-\theta^{\prime}\right)} \mathrm{d} \theta^{\prime}.
\end{aligned}
\end{equation}

This TBA system was obtained in \cite{DT98} for the pure quartic potential using the ODE/IM correspondence. Moreover, it coincides with the  TBA equations for scattering theories of $D_3$. This can be easily verified by comparing it with equations \eqref{eq:zamtba1} and \eqref{eq:d3kernel} for $\mathfrak{g}=D_3$. 

\begin{figure}[htbp]
\centering
 \begin{subfigure}{0.32\linewidth}
        \centering
    \includegraphics[width=1.0\linewidth]{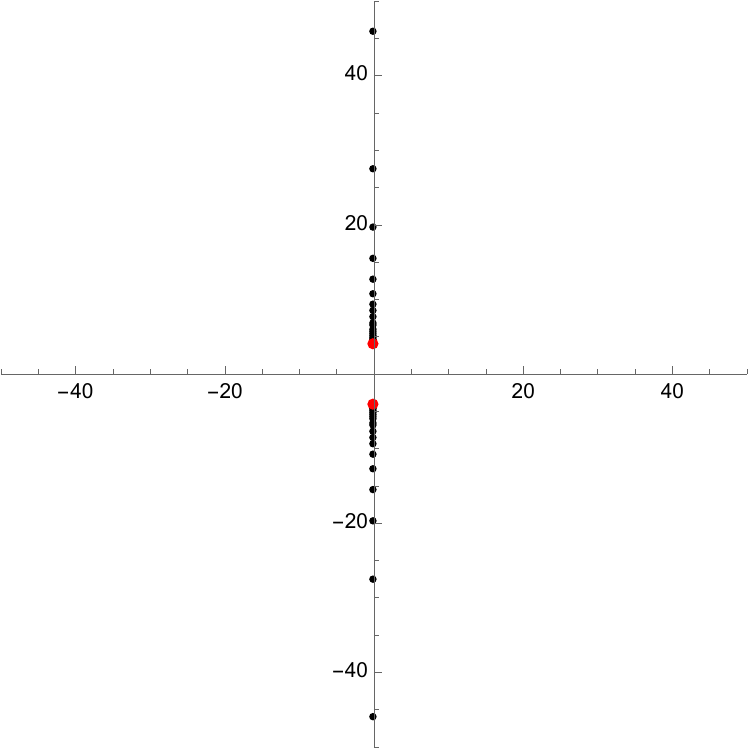}
        \caption{$\Pi_{\gamma_1}$ for the minimal $D_3$}
        \label{fig:d3minpole1}
    \end{subfigure}
     \begin{subfigure}{0.32\linewidth}
        \centering  
    \includegraphics[width=1.0\linewidth]{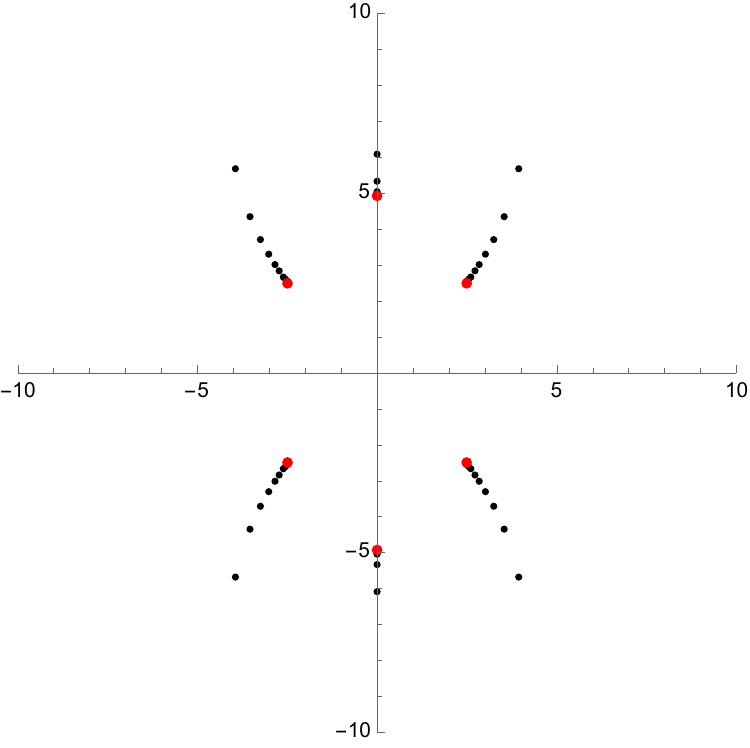}
        \caption{$\Pi_{\gamma_1}$ for the maximal $D_3$}
        \label{fig:d3maxpole1}
    \end{subfigure}
       \begin{subfigure}{0.32\linewidth}
        \centering
    \includegraphics[width=1.0\linewidth]{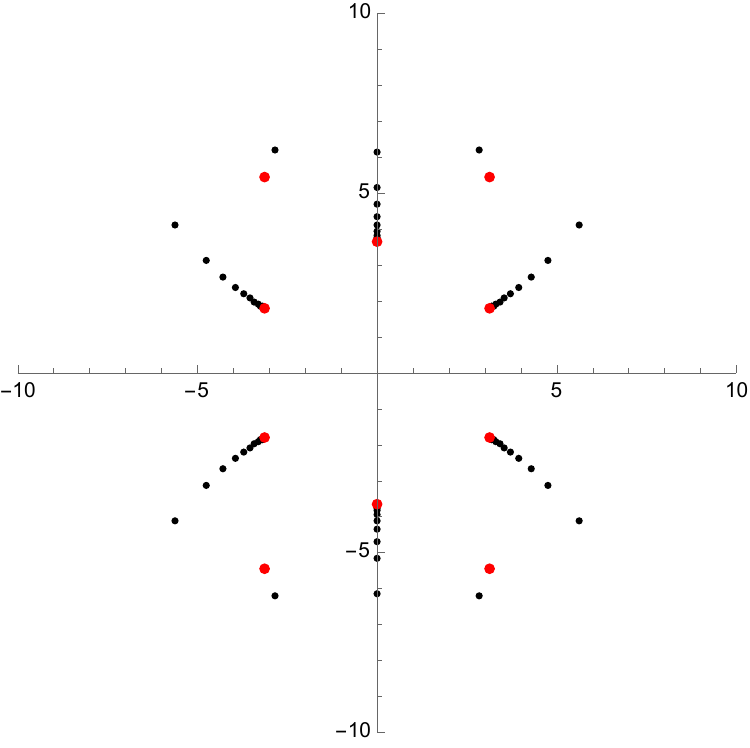}
        \caption{$\Pi_{\gamma_1}$ for the maximal $D_4$}
        \label{fig:d4maxpole1}
    \end{subfigure}
    
    \vspace{1em} 
    
 \begin{subfigure}{0.32\linewidth}
        \centering
    \includegraphics[width=1.0\linewidth]{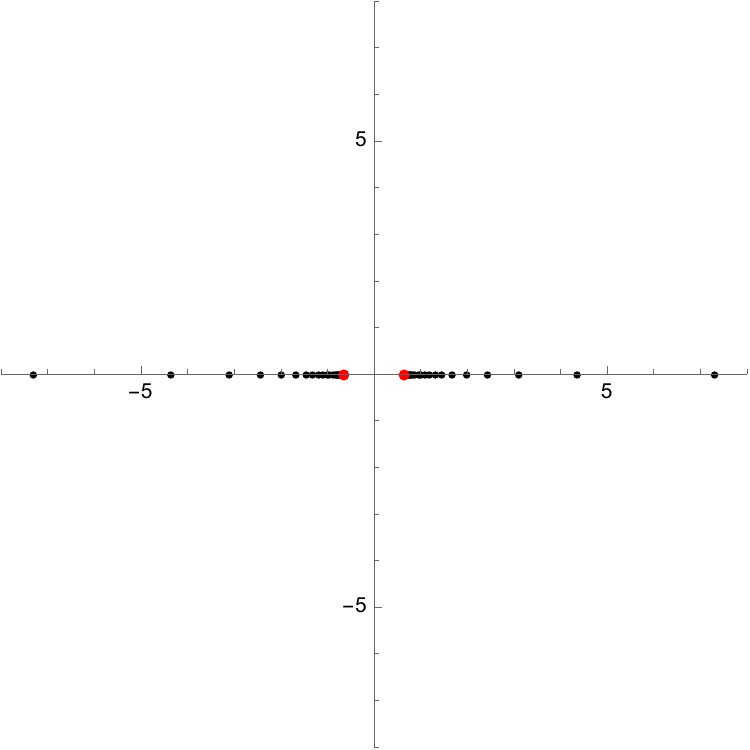}
         \caption{$\Pi_{\gamma_2}$ for the minimal $D_3$}
        \label{fig:d3minpole2}
    \end{subfigure}
    \begin{subfigure}{0.32\linewidth}
        \centering
    \includegraphics[width=1.0\linewidth]{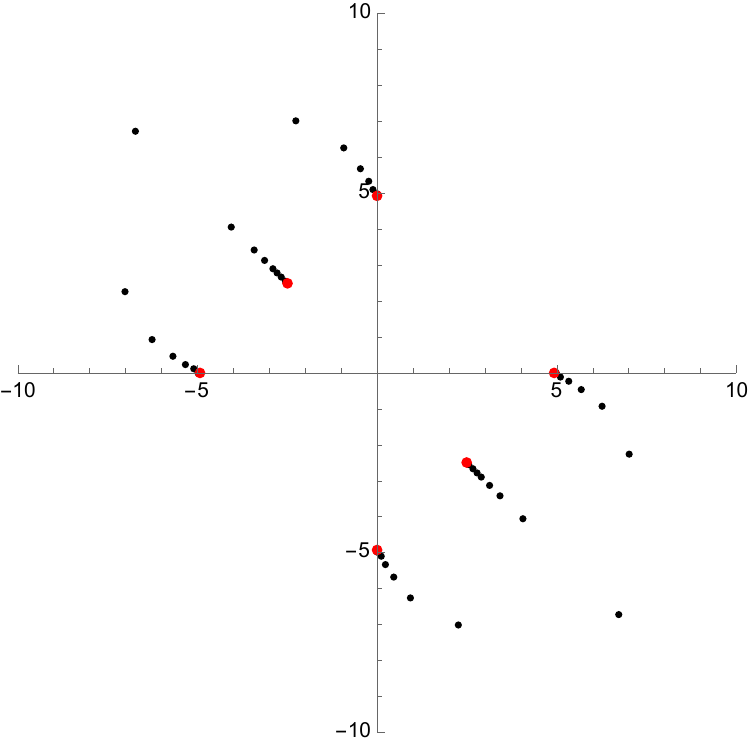}
         \caption{$\Pi_{\gamma_2}$ for the maximal $D_3$}
        \label{fig:d3maxpole2}
    \end{subfigure}
    \begin{subfigure}{0.32\linewidth}
        \centering
    \includegraphics[width=1.0\linewidth]{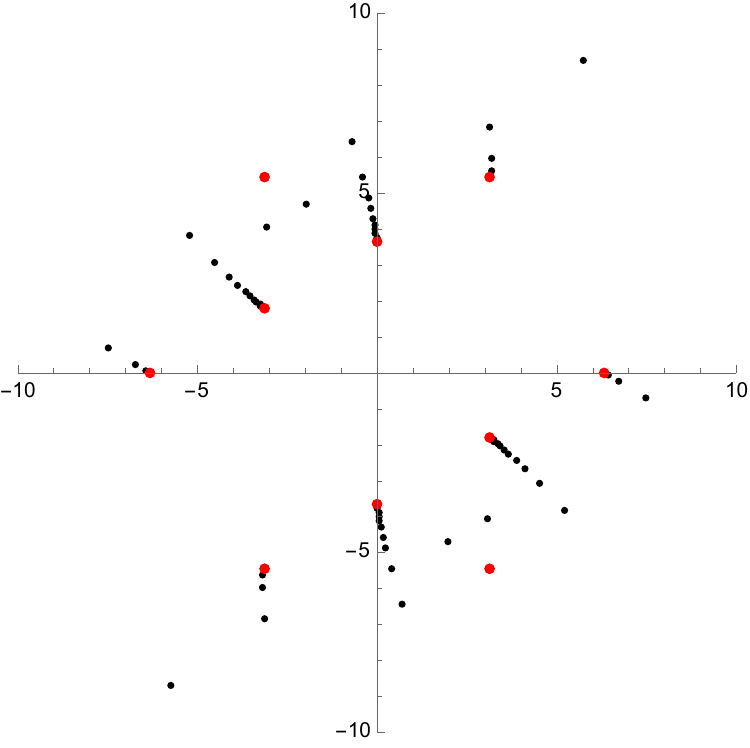}
         \caption{$\Pi_{\gamma_2}$ for the maximal $D_4$}
        \label{fig:d4maxpole2}
    \end{subfigure}

    \caption{Borel-Pad{\'e} singularities of $\Pi_{\gamma_1}$ and $\Pi_{\gamma_2}$ for $(A_1, D_3)$ and $(A_1, D_4)$. The black points represent poles of the Pad{\'e} approximant for the Borel transform of the quantum periods $\Pi_{\gamma_1}$ or $\Pi_{\gamma_2}$. These poles accumulate towards the red points, which denote the central charges that have non-vanishing intersecting numbers with $\gamma_1$ or $\gamma_2$, respectively. We use the first 50 terms of the WKB series for the $D_3$ theory and 70 terms for the $D_4$ theory. Due to computational limitations, some red points in the maximal chamber of the $D_4$ theory do not exhibit apparent pole accumulation. This issue could be addressed by counting for higher-order WKB expansions.}
    \label{fig:d3d4pole}
\end{figure}

These TBA equations are verified against quantum periods for consistency, as shown in Table \ref{tab:d3tbawkb}. Similar to the minimal chamber, the TBA equations in the maximal chamber also predict Borel-Pad{\'e} singularities. For instance, the TBA equation for $\tilde{\epsilon}_{\gamma_1}$ reveals singularities of the convolution kernels at $\arg Z_{\gamma_2}=\frac{\pi}{4}$, $\arg Z_{\gamma_3}=\frac{\pi}{2}$, and $\arg Z_{\gamma_4}=\frac{3\pi}{4}$, which are visible in the poles of the Borel-Pad{\'e} transform of the quantum WKB period $\Pi_{\gamma_1}$, as shown in Figure \ref{fig:d3d4pole}. Based on this analysis, we conclude that $\Pi_{\gamma_1}$ is Borel-summbale, while $\Pi_{\gamma_2}$ is not.

\paragraph{Effective central charge} 

The effective central charge corresponding to the TBA equations \eqref{eq:d3maxtba1} can be computed by analyzing the UV behavior of $\tilde{\epsilon}_{\gamma}(\theta)$:
\begin{equation}
    \tilde{\epsilon}_{\gamma_1}^\star= \tilde{\epsilon}_{\gamma_3}^\star=\log 3, \quad  
    \tilde{\epsilon}_{\gamma_2}^\star=
    \tilde{\epsilon}_{\gamma_4}^\star=\log 2.
\end{equation}
\begin{equation}
\begin{aligned}
    c_{\mathrm{eff}}&=2\left[\frac{6}{\pi^2}\int\left|Z_{\gamma_1}\right|\re^\theta \tilde{L}_1(\theta)\rd \theta+\frac{6}{\pi^2}2\left|Z_{\gamma_2}\right|\int\re^\theta \tilde{L}_2(\theta)\rd \theta\right],\\
    &=2\frac{6}{\pi^2}\left(\mathcal{L}_1\left(\frac{1}{1+\re^{\tilde{\epsilon}_1^\star}}\right)+2\mathcal{L}_1\left(\frac{1}{1+\re^{\tilde{\epsilon}_2^\star}}\right)\right)=2.
    \end{aligned}
\end{equation}
An overall factor 2 is added to account for the contributions from $\gamma_3$ and $\gamma_4$. This result matches the effective central charge in \cite{Zam91} for $D_3$ \footnote{This is the effective central charge associated with the TBA system in \eqref{eq:d3maxsym}, which is straightforwardly evaluated to be 1.}, up to a normalization factor of 2. This computation, along with \eqref{eq:d3mineff}, confirms that the effective central charge remains invariant during the wall-crossing. 

\paragraph{Deformation by $\ell$}

The deformation of the TBA equation given by \eqref{eq:d3mintba3} motivates us to generalize the TBA system \eqref{eq:d3maxtba1} away from $\ell=-\frac{1}{2}$. Based on the experience in the minimal chamber, we separate two doublets and write down the TBA equations for $\ell \in (-1,0)$ as follows: 
\begin{equation}
\begin{aligned}
    \tilde{\epsilon}_{\gamma_1}(\theta)&=\left|Z_{\gamma_1}\right| \mathrm{e}^\theta- K_{1,2}\star \left(\tilde{L}_2^++\tilde{L}_2^-\right)- 2 K_{1,3}\star \tilde{L}_{3}-K_{1,4}\star \left(\tilde{L}_4^++\tilde{L}_4^-\right),\\
     \tilde{\epsilon}_{\gamma_2}(\theta)&=\left|Z_{\gamma_2}\right| \mathrm{e}^\theta+ K_{2,1}\star \tilde{L}_{1}- K_{2,3}\star \tilde{L}_{3}-  K_{2,4}\star \left(\tilde{L}_4^++\tilde{L}_4^-\right),\\
     \tilde{\epsilon}_{\gamma_3}(\theta)&=\left|Z_{\gamma_3}\right| \mathrm{e}^\theta+ 2 K_{3,1}\star \tilde{L}_{1}+ K_{3,2}\star \left(\tilde{L}_2^++\tilde{L}_2^-\right)-K_{3,4}\star \left(\tilde{L}_4^++\tilde{L}_4^-\right),\\
      \tilde{\epsilon}_{\gamma_4}(\theta)&=\left|Z_{\gamma_4}\right| \mathrm{e}^\theta+ K_{4,1}\star \tilde{L}_{1}+K_{4,2}\star \left(\tilde{L}_2^++\tilde{L}_2^-\right)+ K_{4,3}\star \tilde{L}_{3},
\end{aligned}
\end{equation}
where
\begin{equation}
\begin{aligned}
     &\tilde{L}_{2}^+=\log\left(1-e^{2\pi i \ell}e^{-\tilde{\epsilon}_{\gamma_2}}(\theta)\right), \quad \tilde{L}_{2}^-=\log\left(1-e^{-2\pi i \ell}e^{-\tilde{\epsilon}_{\gamma_2}}(\theta)\right), \\
      &\tilde{L}_{4}^+=\log\left(1-e^{2\pi i \ell}e^{-\tilde{\epsilon}_{\gamma_4}}(\theta)\right), \quad \tilde{L}_{4}^-=\log\left(1-e^{-2\pi i \ell}e^{-\tilde{\epsilon}_{\gamma_4}}(\theta)\right).
\end{aligned}
\end{equation}
Numerical computation is performed for $\ell=-\frac{1}{5}$, as shown in Table \ref{tab:d3tbawkb}, which confirms the validity of the $\ell$ deformation in the maximal chamber. It indicates that the analytic continuation on $\ell$ and wall-crossing of the moduli parameters are independent.

In this section, we derive the TBA systems for the $(A_1, D_3)$ theory with SU(2) symmetry. While $D_3$ is identical to $A_3$, whose TBA systems are completely established, our analysis here provides a generalized $\ell$ deformation and offers insight into establishing the TBA equations for generic D-type AD theories, particularly their wall-crossing behavior.

\section{$(A_1,D_4)$ theory}
\label{sc:d4}

In this section, we consider the $(A_1, D_4)$ theory, its Seiberg-Witten differential $\lambda$ is determined by
\begin{equation}
\lambda^2=\left(z^2+u_1 z+u_2+\frac{u_3}{z}+\frac{m^2}{z^2}\right) \mathrm{d} z^2.
\end{equation}
Similar to the $D_3$ case, the residue at $z=0$, denoted by $m$, is associated with an SU(2) flavor symmetry. The residue of the above SW curve at $z=\infty$ is $\frac{1}{8}\left(u_1^2-4u_2\right)$, which corresponds to a U(1) flavor symmetry. We always consider the $m\to 0$ limit. Furthermore, if $\frac{1}{8}\left(u_1^2-4u_2\right)=0$, the flavor symmetry is enhanced to SU(3) from SU(2)$\times$U(1). We will construct TBA equations for the SU(3) case and subsequently generalize them to the SU(2)$\times$U(1) symmetry. The marginal stability curves and chambers are more complicated than the $D_3$ case. We will begin in the minimal chamber and proceed to the maximal chamber through subsequent wall-crossings.

\subsection{TBA equations in the minimal chamber} 

There are four turning points in the limit $m\to 0$, one colliding with the origin. In the minimal chamber, we can order all these turning points in the real axis as
\begin{equation}
    a_3<a_2<a_1<0.
\end{equation}
One can construct three independent cycles, $\gamma_1$ encircling turning points $a_1$ and $a_2$, $\gamma_2$ encircling $a_2$ and $a_3$, $\gamma_3$ encircling $0$ and $a_1$. The orientations of the cycles will be specified in the concrete computations for the given moduli parameters.

\begin{figure}[htbp]
\centering
    \begin{subfigure}{0.45\linewidth}
        \centering
       \includegraphics[width=0.9\linewidth]{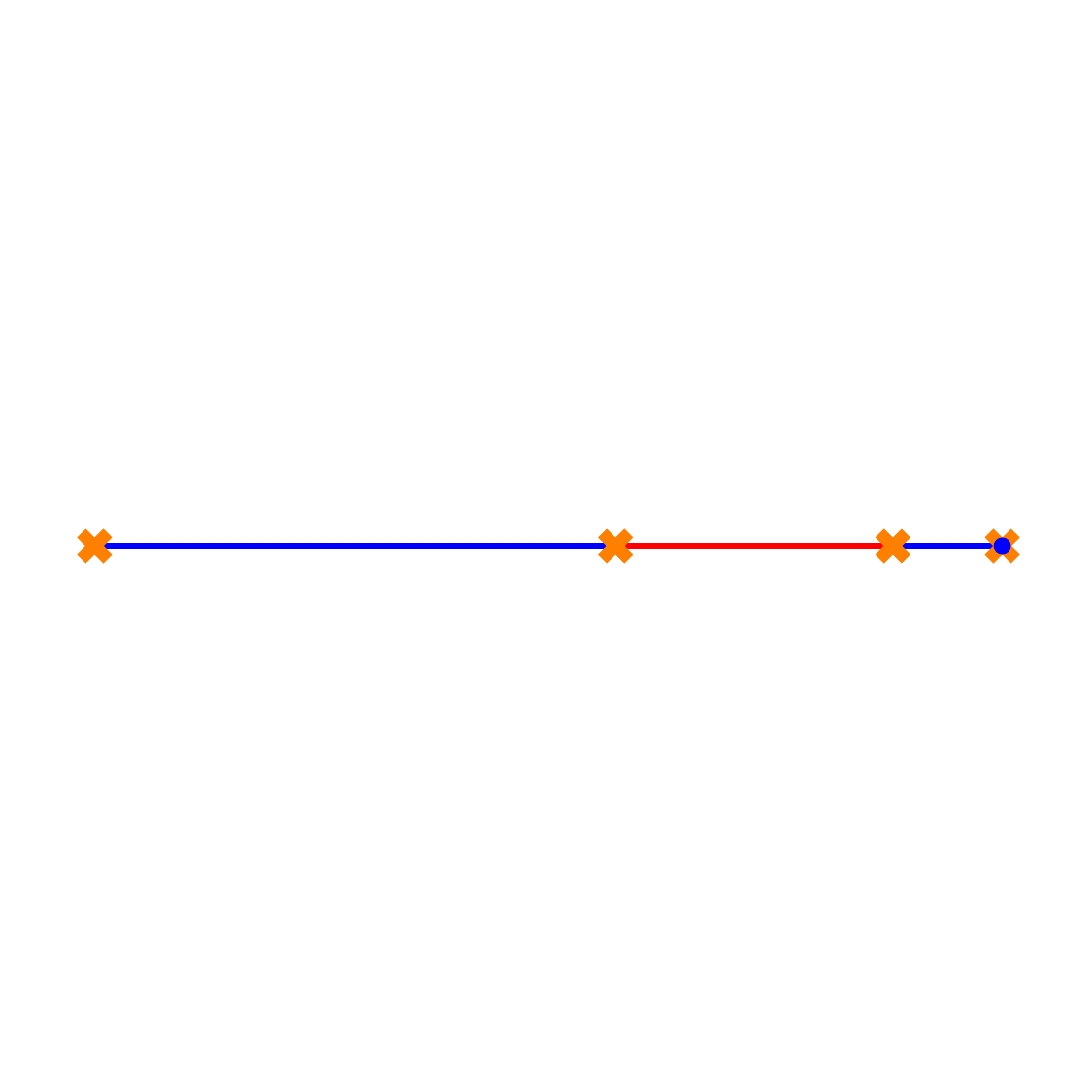}
        \caption{BPS states}
        \label{fig:d4minsu3bpsplot}
    \end{subfigure}
    \begin{subfigure}{0.45\linewidth}
        \centering
       \includegraphics[width=0.9\linewidth]{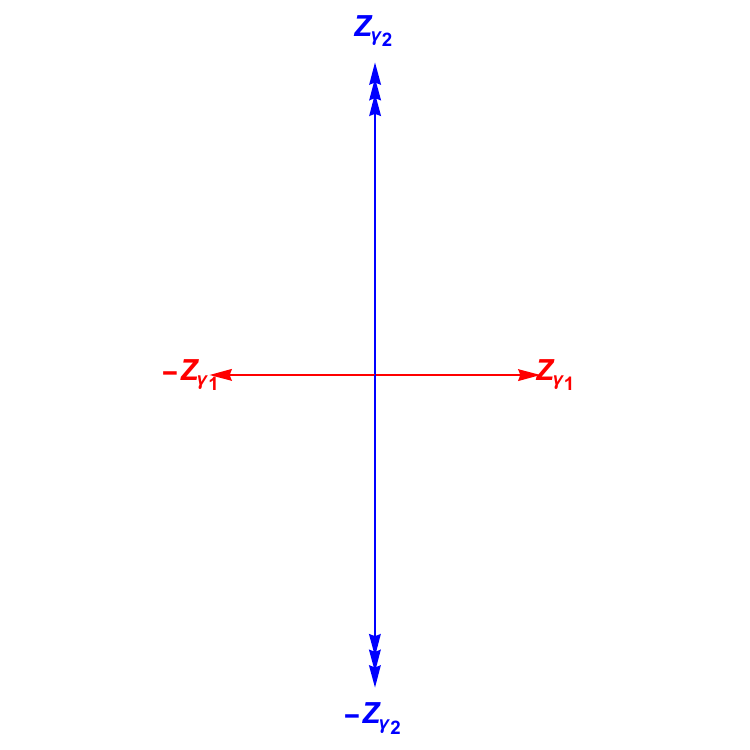}
        \caption{Central charges}
        \label{fig:d4minsu3bpscharge}
    \end{subfigure}

    \caption{BPS spectrum in the minimal chamber for $(A_1,D_4)$ with $SU(3)$ flavor symmetry. The triple arrow indicates a triplet.}
    \label{fig:d4minsu3bps}
\end{figure}

\subsubsection{SU(3) symmetry} 

We choose the parameters such that $u_1^2-4 u_2=0$, which implies $Z_{\gamma_2}=Z_{\gamma_3}$ and enhances the flavor symmetry to SU(3). In this case, $\gamma_2$ and $\gamma_3$ become equivalent, forming a triplet, while $\gamma_1$ remains a singlet. We denote the electromagnetic charges of the singlet $\gamma_1$ and the triplet $\gamma_2$ as 

\begin{equation}
    \gamma_1=(0, 1), \quad   \gamma_2=(1, 0).
\end{equation}

The BPS indices are:

\begin{equation}
\Omega(\gamma)= \begin{cases}
3 & \text { for } \gamma \in
\pm\gamma_2,\\ 
1 & \text { for } \gamma \in
\pm\gamma_1,
 \\
0 & \text{ otherwise.}
\end{cases}
\end{equation}

The TBA equations are expressed as

\begin{equation}
\begin{aligned}
    \tilde{\epsilon}_{\gamma_1}(\theta)&=\left|Z_{\gamma_1}\right| \mathrm{e}^\theta+\frac{1}{2 \pi \mathrm{i}} 3 \int_{\mathbb{R}} \frac{\log \left(1+\mathrm{e}^{-\tilde{\epsilon}_{\gamma_2}\left(\theta^{\prime}\right)}\right)}{\sinh \left(\theta-\theta^{\prime}+\mathrm{i} \phi_{\gamma_2}-\mathrm{i} \phi_{\gamma_1}\right)} \mathrm{d} \theta^{\prime},\\
     \tilde{\epsilon}_{\gamma_2}(\theta)&=\left|Z_{\gamma_2}\right| \mathrm{e}^\theta-\frac{1}{2 \pi \mathrm{i}}  \int_{\mathbb{R}} \frac{\log \left(1+\mathrm{e}^{-\tilde{\epsilon}_{\gamma_1}\left(\theta^{\prime}\right)}\right)}{\sinh \left(\theta-\theta^{\prime}+\mathrm{i} \phi_{\gamma_1}-\mathrm{i} \phi_{\gamma_2}\right)} \mathrm{d} \theta^{\prime}.
\end{aligned}
\end{equation}
The factor $3$ represents the degeneracy of the BPS state $\Omega(\gamma_2)=3$. We select moduli parameters such that $\arg Z_{\gamma_1}=0$ and $\arg Z_{\gamma_2}=\frac{\pi}{2}$, as illustrated in Figure \ref{fig:d4minsu3bps}. The TBA equations are formulated as follows:
\begin{equation}
\label{eq:d4mintba1}
\begin{aligned}
    \tilde{\epsilon}_{\gamma_1}(\theta)&=\left|Z_{\gamma_1}\right| \mathrm{e}^\theta-\frac{1}{2 \pi } 3\int_{\mathbb{R}} \frac{\log \left(1+\mathrm{e}^{-\tilde{\epsilon}_{\gamma_2}\left(\theta^{\prime}\right)}\right)}{\cosh \left(\theta-\theta^{\prime}\right)} \mathrm{d} \theta^{\prime},\\
     \tilde{\epsilon}_{\gamma_2}(\theta)&=\left|Z_{\gamma_2}\right| \mathrm{e}^\theta-\frac{1}{2 \pi } \int_{\mathbb{R}} \frac{\log \left(1+\mathrm{e}^{-\tilde{\epsilon}_{\gamma_1}\left(\theta^{\prime}\right)}\right)}{\cosh \left(\theta-\theta^{\prime}\right)} \mathrm{d} \theta^{\prime}.
\end{aligned}
\end{equation}

Introducing the $\ell$ deformation to the quantum SW curve \eqref{eq:leq} transforms the first TBA equation in \eqref{eq:d4mintba1} into \eqref{eq:d4mintba3}. This indicates that the SU(3) symmetry is preserved only when $\ell=-\frac{1}{2}$. Deviations from $-\frac{1}{2}$ break the equality $\Pi_{\gamma_2}(\hbar)=\Pi_{\gamma_3}(\hbar)$ due to quantum corrections, thereby breaking the TBA equations from SU(3) type to SU(2) type.

\paragraph{Effective central charge}

Similarly, the UV behavior of $\tilde{\epsilon}_{\gamma}(\theta)$ can be described as
\begin{equation}
    \tilde{\epsilon}_{\gamma_1}^\star=-3\log 2, \quad   \tilde{\epsilon}_{\gamma_2}^\star=-\log 3.
\end{equation}
The effective central charge is then evaluated as
\begin{equation}
\label{eq:d4mineff}
\begin{aligned}
    c_{\mathrm{eff}}&=\frac{6}{\pi^2}\int\left|Z_{\gamma_1}\right|\re^\theta \tilde{L}_1(\theta)\rd \theta+\frac{6}{\pi^2}3\left|Z_{\gamma_2}\right|\int\re^\theta \tilde{L}_2(\theta)\rd \theta,\\
    &=\frac{6}{\pi^2}\left(\mathcal{L}_1\left(\frac{1}{1+\re^{\tilde{\epsilon}_{\gamma_1}^\star}}\right)+3\mathcal{L}_1\left(\frac{1}{1+\re^{\tilde{\epsilon}_{\gamma_2}^\star}}\right)\right)=3,
    \end{aligned}
\end{equation}
where the factor 3 accounts for the triplet $\gamma_2$.

\subsubsection{SU(2)$\times$U(1) symmetry}

If we relax the condition $u_1^2-4 u_2=0$, the triplet $\gamma_2$ separates into a singlet $\gamma_2$ (Using the same notation for triplet $\gamma_2$ in the SU(3) case) and a doublet $\gamma_3$, with the following charges:
\begin{equation}
    \gamma_1=(0, 1), \quad   \gamma_2=(1, 0),\quad \gamma_3=(1, 0).
\end{equation}
The BPS indices are:
\begin{equation}
\Omega(\gamma)= \begin{cases}
2 & \text{ for } \gamma \in
\pm\gamma_3,\\ 

1 & \text{ for } \gamma \in
\left\{\pm\gamma_1,\pm\gamma_2\right\},
 \\

0 & \text{ otherwise. }
\end{cases}
\end{equation}
This leads to the TBA equations:
\begin{equation}
\label{eq:d4mintba4}
\begin{aligned}
    \tilde{\epsilon}_{\gamma_1}(\theta)&=\left|Z_{\gamma_1}\right| \mathrm{e}^\theta+\frac{1}{2 \pi \mathrm{i}}  \int_{\mathbb{R}} \frac{\log \left(1+\mathrm{e}^{-\tilde{\epsilon}_{\gamma_2}\left(\theta^{\prime}\right)}\right)}{\sinh \left(\theta-\theta^{\prime}+\mathrm{i} \phi_{\gamma_2}-\mathrm{i} \phi_{\gamma_1}\right)} \mathrm{d} \theta^{\prime}+\frac{1}{2 \pi \mathrm{i}} 2\int_{\mathbb{R}} \frac{\log \left(1+\mathrm{e}^{-\tilde{\epsilon}_{\gamma_3}\left(\theta^{\prime}\right)}\right)}{\sinh \left(\theta-\theta^{\prime}+\mathrm{i} \phi_{\gamma_3}-\mathrm{i} \phi_{\gamma_1}\right)} \mathrm{d} \theta^{\prime},\\
     \tilde{\epsilon}_{\gamma_2}(\theta)&=\left|Z_{\gamma_2}\right| \mathrm{e}^\theta-\frac{1}{2 \pi \mathrm{i}}  \int_{\mathbb{R}} \frac{\log \left(1+\mathrm{e}^{-\tilde{\epsilon}_{\gamma_1}\left(\theta^{\prime}\right)}\right)}{\sinh \left(\theta-\theta^{\prime}+\mathrm{i} \phi_{\gamma_1}-\mathrm{i} \phi_{\gamma_2}\right)} \mathrm{d} \theta^{\prime},\\
      \tilde{\epsilon}_{\gamma_3}(\theta)&=\left|Z_{\gamma_3}\right| \mathrm{e}^\theta-\frac{1}{2 \pi \mathrm{i}}  \int_{\mathbb{R}} \frac{\log \left(1+\mathrm{e}^{-\tilde{\epsilon}_{\gamma_1}\left(\theta^{\prime}\right)}\right)}{\sinh \left(\theta-\theta^{\prime}+\mathrm{i} \phi_{\gamma_1}-\mathrm{i} \phi_{\gamma_3}\right)} \mathrm{d} \theta^{\prime}.
\end{aligned}
\end{equation}
In the minimal chamber, we can choose moduli parameters such that $\arg Z_{\gamma_1}=0$ and $\arg Z_{\gamma_2}=\arg Z_{\gamma_3}=\frac{\pi}{2}$. Under these conditions, the TBA system simplifies to
\begin{equation}
\label{eq:d4mintba2}
\begin{aligned}
    \tilde{\epsilon}_{\gamma_1}(\theta)&=\left|Z_{\gamma_1}\right| \mathrm{e}^\theta-\frac{1}{2 \pi } \int_{\mathbb{R}} \frac{\log \left(1+\mathrm{e}^{-\tilde{\epsilon}_{\gamma_2}\left(\theta^{\prime}\right)}\right)}{\cosh \left(\theta-\theta^{\prime}\right)} \mathrm{d} \theta^{\prime}-\frac{1}{2 \pi } 2\int_{\mathbb{R}} \frac{\log \left(1+\mathrm{e}^{-\tilde{\epsilon}_{\gamma_3}\left(\theta^{\prime}\right)}\right)}{\cosh \left(\theta-\theta^{\prime}\right)} \mathrm{d} \theta^{\prime},\\
     \tilde{\epsilon}_{\gamma_2}(\theta)&=\left|Z_{\gamma_2}\right| \mathrm{e}^\theta-\frac{1}{2 \pi }  \int_{\mathbb{R}} \frac{\log \left(1+\mathrm{e}^{-\tilde{\epsilon}_{\gamma_1}\left(\theta^{\prime}\right)}\right)}{\cosh \left(\theta-\theta^{\prime}\right)} \mathrm{d} \theta^{\prime},\\
     \tilde{\epsilon}_{\gamma_3}(\theta)&=\left|Z_{\gamma_3}\right| \mathrm{e}^\theta-\frac{1}{2 \pi }  \int_{\mathbb{R}} \frac{\log \left(1+\mathrm{e}^{-\tilde{\epsilon}_{\gamma_1}\left(\theta^{\prime}\right)}\right)}{\cosh \left(\theta-\theta^{\prime}\right)} \mathrm{d} \theta^{\prime}.
\end{aligned}
\end{equation}
As in the $(A_1, D_3)$ case, the degeneracy of the doublet $\gamma_3$ can be decomposed through $\ell$ deformation. This results in the first equation splitting into
\begin{equation}
\label{eq:d4mintba3}
\begin{aligned}
     \tilde{\epsilon}_{\gamma_1}(\theta)&=\left|Z_{\gamma_1}\right| \mathrm{e}^\theta-\frac{1}{2 \pi } \int_{\mathbb{R}} \frac{\log \left(1+\mathrm{e}^{-\tilde{\epsilon}_{\gamma_2}\left(\theta^{\prime}\right)}\right)}{\cosh \left(\theta-\theta^{\prime}\right)} \mathrm{d} \theta^{\prime},\\
     &-\frac{1}{2 \pi } \int_{\mathbb{R}} \frac{\log \left(1-\re^{2\pi \ri \ell}\mathrm{e}^{-\tilde{\epsilon}_{\gamma_3}\left(\theta^{\prime}\right)}\right)}{\cosh \left(\theta-\theta^{\prime}\right)} \mathrm{d} \theta^{\prime}-\frac{1}{2 \pi } \int_{\mathbb{R}} \frac{\log \left(1-\re^{-2\pi \ri \ell}\mathrm{e}^{-\tilde{\epsilon}_{\gamma_3}\left(\theta^{\prime}\right)}\right)}{\cosh \left(\theta-\theta^{\prime}\right)} \mathrm{d} \theta^{\prime}.
     \end{aligned}
\end{equation}
This corresponds to the TBA equations in \cite{IS19} for $r=1$. The TBA system is related to the Dynkin diagram of the $D_4$ Lie algebra, as shown in Figure \ref{fig:dynkin} for $N=2$. It is evident that the effective central charge for TBA system \eqref{eq:d4mintba4} is $c_{\mathrm{eff}}=3$, which is the same with \eqref{eq:d4mineff}.

\subsection{TBA equations in the intermediate chambers}
\begin{figure}[htbp]
\centering
    \begin{subfigure}{0.24\linewidth}
        \centering
        \includegraphics[width=1.0\linewidth]{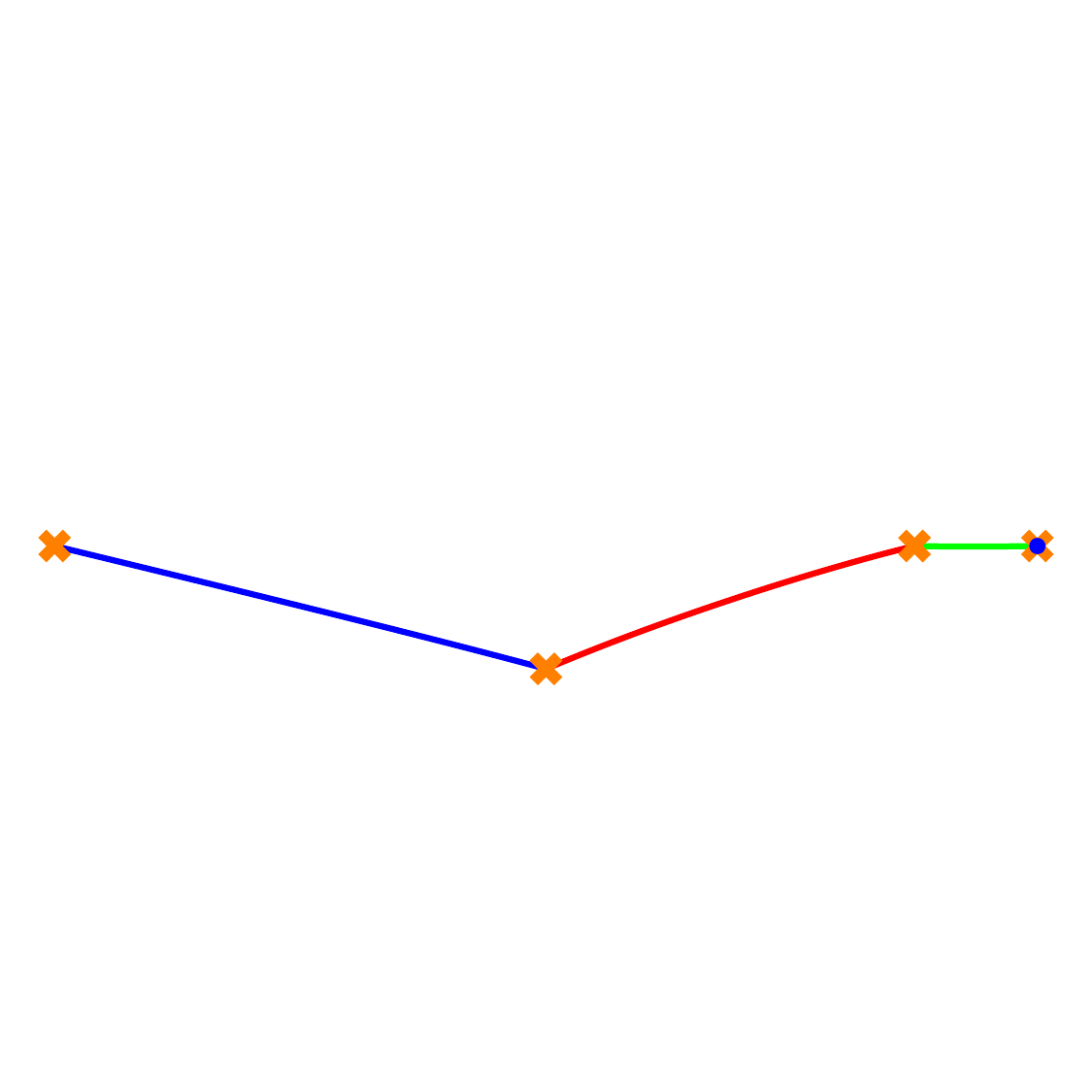}
        \caption{Minimal chamber}
        \label{fig:d4cham1bpsplot}
    \end{subfigure}
    \begin{subfigure}{0.24\linewidth}
        \centering
        \includegraphics[width=1.0\linewidth]{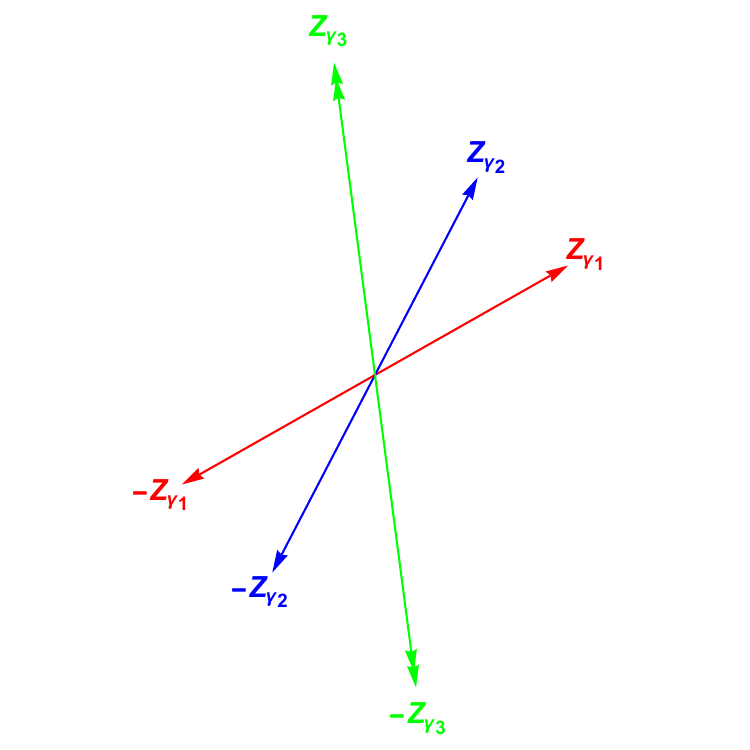}
        \caption{Minimal chamber}
        \label{fig:d4cham1bpscharge}
    \end{subfigure}
     \begin{subfigure}{0.24\linewidth}
        \centering
        \includegraphics[width=1.0\linewidth]{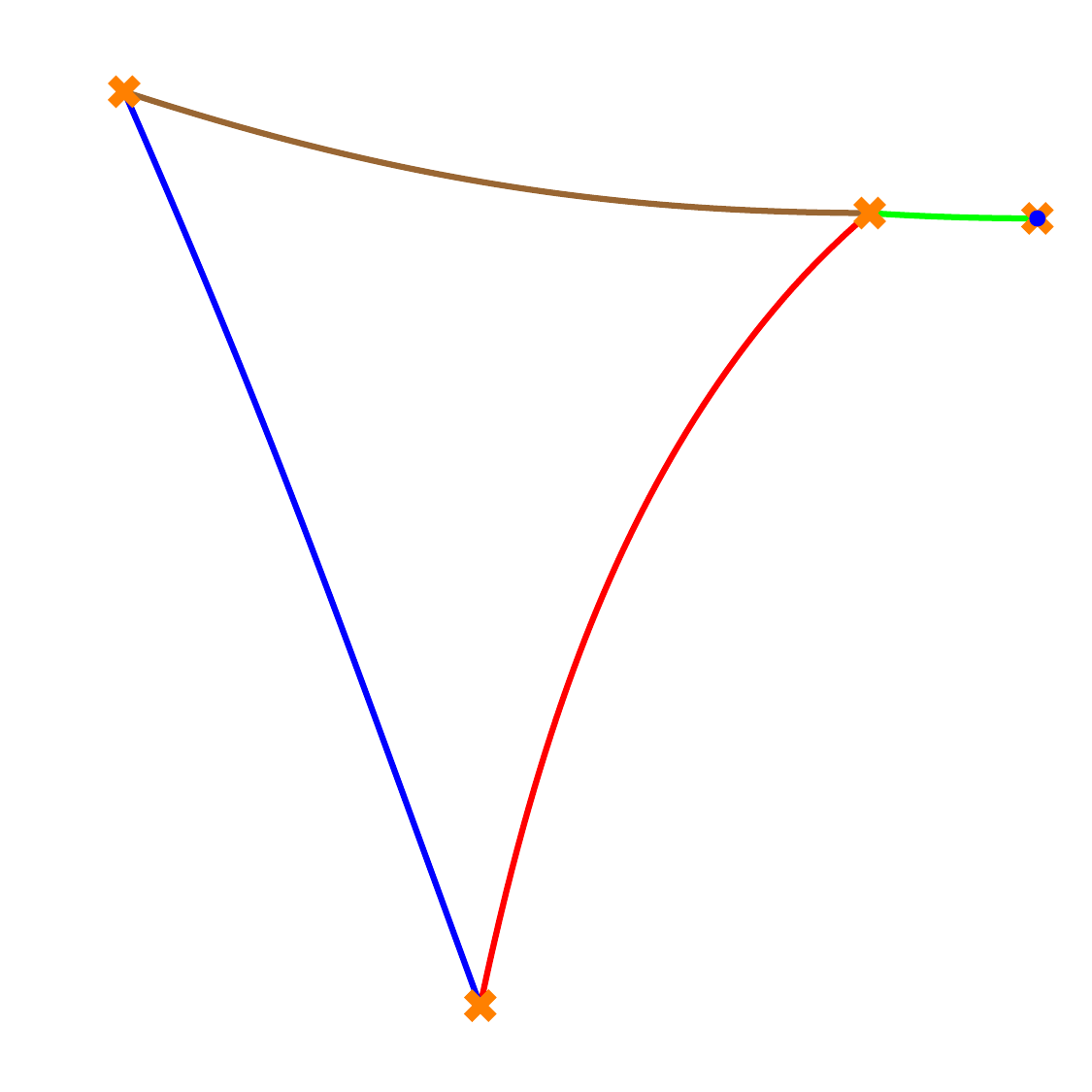}
        \caption{Chamber \RN{1}}
        \label{fig:d4cham2bpsplot}
    \end{subfigure}
    \begin{subfigure}{0.24\linewidth}
        \centering
        \includegraphics[width=1.0\linewidth]{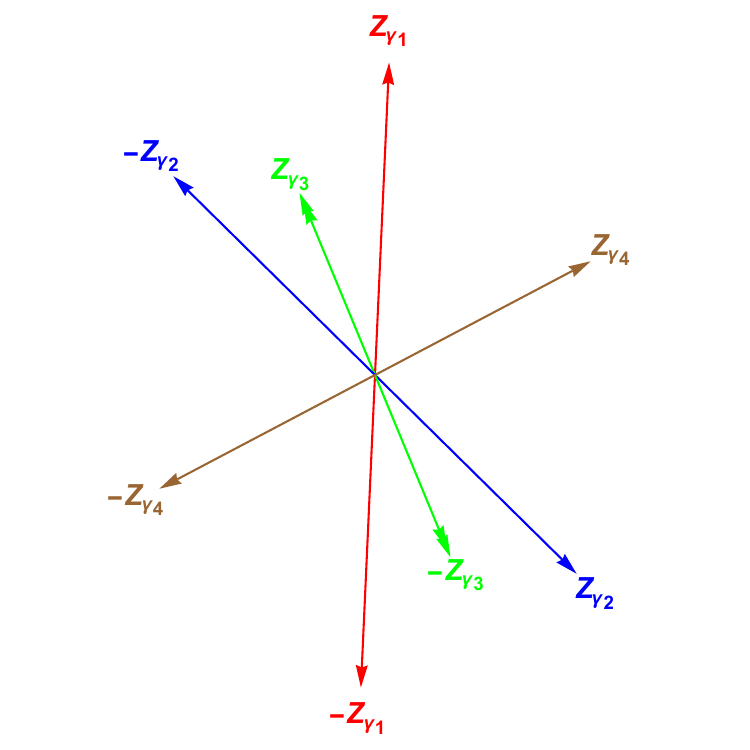}
        \caption{Chamber \RN{1}}
        \label{fig:d4cham2bpscharge}
    \end{subfigure}

     \begin{subfigure}{0.24\linewidth}
        \centering
        \includegraphics[width=1.0\linewidth]{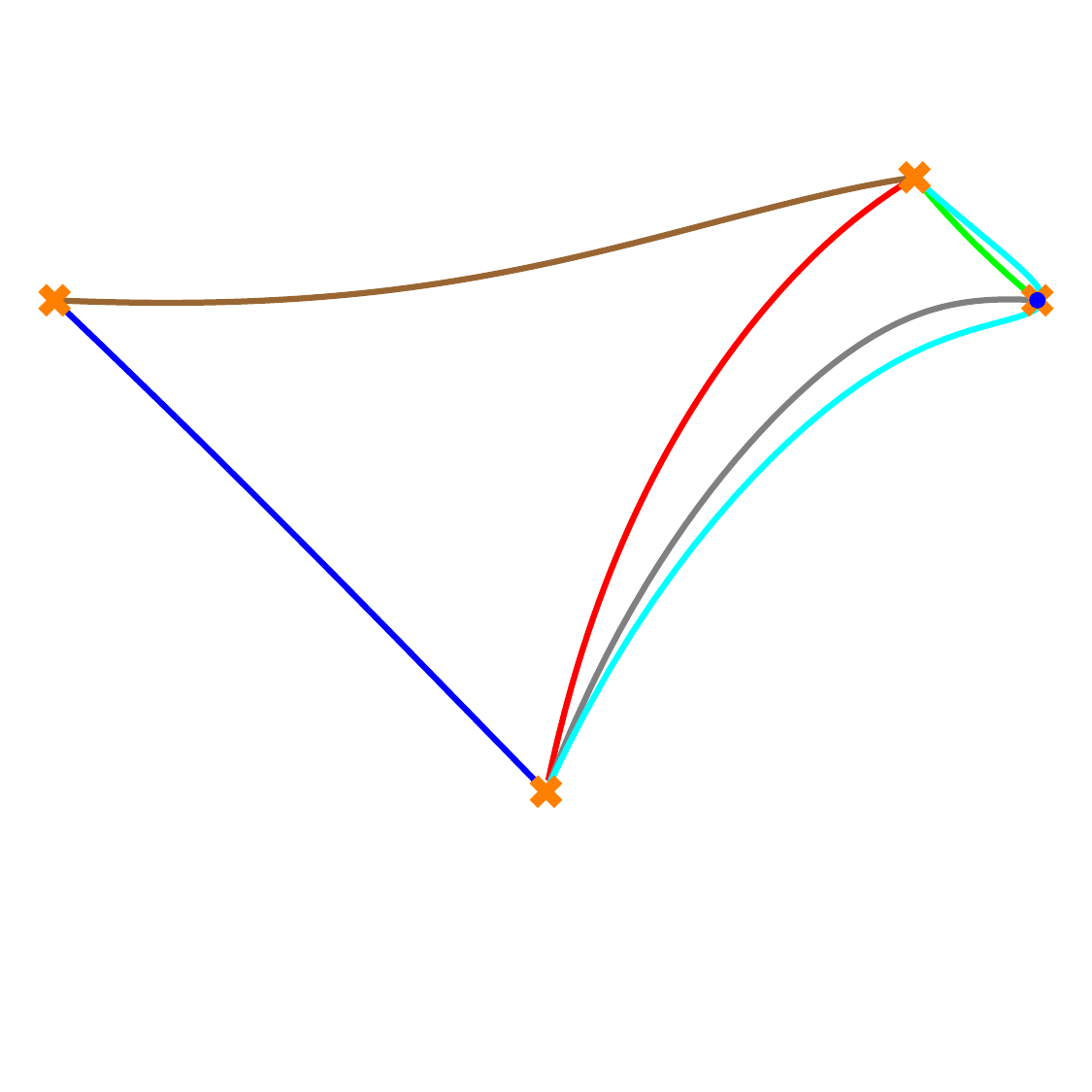}
        \caption{Chamber \RN{2}}
        \label{fig:d4cham3bpsplot}
    \end{subfigure}
    \begin{subfigure}{0.24\linewidth}
        \centering
        \includegraphics[width=1.0\linewidth]{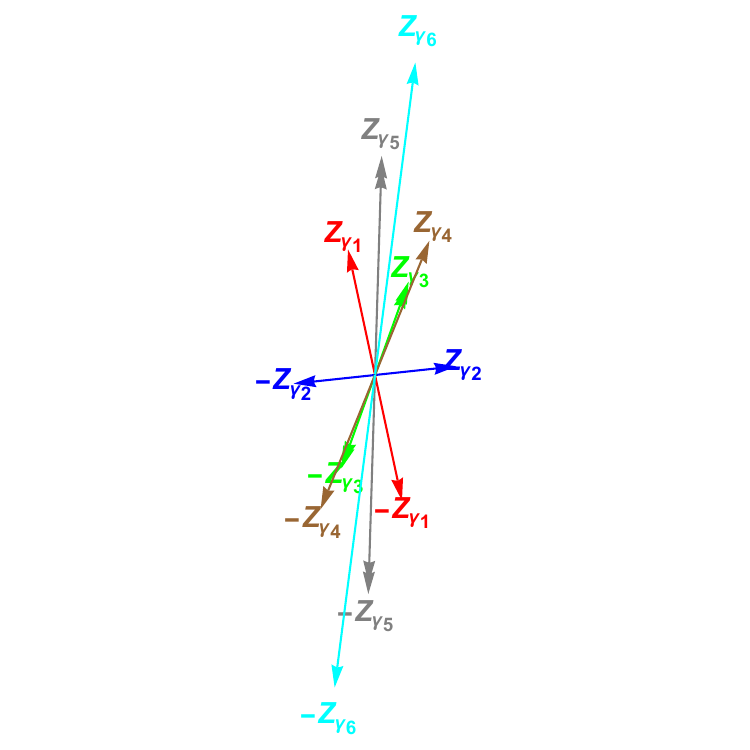}
        \caption{Chamber \RN{2}}
        \label{fig:d4cham3bpscharge}
    \end{subfigure}
     \begin{subfigure}{0.24\linewidth}
        \centering
        \includegraphics[width=1.0\linewidth]{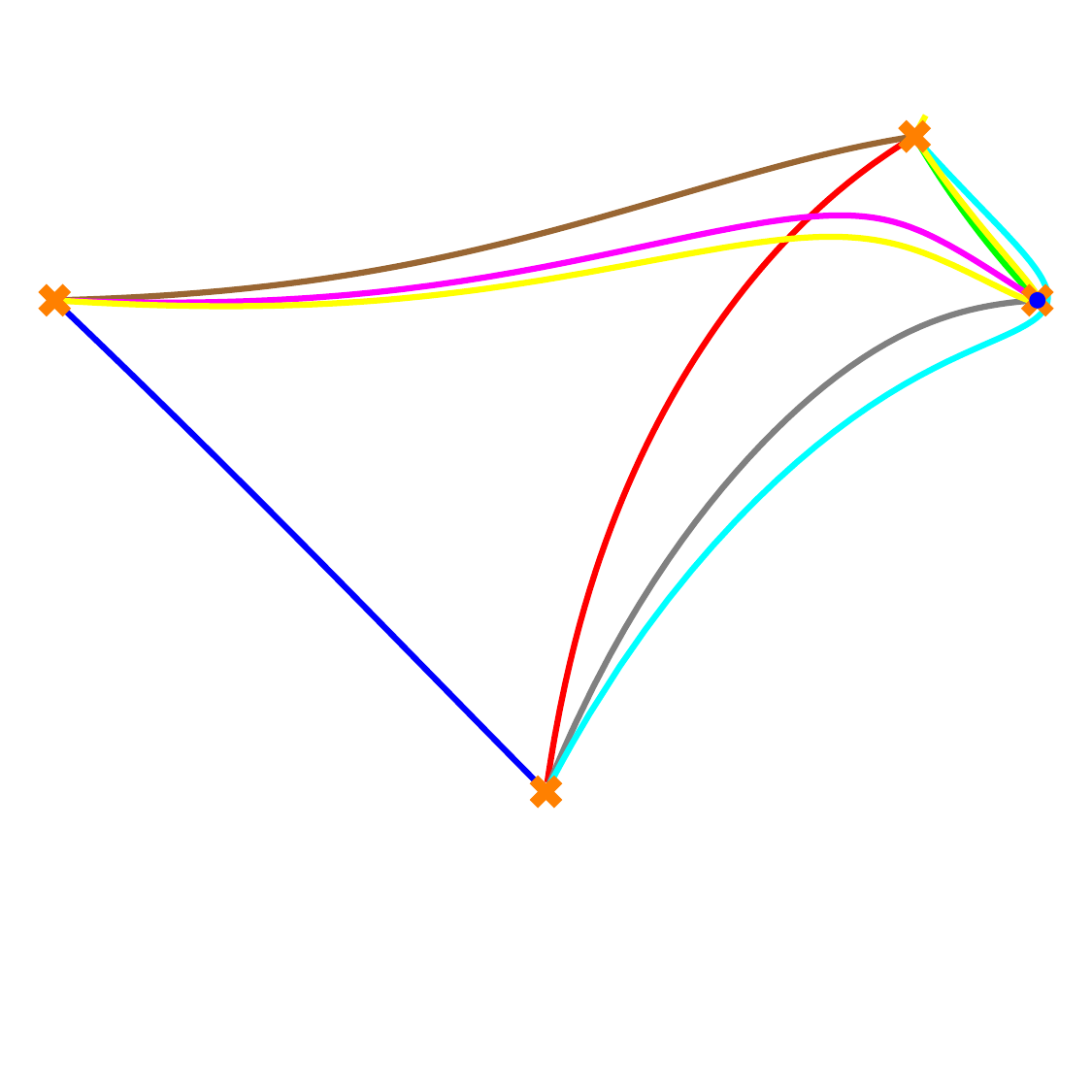}
        \caption{Chamber \RN{3}}
        \label{fig:d4cham4bpsplot}
    \end{subfigure}
    \begin{subfigure}{0.24\linewidth}
        \centering
        \includegraphics[width=1.0\linewidth]{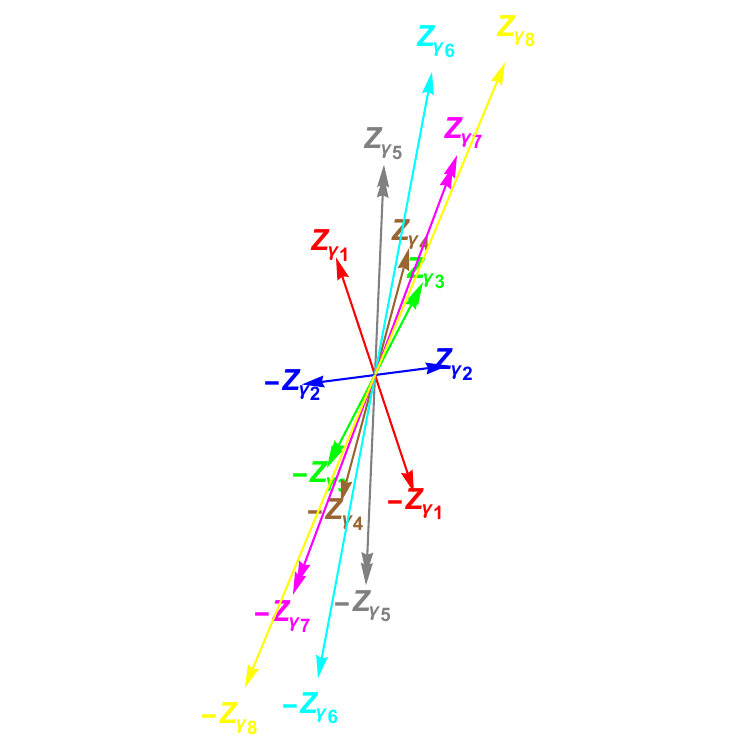}
        \caption{Chamber \RN{3}}
        \label{fig:d4cham4bpscharge}
    \end{subfigure}

     \begin{subfigure}{0.24\linewidth}
        \centering
        \includegraphics[width=1.0\linewidth]{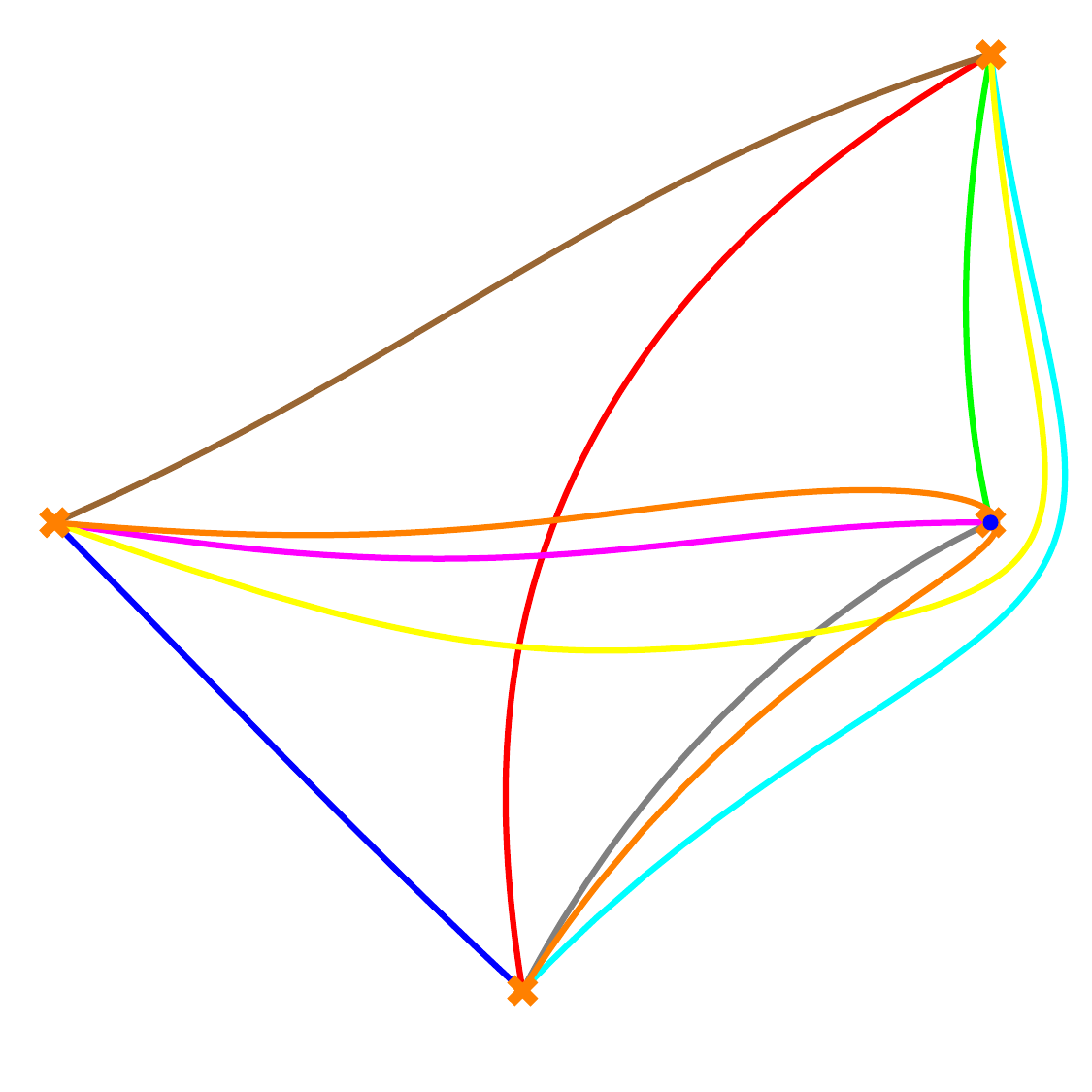}
        \caption{Maximal chamber}
        \label{fig:d4cham5bpsplot}
    \end{subfigure}
    \begin{subfigure}{0.24\linewidth}
        \centering
        \includegraphics[width=1.0\linewidth]{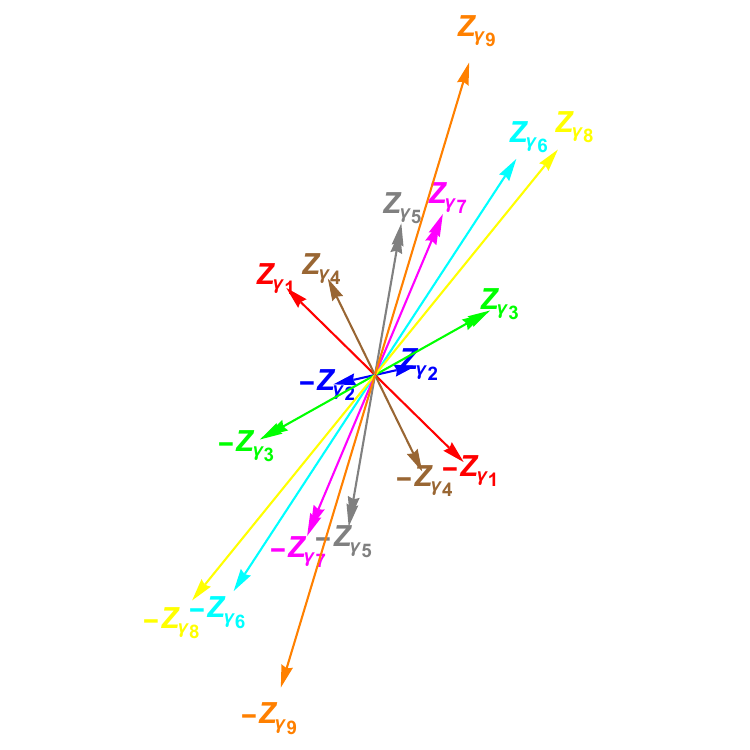}
        \caption{Maximal chamber}
        \label{fig:d4cham5bpscharge}
    \end{subfigure}

    \caption{BPS spectrum and wall-crossing of $(A_1,D_4)$ with SU(2)$\times$U(1) symmetry.}
    \label{fig:d4chambps}
\end{figure}

We consider the wall-crossing along a path from the minimal chamber to the maximal chamber, with a typical process illustrated in Figure \ref{fig:d4chambps}. We start with the BPS spectrum in the minimal chamber with SU(2)$\times$U(1) symmetry as depicted in Figures \ref{fig:d4cham1bpsplot} and \ref{fig:d4cham1bpscharge}\footnote{For SU(3) symmetry, there are only minimal chamber and maximal chambers, with no intermediate chamber. Hence, this discussion is limited to SU(2)$\times$U(1) symmetry.}. As we move through the moduli space, the phase of central charges changes smoothly. We adjust the moduli parameters so that the phase of $Z_{\gamma_1}$ increases while the phase of $Z_{\gamma_2}$ decreases. We encounter a marginal stability curve when $\arg Z_{\gamma_1}=\arg Z_{\gamma_2}$, leading to a type A wall-crossing. As we continue, we reach an intermediate chamber (chamber \RN{1}), where $\arg Z_{\gamma_1}>\arg Z_{\gamma_2}$, and
a new singlet $\gamma_4=(1,1)$ appears as shown in Figures \ref{fig:d4cham2bpsplot} and \ref{fig:d4cham2bpscharge}. From the chamber \RN{1}, we proceed to another chamber (Chamber \RN{2}), where a doublet $\gamma_5=(1,1)$ and a singlet $\gamma_6=(2,1)$ are created, as indicated in Figures \ref{fig:d4cham3bpsplot} and \ref{fig:d4cham3bpscharge}. We then arrive at the intermediate chamber \RN{3} where a doublet $\gamma_7=(2,1)$ and a singlet $\gamma_8=(3,1)$ are created, as depicted in Figures \ref{fig:d4cham4bpsplot} and \ref{fig:d4cham4bpscharge}. Through another wall-crossing, we reach the maximal chamber, where an extra singlet $\gamma_9=(3,2)$ appears. 

\begin{figure}[htbp]
\centering
    \begin{subfigure}{0.45\linewidth}
        \centering
        \includegraphics[width=0.9\linewidth]{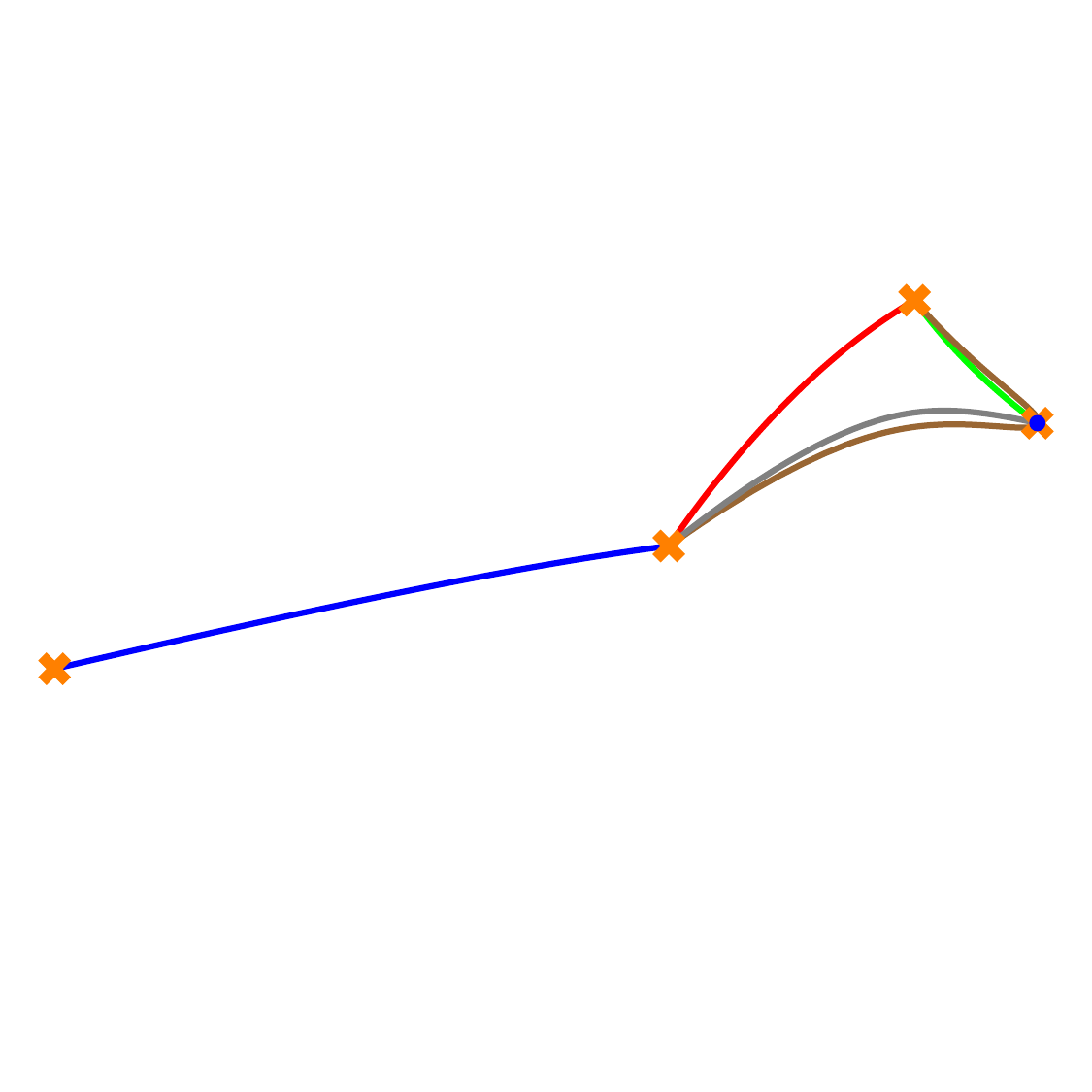}
        \caption{BPS states}
        \label{fig:d4cham6bpsplot}
    \end{subfigure}
    \begin{subfigure}{0.45\linewidth}
        \centering
        \includegraphics[width=0.9\linewidth]{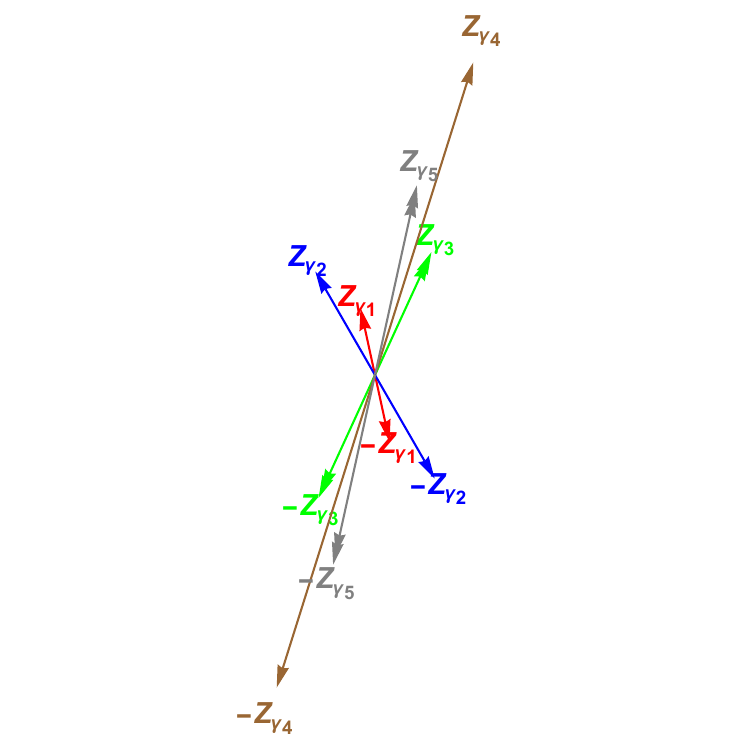}
        \caption{Central charges}
        \label{fig:d4cham6bpscharge}
    \end{subfigure}
    \caption{BPS spectrum in the intermediate chamber \RN{4} of $(A_1,D_4)$ with U(1)$\times$SU(2) flavor symmetry.}
    \label{fig:d4midbps}
\end{figure}

However, these are not all the chambers for $D_4$ theory. An additional example is shown in Figure \ref{fig:d4midbps}. We reach this additional chamber (chamber \RN{4}) through a type D wall-crossing from the minimal chamber, as $\arg Z_{\gamma_1}$ and $\arg Z_{\gamma_3}$ align. This wall-crossing creates two new BPS states, a doublet $\gamma_5=(1,1)$ and a singlet $\gamma_4=(2,1)=\gamma_3+\gamma_5$. The phases of their central charges lie between those of $\gamma_1$ and $\gamma_3$.

Here we explicitly write down the TBA equations for the intermediate chamber \RN{1} and \RN{4}. The TBA equations for the other intermediate chambers can be easily deduced from their charges and BPS indices.
In chamber \RN{1}, there are four BPS states, with three singlets and one doublet:
\begin{equation}
    \gamma_1=(0, 1), \quad   \gamma_2=(1, 0),\quad \gamma_3=(1, 0), \quad   \gamma_4=(1, 1).
\end{equation}
The BPS indices are:
\begin{equation}
\Omega(\gamma)= \begin{cases}
2 & \text { for } \gamma \in
\pm\gamma_3,\\ 

1 & \text{ for } \gamma \in
\left\{\pm\gamma_1,\pm\gamma_2,\pm\gamma_4\right\},
 \\

0 & \text{ otherwise. }
\end{cases}
\end{equation}
TBA equations are written as
\begin{equation}
\begin{aligned}
    \tilde{\epsilon}_{\gamma_1}(\theta)&=\left|Z_{\gamma_1}\right| \mathrm{e}^\theta+ K_{1,2}\star \tilde{L}_2+2 K_{1,3}\star \tilde{L}_3+ K_{1,4}\star \tilde{L}_4,\\
     \tilde{\epsilon}_{\gamma_2}(\theta)&=\left|Z_{\gamma_2}\right| \mathrm{e}^\theta- K_{2,1}\star \tilde{L}_1-  K_{2,4}\star \tilde{L}_4,\\
      \tilde{\epsilon}_{\gamma_3}(\theta)&=\left|Z_{\gamma_3}\right| \mathrm{e}^\theta- K_{3,1}\star \tilde{L}_1- K_{3,4}\star \tilde{L}_4,\\
       \tilde{\epsilon}_{\gamma_4}(\theta)&=\left|Z_{\gamma_4}\right| \mathrm{e}^\theta- K_{4,1}\star \tilde{L}_1+K_{4,2}\star \tilde{L}_2+ 2 K_{4,3}\star \tilde{L}_3.
\end{aligned}
\end{equation}
For chamber \RN{4}, there are five BPS states: three singlets ($\gamma_1$, $\gamma_2$, and $\gamma_4$) and two doublets ($\gamma_3$ and $\gamma_5$). Their BPS indices are:
\begin{equation}
\Omega(\gamma)= \begin{cases}
2 & \text { for } \gamma \in \left\{
\pm\gamma_3, \pm \gamma_5\right\},\\ 

1 & \text{ for } \gamma \in
\left\{\pm\gamma_1,\pm\gamma_2,\pm\gamma_4\right\},
 \\

0 & \text{ otherwise. }
\end{cases}
\end{equation} 

TBA system in this chamber is written as
\begin{equation}
\begin{aligned}
    \tilde{\epsilon}_{\gamma_1}(\theta)&=\left|Z_{\gamma_1}\right| \mathrm{e}^\theta+ K_{1,2}\star \tilde{L}_2+2 K_{1,3}\star \tilde{L}_3+2 K_{1,4}\star \tilde{L}_4+2 K_{1,5}\star \tilde{L}_5,\\
     \tilde{\epsilon}_{\gamma_2}(\theta)&=\left|Z_{\gamma_2}\right| \mathrm{e}^\theta- K_{2,1}\star \tilde{L}_1-  K_{2,4}\star \tilde{L}_4-2 K_{2,5}\star \tilde{L}_5,\\
      \tilde{\epsilon}_{\gamma_3}(\theta)&=\left|Z_{\gamma_3}\right| \mathrm{e}^\theta- K_{3,1}\star \tilde{L}_1- K_{3,4}\star \tilde{L}_4-2 K_{3,5}\star \tilde{L}_5,\\
       \tilde{\epsilon}_{\gamma_4}(\theta)&=\left|Z_{\gamma_4}\right| \mathrm{e}^\theta-2 K_{4,1}\star \tilde{L}_1+K_{4,2}\star \tilde{L}_2+2  K_{4,3}\star \tilde{L}_3-2 K_{4,5}\star \tilde{L}_5,\\
        \tilde{\epsilon}_{\gamma_5}(\theta)&=\left|Z_{\gamma_5}\right| \mathrm{e}^\theta- K_{5,1}\star \tilde{L}_1+K_{5,2}\star \tilde{L}_2+2 K_{5,3}\star \tilde{L}_3+ K_{5,4}\star \tilde{L}_4.
\end{aligned}
\end{equation}
We perform numerical calculations for these two 
intermediate chambers with $\ell=-\frac{1}{2}$, as shown in Table \ref{tab:d4tbawkb}. It is straightforward to consider their $\ell$ deformation by splitting the $L$-functions corresponding to the doublets.

\subsection{TBA equations in the maximal chamber}
\begin{figure}[htbp]
\centering
    \begin{subfigure}{0.45\linewidth}
        \centering
        \includegraphics[width=0.9\linewidth]{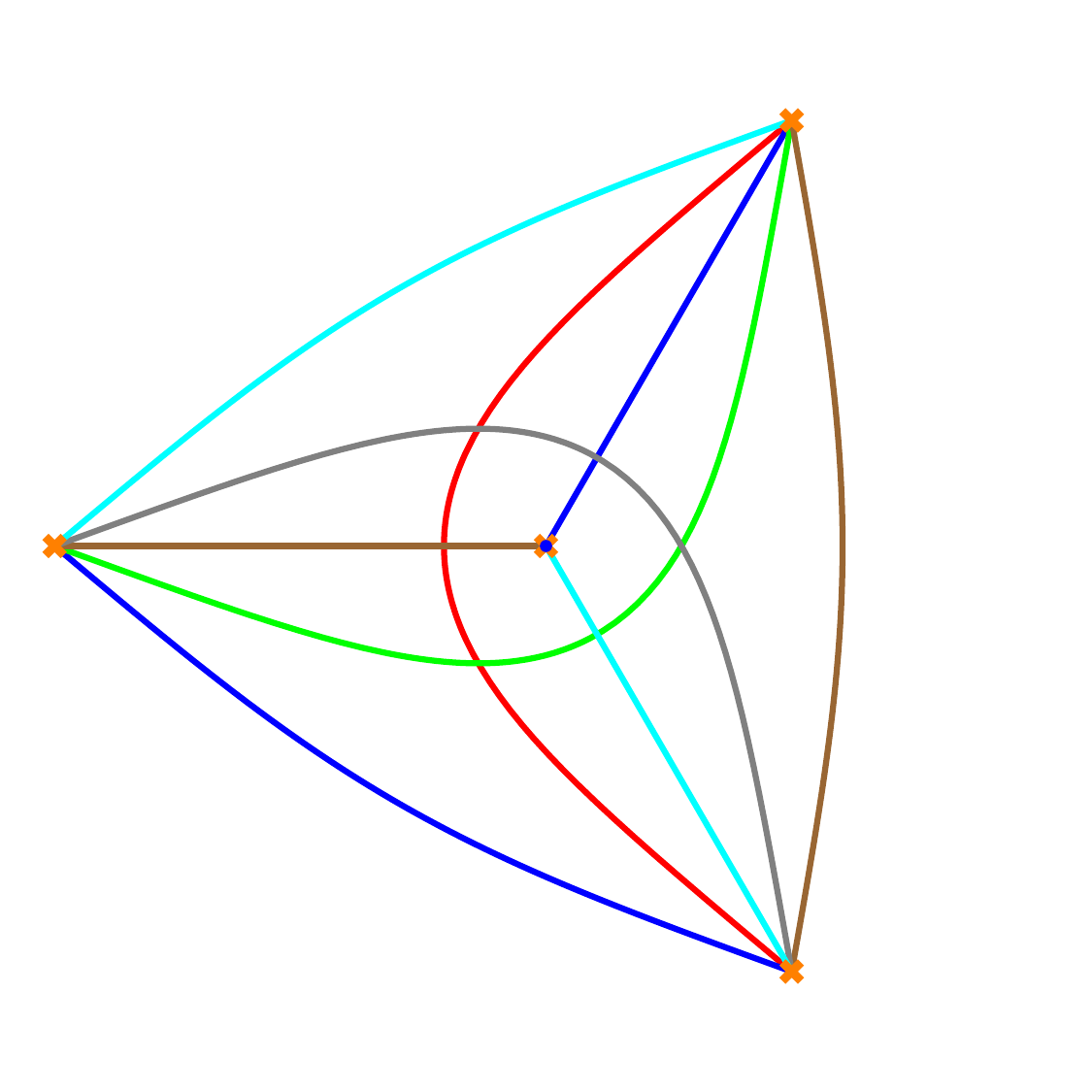}
        \caption{BPS states}
        \label{fig:d4maxsu3bpsplot}
    \end{subfigure}
    \begin{subfigure}{0.45\linewidth}
        \centering
        \includegraphics[width=0.9\linewidth]{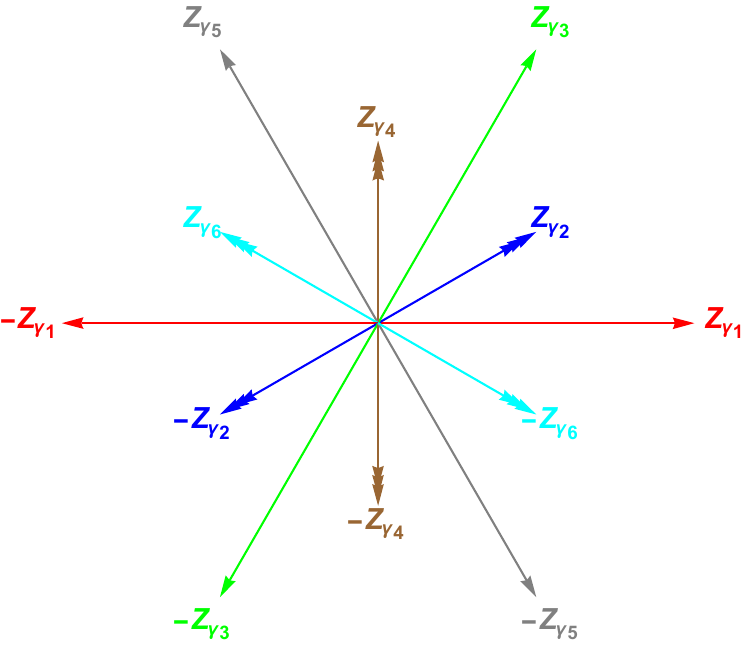}
        \caption{Central charges}
        \label{fig:d4maxsu3bpscharge}
    \end{subfigure}
    \caption{BPS spectrum in the maximal chamber for $(A_1,D_4)$ with SU(3) flavor symmetry.}
    \label{fig:d4maxbps}
\end{figure}

We are particularly interested in the TBA equations in the maximal chamber; the BPS spectrum for SU(2)$\times$U(1) symmetry is shown in Figures \ref{fig:d4cham5bpsplot} and \ref{fig:d4cham5bpscharge}. The spectrum includes three doublets and six singlets. One can derive the TBA equations in the maximal chamber directly from these BPS states. However, in this part, we will first consider the TBA equations for the SU(3) case and then generalize them to SU(2)$\times$U(1) symmetry.

\subsubsection{SU(3) symmetry}

For the SU(3) flavor symmetry case with $m=0$ and $u_1^2-4u_2=0$, the central charges of three of six singlets are the same as that of three doublets, forming triplets. This results in three triplets and three singlets, as shown in Figure \ref{fig:d4maxbps}. Their electromagnetic charges are
\begin{equation}
\begin{aligned}
      &\gamma_1=(1,0), \quad  \gamma_3=(2,3),\quad \gamma_5=(1,3), \\
     &\gamma_2=(1, 1), \quad  \gamma_4=(1,2),\quad \gamma_6=(0,1).
\end{aligned}
\end{equation}
Three singlets can be represented as a combination of the three triplets:
\begin{equation}
     \gamma_1=\gamma_2-\gamma_6, \quad \gamma_3=\gamma_2+\gamma_4, \quad \gamma_5=\gamma_4+\gamma_6.
\end{equation}
As mentioned earlier, the conventions for the maximal chamber with SU(3) symmetry differ from those used in the minimal and intermediate chambers. Therefore, the notation $\gamma_a$ in this part may not correspond to the same state as $\gamma_a$ in the other chambers. However, this discrepancy is merely a reordering of BPS states or an exchange between BPS states and their anti-states. The resulting TBA equations should remain consistent regardless of this reordering. The BPS indices of these states are:
\begin{equation}
\Omega(\gamma)= \begin{cases}
3 & \text { for } \gamma \in\left\{\pm \gamma_2,\pm \gamma_4,\pm \gamma_6\right\}, \\ 
1 & \text { for } \gamma \in\left\{\pm \gamma_1,\pm \gamma_3,\pm \gamma_5\right\}, \\
0 & \text { otherwise. }
\end{cases}
\end{equation}
The TBA equations are written as:
\begin{equation}
\label{eq:d4maxtba1}
\begin{aligned}
    \tilde{\epsilon}_{\gamma_1}(\theta)&=\left|Z_{\gamma_1}\right| \mathrm{e}^\theta-3 K_{1,2}\star \tilde{L}_2-3 K_{1,3}\star \tilde{L}_3-6 K_{1,4}\star \tilde{L}_4-3 K_{1,5}\star \tilde{L}_5-3 K_{1,6}\star \tilde{L}_6,\\
     \tilde{\epsilon}_{\gamma_2}(\theta)&=\left|Z_{\gamma_2}\right| \mathrm{e}^\theta+ K_{2,1}\star \tilde{L}_1- K_{2,3}\star \tilde{L}_3-3  K_{2,4}\star \tilde{L}_4-2 K_{2,5}\star \tilde{L}_5-3 K_{2,6}\star \tilde{L}_6,\\
    \tilde{\epsilon}_{\gamma_3}(\theta)&=\left|Z_{\gamma_3}\right| \mathrm{e}^\theta+3 K_{3,1}\star \tilde{L}_1+3 K_{3,2}\star \tilde{L}_2-3 K_{3,4}\star \tilde{L}_4- 3 K_{3,5}\star \tilde{L}_5-6 K_{3,6}\star \tilde{L}_6,\\
       \tilde{\epsilon}_{\gamma_4}(\theta)&=\left|Z_{\gamma_4}\right| \mathrm{e}^\theta+2 K_{4,1}\star \tilde{L}_1+3 K_{4,2}\star \tilde{L}_2+ K_{4,3}\star \tilde{L}_3- K_{4,5}\star \tilde{L}_5-3 K_{4,6}\star \tilde{L}_6,\\
       \tilde{\epsilon}_{\gamma_5}(\theta)&=\left|Z_{\gamma_5}\right| \mathrm{e}^\theta+3 K_{5,1}\star \tilde{L}_1+6 K_{5,2}\star \tilde{L}_2+3 K_{5,3}\star \tilde{L}_3+3 K_{5,4}\star \tilde{L}_4-3 K_{5,6}\star \tilde{L}_6,\\
       \tilde{\epsilon}_{\gamma_6}(\theta)&=\left|Z_{\gamma_6}\right| \mathrm{e}^\theta+ K_{6,1}\star \tilde{L}_1+3 K_{6,2}\star \tilde{L}_2+2 K_{6,3}\star \tilde{L}_3+3 K_{6,4}\star \tilde{L}_4+K_{6,5}\star \tilde{L}_5,
\end{aligned}
\end{equation}

\paragraph{Maximally symmetric point}

If we choose the moduli parameters at the maximally symmetric point: $u_1=0, u_2=0, u_3=1$, an extra $\mathbb{Z}_3$ symmetry appears. The central charges of the BPS states are given by
\begin{equation}
    \begin{aligned}
        &\left|Z_{\gamma_1}\right|= \left|Z_{\gamma_3}\right|= \left|Z_{\gamma_5}\right|=\frac{2\sqrt{3\pi}\Gamma\left(\frac{7}{6}\right)}{\Gamma\left(\frac{5}{3}\right)},\\
       &\left|Z_{\gamma_2}\right|= \left|Z_{\gamma_4}\right|= \left|Z_{\gamma_6}\right|=\frac{2\sqrt{\pi}\Gamma\left(\frac{7}{6}\right)}{\Gamma\left(\frac{5}{3}\right)}.
    \end{aligned}
\end{equation}
\begin{equation}
     \arg Z_{\gamma_i}=\frac{\pi}{6}(i-1).
\end{equation}
One can identify $\tilde{\epsilon}_{\gamma_1}=\tilde{\epsilon}_{\gamma_3}=\tilde{\epsilon}_{\gamma_5}$ and $\tilde{\epsilon}_{\gamma_2}=\tilde{\epsilon}_{\gamma_4}=\tilde{\epsilon}_{\gamma_6}$, which simplifies the TBA equation further and reduced TBA system \eqref{eq:d4maxtba1} to two equations as follows:
\begin{equation}
\label{eq:d4maxtba2}
\begin{aligned}
    \tilde{\epsilon}_{\gamma_1}(\theta)&=\left|Z_{\gamma_1}\right| \mathrm{e}^\theta+3K_1\star\tilde{L}_1+3K_2\star\tilde{L}_2,\\
     \tilde{\epsilon}_{\gamma_2}(\theta)&=\left|Z_{\gamma_2}\right| \mathrm{e}^\theta+K_2\star\tilde{L}_1+3 K_1\star\tilde{L}_2,
\end{aligned}
\end{equation}
with convolutional kernels
\begin{equation}
     K_1=\frac{2\sqrt{3}\cosh\theta}{\pi+2\pi \cosh2\theta},\quad K_2=\frac{3\cosh2\theta}{\pi\cosh3\theta}.
\end{equation}
We compute the large $\theta$ expansions from this TBA system and compare them with the WKB expansions, finding consistency, as shown in Table \ref{tab:d4tbawkb} for $\ell=-\frac{1}{2}$. The Borel-Pad{\'e} poles for $\Pi_{\gamma_1}$ and $\Pi_{\gamma_2}$ are illustrated in Figure \ref{fig:d3d4pole}, indicating that the WKB series of $\Pi_{\gamma_1}$ is Borel-summable, while that of $\Pi_{\gamma_2}$ is not. This is consistent with the prediction of the TBA equations \eqref{eq:d4maxtba1} or \eqref{eq:d4maxtba2}. 

Furthermore, the TBA equations given in \eqref{eq:d4maxtba2} coincide with TBA for the scattering theory with $\mathfrak{g}=D_4$; this can be verified by examining the kernel in\eqref{eq:d4kernel}.

\paragraph{Effective central charge} 

One finds the following limit behavior of $\tilde{\epsilon}_{\gamma}(\theta)$ for TBA equations \eqref{eq:d4maxtba1} at $\theta\to -\infty$:
\begin{equation}
    \tilde{\epsilon}_{\gamma_1}^\star=
     \tilde{\epsilon}_{\gamma_3}^\star=
      \tilde{\epsilon}_{\gamma_5}^\star=3\log 2, \quad   \tilde{\epsilon}_{\gamma_2}^\star=
       \tilde{\epsilon}_{\gamma_4}^\star=
        \tilde{\epsilon}_{\gamma_6}^\star=\log 3.
\end{equation}
The effective central charge for \eqref{eq:d4maxtba1} is evaluated as
\begin{equation}
\label{eq:d4maxeff}
\begin{aligned}
    c_{\mathrm{eff}}&=3\left[\frac{6}{\pi^2}\int\left|Z_{\gamma_1}\right|\re^\theta \tilde{L}_1(\theta)\rd \theta+\frac{6}{\pi^2}3\left|Z_{\gamma_2}\right|\int\re^\theta \tilde{L}_2(\theta)\rd \theta\right],\\
    &=3\frac{6}{\pi^2}\left(\mathcal{L}_1\left(\frac{1}{1+\re^{\tilde{\epsilon}_1^\star}}\right)+2\mathcal{L}_1\left(\frac{1}{1+\re^{\tilde{\epsilon}_2^\star}}\right)\right)=3.
    \end{aligned}
\end{equation}
An overall factor of 3 accounts for the contributions from $\gamma_3$, $\gamma_4$, $\gamma_5$, and $\gamma_6$. This result matches the effective central charge in \cite{Zam91} for $D_4$, except for a normalization factor of 3. \eqref{eq:d4mineff} and \eqref{eq:d4maxeff} further demonstrate that the effective central charge remains invariant during the wall-crossing. 

\subsubsection{SU(2)$\times$U(1) symmetry}

By relaxing the constriant $u_1^2-4 u_2=0$, the symmetry is broken from SU(3) to SU(2)$\times$U(1). The three triplets ($\gamma_2$, $\gamma_4$, and $\gamma_6$) are split into three doublets (using the same notation for these doublets) and three singlets with the same charges:\footnote{These labelings differ from those used in Figures \ref{fig:d4cham5bpsplot} and \ref{fig:d4cham5bpscharge}.}
\begin{equation}
    \gamma_7=(1,1), \quad \gamma_8=(1,2),\quad \gamma_9=(0,1).
\end{equation}

We could write down the corresponding TBA equations straightforwardly as above. However, since we already know the TBA equations for the SU(3) case, it is easier to obtain the TBA equations for SU(2)$\times$U(1) by modifying \eqref{eq:d4maxtba1}. The TBA equations corresponding to $\gamma_a$ (where $a$ ranges from 1 to 6) for SU(2)$\times$U(1) case can be obtained by the following simple substitution from the SU(3) TBA \eqref{eq:d4maxtba1}: 
\begin{equation}
\begin{aligned}
      &3 K_{a,2}\star \tilde{L}_2 \to 2 K_{a,2}\star \tilde{L}_2+K_{a,7}\star \tilde{L}_7,\\
       &3 K_{a,4}\star \tilde{L}_4 \to 2 K_{a,4}\star \tilde{L}_4+K_{a,8}\star \tilde{L}_8,\\
        &3 K_{a,6}\star \tilde{L}_6 \to 2 K_{a,6}\star \tilde{L}_6+K_{a,9}\star \tilde{L}_9,\\
\end{aligned}
\end{equation}
The TBA equations for $\gamma_7$, $\gamma_8$ and $\gamma_9$ are given by
\begin{equation}
\begin{aligned}
       \tilde{\epsilon}_{\gamma_7}(\theta)&=\left|Z_{\gamma_7}\right| \mathrm{e}^\theta+ K_{7,1}\star \tilde{L}_1- K_{7,3}\star \tilde{L}_3-2 K_{7,4}\star \tilde{L}_4-2 K_{7,5}\star \tilde{L}_5-2 K_{7,6}\star \tilde{L}_6
       -K_{7,8}\star \tilde{L}_8-K_{7,9}\star \tilde{L}_9,\\
       \tilde{\epsilon}_{\gamma_8}(\theta)&=\left|Z_{\gamma_8}\right| \mathrm{e}^\theta+ 2 K_{8,1}\star \tilde{L}_1+2 K_{8,2}\star \tilde{L}_2+K_{8,3}\star \tilde{L}_3- K_{8,5}\star \tilde{L}_5-2 K_{8,6}\star \tilde{L}_6+K_{8,7}\star \tilde{L}_7- K_{8,9}\star \tilde{L}_9,\\
        \tilde{\epsilon}_{\gamma_9}(\theta)&=\left|Z_{\gamma_9}\right| \mathrm{e}^\theta+ K_{9,1}\star \tilde{L}_1+2 K_{9,2}\star \tilde{L}_2+2 K_{9,3}\star \tilde{L}_3+2 K_{9,4}\star \tilde{L}_4+ K_{9,5}\star \tilde{L}_5+K_{9,7}\star \tilde{L}_7+K_{9,8}\star \tilde{L}_8.
\end{aligned}
\end{equation}
Note that there are no terms like $K_{7,2}$, $K_{8,4}$, or $K_{9,6}$ since the corresponding electromagnetic central charges are the same and the intersection parings vanish.

\paragraph{Deformation by $\ell$}
The above TBA equations for $\ell=-\frac{1}{2}$ can be directly generalized to $\ell \in (-1,0)$. This can be achieved by replacing
\begin{equation}
    \begin{aligned}
        2\tilde{L}_2\to \tilde{L}_2^{+}+ \tilde{L}_2^{-}, \quad     2\tilde{L}_4\to \tilde{L}_4^{+}+ \tilde{L}_4^{-}, \quad     2\tilde{L}_6\to \tilde{L}_6^{+}+ \tilde{L}_6^{-}, 
    \end{aligned}
\end{equation}
where 
\begin{equation}
\begin{aligned}
     &\tilde{L}_{s}^{\pm}=\log\left(1-\re^{\pm 2\pi i \ell}\re^{-\tilde{\epsilon}_{\gamma_s}}(\theta)\right), \quad s=2,4,6.
\end{aligned}
\end{equation}

This set of TBA equations with $\ell$ deformation also applies to the SU(3) case, yet the triplets are split by quantum corrections. We perform numerical calculations for $\ell=-\frac{1}{5}$ at the maximally symmetric point. The WKB expansions of the quantum periods and large $\theta$ expansions of the pseudo-energies match well, as detailed in Table \ref{tab:d4tbawkb}.

In this section, we establish the TBA systems for the $(A_1, D_4)$ theory with SU(3) and SU(2)$\times$U(1) symmetry, respectively. We explicitly present the TBA equations in the minimal, maximal, and two intermediate chambers. We discuss the $\ell$ deformation of the TBA systems and demonstrate the relation between the TBA equations for the $D_4$ scattering theory and the $(A_1, D_4)$ GMN TBA at the maximally symmetric point. TBA equations for any chamber can be similarly constructed.

\section{$(A_1,D_{N+2})$ theory}
\label{sc:dn}

In Sections $\ref{sc:d3}$ and $\ref{sc:d4}$, we discuss the TBA equations across the moduli space for the $(A_1, D_3)$ and $(A_1, D_4)$ theories. The examples presented illustrate all features of the D-type TBA equations. This procedure can be readily generalized to $(A_1, D_{N+2})$ theories with $N>2$. We focus particularly on the TBA systems in the minimal chamber and at the maximally symmetric point of the maximal chamber, where the TBA equations can be systematically derived following a set of simple rules.

\subsection{TBA equations in the minimal chamber}

\begin{figure}[h]
    \centering
    \begin{tikzpicture}
    
        \foreach \x in { 0, 2, 4, 6, 8, 10} {
            \draw[line width=2pt, orange] (\x-0.1, -0.1) -- (\x+0.1, 0.1);
            \draw[line width=2pt, orange] (\x-0.1, 0.1) -- (\x+0.1, -0.1);
        }

        \fill[blue] (10,0) circle (0.06cm);

        \node at (3, 0) {\tikz \fill (0,0) circle (0.05cm); \hspace{0.1cm} \tikz \fill (0,0) circle (0.05cm); \hspace{0.1cm} \tikz \fill (0,0) circle (0.05cm);};
        
        \draw[line width=0.5mm, black] (1, 0) ellipse (1.3cm and 0.4cm);
        \draw[line width=0.5mm, blue] (5, 0) ellipse (1.3cm and 0.4cm);
        \draw[line width=0.5mm, red] (7, 0) ellipse (1.3cm and 0.4cm);
        \draw[line width=0.5mm, blue] (9, 0) ellipse (1.3cm and 0.4cm);
        
        \node[below] at (0, -0.5) {$a_{N+1}$};
        \node[below] at (2, -0.5) {$a_{N}$};
        \node[below] at (4, -0.5) {$a_3$};
        \node[below] at (6, -0.5) {$a_2$};
        \node[below] at (8, -0.5) {$a_1$};
        \node[below] at (10, -0.5) {$0$};
        
        \node[above] at (1, 0.5) {$\gamma_N$};
        \node[above] at (5, 0.5) {$\gamma_2$};
        \node[above] at (7, 0.5) {$\gamma_1$};
        \node[above] at (9, 0.5) {$\gamma_{N+1}$};
    \end{tikzpicture}
    \caption{Cycles corresponding to the BPS states in the minimal chamber of $(A_1, D_{N+2})$. The red cycles represent classically allowed regions, while the blue cycles represent forbidden regions. The black cycle for $\gamma_N$ represents an allowed region for odd $N$ and a forbidden region for even $N$.}
    \label{fig:elliptical_cycles}
\end{figure}
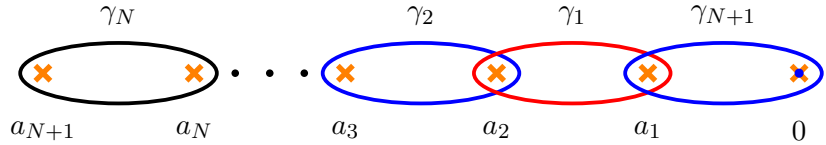

The Seiberg-Witten differential for generic $(A_1, D_{N+2})$ theory is determined by \eqref{eq:dsw}. We consider the limit $m\to 0$ when there are $N+2$ branch points, including $0$. The moduli parameters are chosen so that these $N+2$ branch points are ordered in the real axis as: 
\begin{equation}
    a_{N+1}<a_{N}<\cdots<a_{1}<0.
\end{equation}
These turning points divide the negative real axis into $N+1$ adjacent regions. We classify them as classically allowed or forbidden regions when $\lambda^2$ is positive or negative, respectively. We denote $\gamma_s$ (singlet) the one-cycle encircling branch points $a_s$ and $a_{s+1}$ for $s=1,\cdots,N$, and $\gamma_{N+1}$ (doublet) as the one-cycle around $0$ and $a_1$. The orientation of the cycles is chosen so that the central charges for classically allowed regions are real and positive, and those for the forbidden regions are purely imaginary. We assign electromagnetic charges for BPS states corresponding to these one-cycles as
\begin{equation}
 \gamma_{2i-1}=(0,1), \quad \gamma_{2i}=(1,0), \quad \gamma_{N+1}=(1,0).
\end{equation}
Their BPS indices are:
\begin{equation}
\Omega(\gamma)= \begin{cases}
2 & \text { for } \gamma \in
\pm\gamma_{N+1},\\ 
1 & \text { for } \gamma \in
\pm\gamma_s,\; s=1,\cdots,N,
 \\
0 & \text{ otherwise. }
\end{cases}
\end{equation}
Each BPS state $\gamma_s$ only intersects with its neighboring states $\gamma_{s-1}$ and $\gamma_{s+1}$.\footnote{Due to our notation, $\gamma_1$ and $\gamma_{N+1}$ are adjacent to each other.} The TBA equations for these $N$ singlets and one doublet in the minimal chamber are written as 
\begin{equation}
\begin{aligned}
    \tilde{\epsilon}_{\gamma_1}(\theta)&=\left|Z_{\gamma_1}\right| \mathrm{e}^\theta+K_{1,2}\star\tilde{L}_{2}+2 K_{1,N+1}\star\tilde{L}_{N+1},\\
     \tilde{\epsilon}_{\gamma_s}(\theta)&=\left|Z_{\gamma_s}\right| \mathrm{e}^\theta-(-1)^s K_{s,s-1}\star\tilde{L}_{s-1}-(-1)^s K_{s,s+1}\star\tilde{L}_{s+1}, \quad s=2,\cdots,N-1,\\
       \tilde{\epsilon}_{\gamma_{N}}(\theta)&=\left|Z_{\gamma_{N}}\right| \mathrm{e}^\theta-(-1)^N K_{N,N-1}\star\tilde{L}_{N-1},\\
      \tilde{\epsilon}_{\gamma_{N+1}}(\theta)&=\left|Z_{\gamma_{N+1}}\right| \mathrm{e}^\theta- K_{N+1,1}\star\tilde{L}_{1}.
\end{aligned}
\end{equation}
When all the turning points are real, the phases of the central charges are:
\begin{equation}
    \arg Z_{\gamma_{2i-1}}=0, \quad \arg Z_{\gamma_{2i}}=\arg Z_{\gamma_{N+1}}=\frac{\pi}{2}.
\end{equation}
The TBA system simplifies to:
\begin{equation}
\label{eq:dnmintba2}
\begin{aligned}
    \tilde{\epsilon}_{\gamma_1}(\theta)&=\left|Z_{\gamma_1}\right| \mathrm{e}^\theta-\frac{1}{2 \pi } \int_{\mathbb{R}} \frac{\log \left(1+\mathrm{e}^{-\tilde{\epsilon}_{\gamma_2}\left(\theta^{\prime}\right)}\right)}{\cosh \left(\theta-\theta^{\prime}\right)} \mathrm{d} \theta^{\prime}-\frac{1}{2 \pi } 2\int_{\mathbb{R}} \frac{\log \left(1+\mathrm{e}^{-\tilde{\epsilon}_{\gamma_{N+1}}\left(\theta^{\prime}\right)}\right)}{\cosh \left(\theta-\theta^{\prime}\right)} \mathrm{d} \theta^{\prime},\\
     \tilde{\epsilon}_{\gamma_s}(\theta)&=\left|Z_{\gamma_s}\right| \mathrm{e}^\theta-\frac{1}{2 \pi }  \int_{\mathbb{R}} \frac{\log \left(1+\mathrm{e}^{-\tilde{\epsilon}_{\gamma_{s-1}}\left(\theta^{\prime}\right)}\right)}{\cosh \left(\theta-\theta^{\prime}\right)} \mathrm{d} \theta^{\prime}-\frac{1}{2 \pi }  \int_{\mathbb{R}} \frac{\log \left(1+\mathrm{e}^{-\tilde{\epsilon}_{\gamma_{s+1}}\left(\theta^{\prime}\right)}\right)}{\cosh \left(\theta-\theta^{\prime}\right)} \mathrm{d} \theta^{\prime},\quad s=2,\cdots,N-1,\\
      \tilde{\epsilon}_{\gamma_{N}}(\theta)&=\left|Z_{\gamma_{N}}\right| \mathrm{e}^\theta-\frac{1}{2 \pi }  \int_{\mathbb{R}} \frac{\log \left(1+\mathrm{e}^{-\tilde{\epsilon}_{\gamma_{N-1}}\left(\theta^{\prime}\right)}\right)}{\cosh \left(\theta-\theta^{\prime}\right)} \mathrm{d} \theta^{\prime},\\
     \tilde{\epsilon}_{\gamma_{N+1}}(\theta)&=\left|Z_{\gamma_{N+1}}\right| \mathrm{e}^\theta-\frac{1}{2 \pi }  \int_{\mathbb{R}} \frac{\log \left(1+\mathrm{e}^{-\tilde{\epsilon}_{\gamma_1}\left(\theta^{\prime}\right)}\right)}{\cosh \left(\theta-\theta^{\prime}\right)} \mathrm{d} \theta^{\prime}.
\end{aligned}
\end{equation}

\begin{figure}[htbp]
    \centering
    \begin{tikzpicture}
        \node[circle, draw, fill=black, minimum size=0.3cm] (A0) at (0,0) {};
        \node[circle, draw, fill=black, minimum size=0.3cm] (A1) at (1.5,0) {};
        \node[circle, draw, fill=black, minimum size=0.3cm] (A2) at (3,0) {};
        \node[circle, draw, fill=black, minimum size=0.3cm] (A3) at (4.5,0)  {};
        
        \node[circle, draw, fill=black, minimum size=0.3cm] (B1) at (6,1) {};
        \node[circle, draw, fill=black, minimum size=0.3cm] (B2) at (6,-1) {};

        \draw[line width=0.3mm] (A0) -- (A1);
        \draw[densely dashed, line width=0.3mm] (A1) -- (A2);
        \draw[line width=0.3mm] (A2) -- (A3);

        \draw[line width=0.3mm] (A3) -- (B1);
        \draw[line width=0.3mm] (A3) -- (B2);

        \node at (0,-0.5) {$\gamma_N$};
        \node at (1.5,-0.5) {$\gamma_{N-1}$};
        \node at (3,-0.5) {$\gamma_2$};
        \node at (4.5,-0.5) {$\gamma_1$};
        \node at (6.8,1) {$\gamma_{N+1}$};
        \node at (6.8,-1) {$\gamma_{N+1}$};
    \end{tikzpicture}
    \caption{The corresponding Dynkin diagram of the TBA system for $(A_1, D_{N+2})$ in the minimal chamber.}
    \label{fig:dynkin}
\end{figure}
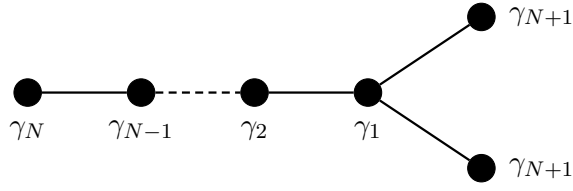

This set of TBA equations corresponds to the Dynkin diagrams for the $D_{N+2}$ Lie algebra, as shown in Figure \ref{fig:dynkin}. They apply to generic $N$ and return to \eqref{eq:d3mintba1} and \eqref{eq:d4mintba4} for $N=1$ and $2$, respectively. Introducing the $\ell$ deformation and its analytic continuation to the above TBA equations is straightforward by splitting the doublet $\gamma_{N+2}$, as demonstrated in sections \ref{sc:d3} and \ref{sc:d4}. We will not repeat the explicit formulas here.

\paragraph{Effective central charge}

We can find the limit behaviors of the above $\tilde{\epsilon}_{\gamma}(\theta)$ at $\theta\to-\infty$ as follows:
\begin{equation}
    \tilde{\epsilon}_{\gamma_s}^\star=-\log \left[\left(N+1-s\right)\left(N+3-s\right)\right], \quad   \tilde{\epsilon}_{\gamma_{N+1}}^\star=-\log \left(N+1\right), \quad s=1, \cdots, N.
\end{equation}
The effective central charge is then evaluated as
\begin{equation}
\label{eq:dnmineff}
\begin{aligned}
    c_{\mathrm{eff}}&=\frac{6}{\pi^2}\sum_{s=1}^{N}\int\left|Z_{\gamma_s}\right|\re^\theta \tilde{L}_s(\theta)\rd \theta+\frac{6}{\pi^2}2\int \left|Z_{\gamma_{N+1}}\right| \re^\theta \tilde{L}_{N+1}(\theta)\rd \theta,\\
    &=\frac{6}{\pi^2}\left(\sum_{s=1}^{N}\mathcal{L}_1\left(\frac{1}{1+\re^{\tilde{\epsilon}_{\gamma_s}^\star}}\right)+2\mathcal{L}_1\left(\frac{1}{1+\re^{\tilde{\epsilon}_{\gamma_{N+1}}^\star}}\right)\right)=N+1,
    \end{aligned}
\end{equation}
where the factor 2 accounts for the doublet $\gamma_{N+1}$. This reduces to \eqref{eq:d3mineff} and \eqref{eq:d4mineff} when $N=1$ and $2$, respectively.

\subsection{TBA equations at the maximally symmetric point}

\begin{figure}[htbp]
\centering
    \begin{subfigure}{0.45\linewidth}
        \centering
        \includegraphics[width=0.9\linewidth]{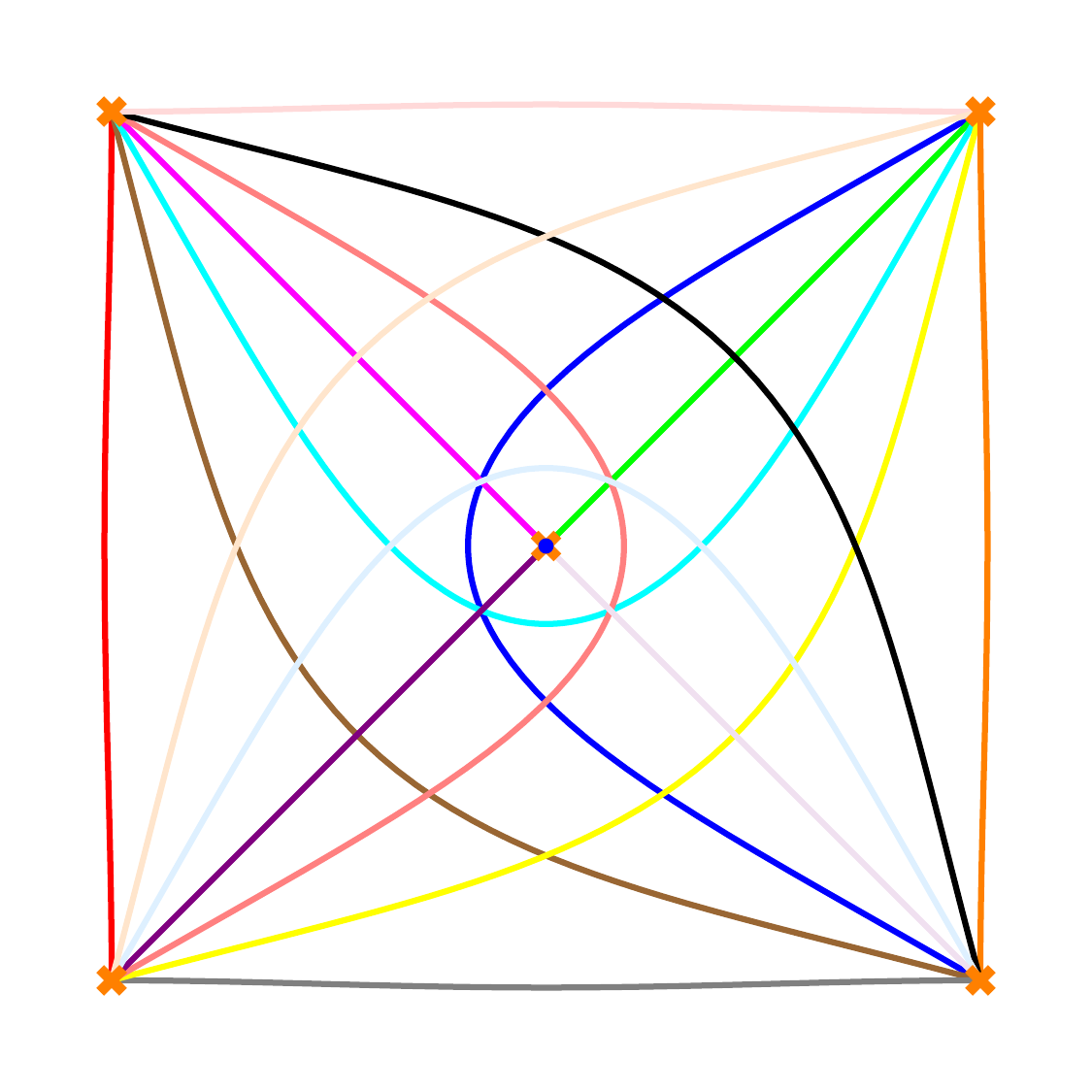}
        \caption{BPS states}
        \label{fig:d5maxbpsplot}
    \end{subfigure}
    \begin{subfigure}{0.45\linewidth}
        \centering
        \includegraphics[width=0.9\linewidth]{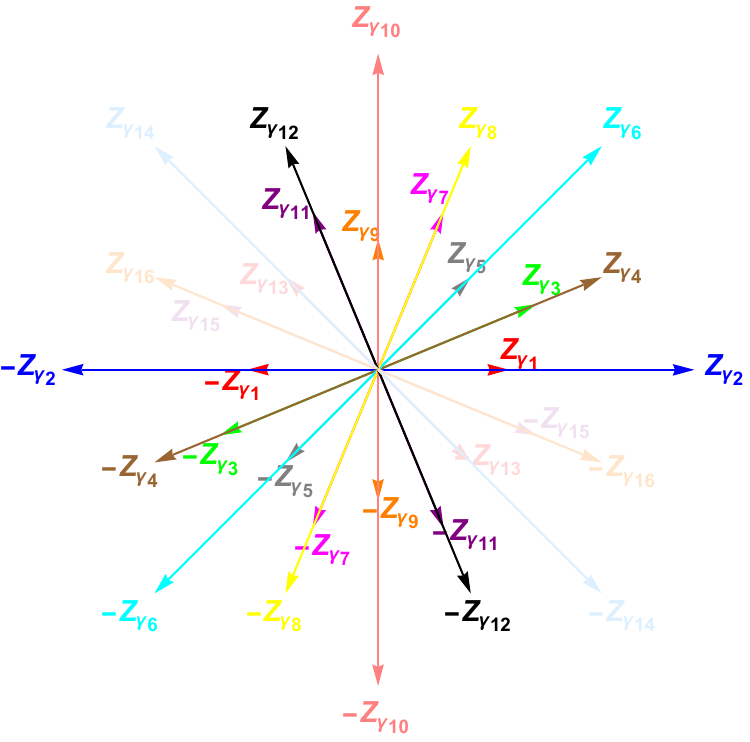}
        \caption{Central charges}
        \label{fig:d5maxbpscharge}
    \end{subfigure}
    \caption{BPS spectrum for $(A_1,D_5)$ at the maximally symmetric point.}
    \label{fig:d5maxbps}
\end{figure}

Let us now consider the TBA equations at the maximally symmetric point of the maximal chamber for generic $N$. We still work in the limit $m\to 0$, where there are $(N+1)^2$ BPS states, comprising $N+1$ doublets and $N(N+1)$ singlets. We find these TBA equations can be directly derived from the diagrammatic representation of the BPS spectrum using a few simple rules:

\begin{enumerate}
    \item BPS states: We draw the variety of Stokes graphs for $\vartheta\in[0,\pi)$, identifying all BPS states as the saddle connections (lines) that link two branch points.

    \item BPS indices: A line (corresponding to a BPS state) connecting to the origin represents a doublet with its BPS index $\Omega(\gamma)=2$, while the other lines represent singlets with $\Omega(\gamma)=1$.

    \item Dirac pairings: A common endpoint (which is a branch point) of two lines representing BPS states $\gamma_a$ and $\gamma_b$ contributes a factor of $1$ to $\left\langle\gamma_a, \gamma_b\right\rangle$, up to an overall sign, while an intersection point contributes a factor of $2$, also up to an overall sign. The intersection number is the sum of the contributions from all common endpoints and intersecting points.

    \item Signs: $\left\langle\gamma_a, \gamma_b\right\rangle$ is positive whenever $a<b$, and negative whenever $a>b$, if the central charges are ordered according to their phase from smallest to largest, $\arg Z_{\gamma_1}<\arg Z_{\gamma_2}<\cdots<\arg Z_{\gamma_{(N+1)^2}}$.
\end{enumerate}

The first three rules are valid for any chamber, which means that their TBA systems can be determined up to $\pm$ signs. The determination of the signs in generic chambers depends on the conventions for branch cuts and orientations, which are offset by the signs of the central charges.

\paragraph{$(A_1,D_5)$ TBA at the maximally symmetric point}

We use the $(A_1, D_5)$ theory as an example to illustrate the application of the above rules. At the maximally symmetric point, where $u_1=u_2=u_3=0$ and $u_4=1$, the BPS spectrum and the central charges are depicted in Figure \ref{fig:d5maxbps}. The spectrum includes four doublets: $\gamma_3$, $\gamma_7$, $\gamma_{11}$, and $\gamma_{15}$, along with 12 singlets. 

From Figure \ref{fig:d5maxbps}, we can directly determine the Dirac pairings. For instance, there are no common endpoints or intersections between the lines for $\gamma_1$ and $\gamma_2$, indicating that $\left\langle\gamma_1, \gamma_2\right\rangle$=0. The states $\gamma_2$ and $\gamma_3$ share one common endpoint, yielding $\left\langle\gamma_2, \gamma_3\right\rangle=-\left\langle\gamma_3, \gamma_2\right\rangle=1$. In contrast, $\gamma_2$ and $\gamma_9$ have two common endpoints, yielding $\left\langle\gamma_2, \gamma_9\right\rangle=2$. For $\gamma_2$ and $\gamma_{12}$, which have one common endpoint and one intersection point, the pairing is $\left\langle\gamma_2, \gamma_{12}\right\rangle=1+1\times2=3$. $\gamma_2$ and $\gamma_{10}$ have two intersection points, resulting in $\left\langle\gamma_2, \gamma_{10}\right\rangle=2\times2=4$. Other Dirac pairings can be determined analogously.

\begin{figure}[htbp]
    \centering
    \includegraphics[width=0.4\textwidth]{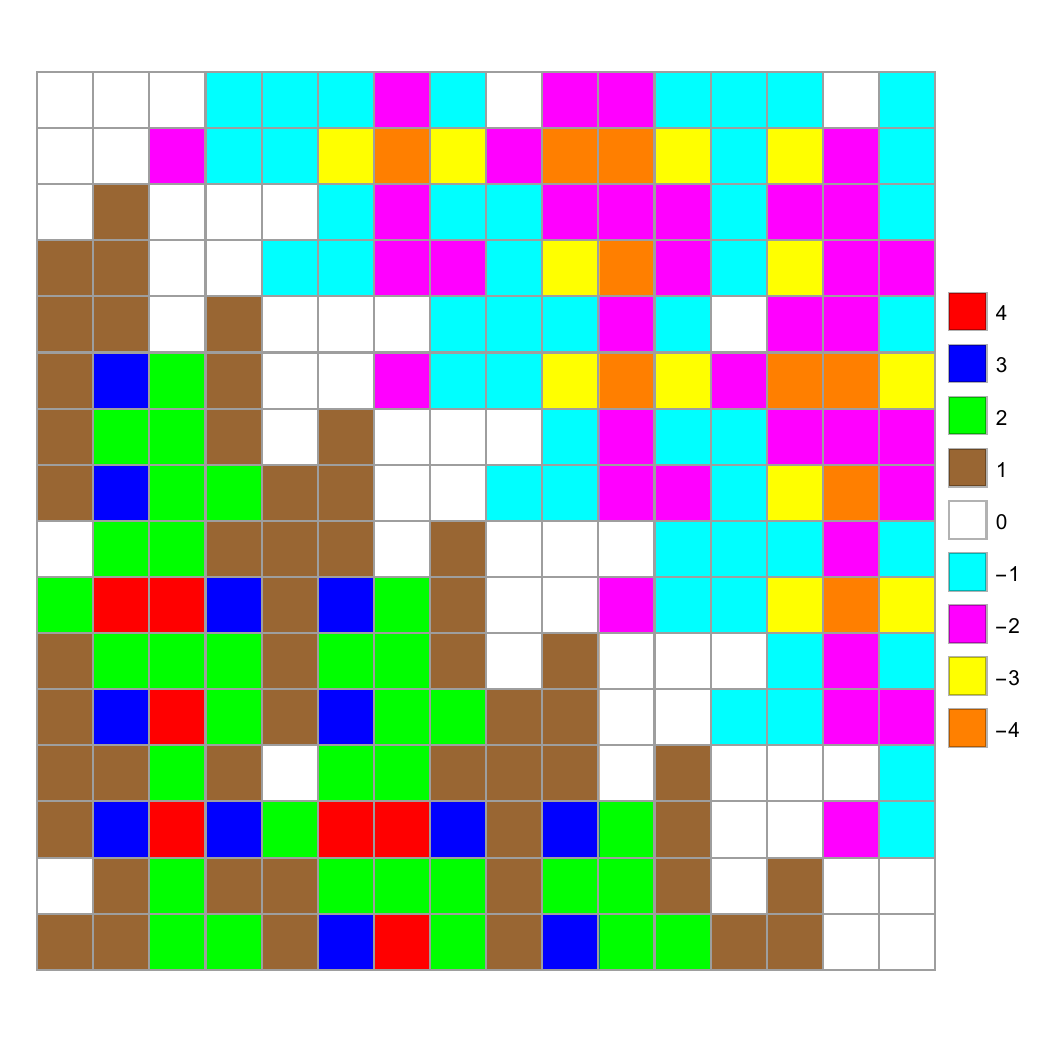}
    \caption{Intersection matrix $S_{a,b}$ for $(A_1,D_5)$ with an SU(2) flavor symmetry.}
    \label{fig:d5maxintsec}
\end{figure}

The TBA equations for the $D_5$ theory at the maximally symmetric point can be readily written down using the above rules. These equations can be organized as

\begin{equation}
\label{eq:dnmaxtba1}
\tilde{\epsilon}_{\gamma_a}(\theta)=\left|Z_{\gamma_a}\right| \mathrm{e}^\theta+\sum_{b\neq a} S_{a, b} K_{a, b} \star \tilde{L}_b, \quad a=1,2,\cdots,16,
\end{equation}
where $S_{a, b}=-\left\langle\gamma_a, \gamma_b\right\rangle \Omega\left(\gamma_b\right)$ is the intersection matrix, represented in Figure \ref{fig:d5maxintsec}. Note that it is not antisymmetric because this matrix is dressed by the BPS indices.

At the maximally symmetric point, the central charges are evaluated as
\begin{equation}
    \begin{aligned}
        &\left|Z_{\gamma_1}\right|= \left|Z_{\gamma_5}\right|= \left|Z_{\gamma_9}\right|=
        \left|Z_{\gamma_{13}}\right|=\frac{2\sqrt{\left(2-\sqrt{2}\right)\pi}\Gamma\left(\frac{1}{8}\right)}{5\Gamma\left(\frac{5}{8}\right)},\\
       &\left|Z_{\gamma_2}\right|= \left|Z_{\gamma_6}\right|= \left|Z_{\gamma_{10}}\right|=
       \left|Z_{\gamma_{14}}\right|=\frac{2\sqrt{\left(2+\sqrt{2}\right)\pi}\Gamma\left(\frac{1}{8}\right)}{5\Gamma\left(\frac{5}{8}\right)},\\
       &\left|Z_{\gamma_3}\right|= \left|Z_{\gamma_7}\right|= \left|Z_{\gamma_{11}}\right|=
       \left|Z_{\gamma_{15}}\right|=\frac{2\sqrt{\pi}\Gamma\left(\frac{1}{8}\right)}{5\Gamma\left(\frac{5}{8}\right)},\\
       &\left|Z_{\gamma_4}\right|= \left|Z_{\gamma_8}\right|= \left|Z_{\gamma_{12}}\right|=
       \left|Z_{\gamma_{16}}\right|=\frac{2\sqrt{2\pi}\Gamma\left(\frac{9}{8}\right)}{\Gamma\left(\frac{13}{8}\right)}.
    \end{aligned}
\end{equation}
\begin{equation}
     \arg Z_{\gamma_{2i-1}}= \arg Z_{\gamma_{2i}}=\frac{\pi}{8}(i-1).
\end{equation}
At this point in the moduli space, 
 an additional $\mathbb{Z}_4$ symmetry emerges. Specifically, we can identify $\tilde{\epsilon}_{\gamma_1}=\tilde{\epsilon}_{\gamma_5}=\tilde{\epsilon}_{\gamma_9}=\tilde{\epsilon}_{\gamma_{13}}$, $\tilde{\epsilon}_{\gamma_2}=\tilde{\epsilon}_{\gamma_6}=\tilde{\epsilon}_{\gamma_{10}}=\tilde{\epsilon}_{\gamma_{14}}$, $\tilde{\epsilon}_{\gamma_3}=\tilde{\epsilon}_{\gamma_7}=\tilde{\epsilon}_{\gamma_{11}}=\tilde{\epsilon}_{\gamma_{15}}$, $\tilde{\epsilon}_{\gamma_4}=\tilde{\epsilon}_{\gamma_8}=\tilde{\epsilon}_{\gamma_{12}}=\tilde{\epsilon}_{\gamma_{16}}$, which simplify the TBA equations, reducing the TBA system to four distinct equations as follows:
\begin{equation}
\label{eq:dnmaxtba2}
\begin{aligned}
    \tilde{\epsilon}_{\gamma_1}(\theta)&=\left|Z_{\gamma_1}\right| \mathrm{e}^\theta+K_{11}\star\tilde{L}_1+K_{12}\star\tilde{L}_2+K_{13}\star\tilde{L}_3+K_{14}\star\tilde{L}_4,\\
     \tilde{\epsilon}_{\gamma_2}(\theta)&=\left|Z_{\gamma_2}\right| \mathrm{e}^\theta+K_{21}\star\tilde{L}_1+K_{22}\star\tilde{L}_2+K_{23}\star\tilde{L}_3+K_{24}\star\tilde{L}_4,\\
      \tilde{\epsilon}_{\gamma_3}(\theta)&=\left|Z_{\gamma_3}\right| \mathrm{e}^\theta+K_{31}\star\tilde{L}_1+K_{32}\star\tilde{L}_2+K_{33}\star\tilde{L}_3+K_{34}\star\tilde{L}_4,\\
     \tilde{\epsilon}_{\gamma_4}(\theta)&=\left|Z_{\gamma_4}\right| \mathrm{e}^\theta+K_{41}\star\tilde{L}_1+K_{42}\star\tilde{L}_2+K_{43}\star\tilde{L}_3+K_{44}\star\tilde{L}_4.
\end{aligned}
\end{equation}
The kernels are represented in terms of 
\begin{equation}
\begin{aligned}
     &K_1=\frac{1}{\pi \cosh\theta}, &&K_2=\frac{\sqrt{2}\cosh\theta}{\pi\cosh2\theta},\\
     &K_3=\frac{\sqrt{4+2\sqrt{2}}\cosh\theta}{\sqrt{2}\pi\cosh2\theta+\pi}, &&K_4=\frac{\sqrt{4-2\sqrt{2}}\cosh\theta}{\sqrt{2}\pi\cosh2\theta-\pi},
     \end{aligned}
\end{equation}
as follows:
\begin{equation}
\begin{aligned}
    &K_{11} = K_2,           &&K_{12} = K_1 + K_2,    &&K_{13} = 2 K_3,         &&K_{14} = K_3 + K_4, \\
    &K_{21} = K_1 + K_2,     &&K_{22} = 2 K_1 + 3 K_2, &&K_{23} = 4 K_3 + 2 K_4,  &&K_{24} = 3 K_3 + K_4, \\
    &K_{31} = K_3,           &&K_{32} = 2 K_3 + K_4,  &&K_{33} = K_1 + 2 K_2,   &&K_{34} = K_1 + K_2, \\
    &K_{41} = K_3 + K_4,     &&K_{42} = 3 K_3 + K_4,  &&K_{43} = 2 K_1 + 2 K_2, &&K_{44} = K_1 + 2 K_2.
\end{aligned}
\end{equation}

We solve these TBA equations \eqref{eq:dnmaxtba2} numerically and compare the results with the WKB expansions, as shown in Table \ref{tab:d5tbawkb}.
The TBA \eqref{eq:dnmaxtba2} coincides with the TBA system for the scattering theory with $\mathfrak{g}=D_5$ \eqref{eq:zamtba1}. The effective central charge for \eqref{eq:dnmaxtba1} is $4$, the same as \eqref{eq:dnmineff} when $N=3$. Additionally, the effective central charge for \eqref{eq:dnmaxtba2} is $1$, consistent with the result reported in \cite{Zam91}. It is straightforward to introduce $\ell$ deformations into the TBA equations \eqref{eq:dnmaxtba1}. However, we will not present these calculations explicitly here. 

\section{Conclusion and discussion}
\label{sc:concl}

In this paper, we have analyzed the BPS spectrum for D-type AD theories using the spectral network and constructed the corresponding TBA systems for each chamber in the moduli space. Our focus has been particularly on the TBA equations in the minimal and maximal chambers. These TBA equations have been validated by comparing them with quantum periods obtained via the exact WKB method of the Schr{\"o}dinger equation including a centrifugal correction. The effective central charges associated with the TBA equations for different chambers have been evaluated, confirming their invariance under wall-crossing. Notably, we have found that the TBA system at the maximally symmetric point aligns with TBA equations for D scattering theories. 

Our analysis provides a general procedure for deriving TBA equations for any chamber of a $(A_1, D_r)$ theory, which precisely governs the quantum periods. This approach can be extended to a broader class of four-dimensional theories in the NS limit of the $\Omega$ background, allowing for a non-perturbative determination of quantum deformations. A natural extension of this work involves AD theories that result from compactifying 6d $A_{N-1}$ theories with $N>2$, which are associated with an $N$-th order differential equation. For a third-order ODE with a cubic potential, the $D_4$-type TBA appears at the maximally symmetric point, suggesting the duality of the AD theories $(A_1, D_4)=(A_2, A_2)$. In the case of a third-order ODE with a quartic potential, the $E_6$-type TBA appears at the maximally symmetric point \cite{IKKS21, IKS21}. It is interesting to explore ODEs corresponding to other $E$-type TBAs \cite{Cecotti:2014zga,Ito:2017ypt}. The BPS spectrum of these theories has been studied in \cite{MPY13}, and the TBA equations for specific cases have been derived from the ODE/IM correspondence in \cite{IKKS21, IKS21}. However, a comprehensive generalization is still needed.

An intriguing issue not addressed in our current discussion is the case for the $m\neq 0$ case, when the Stokes graphs exhibit closed loops, leading to a more complicated BPS spectrum compared to the $m=0$ case. These types of Stokes graphs are also observed in the weak coupling regime of pure SU(2) gauge theory \cite{GGM19} and in SQCD with one flavor \cite{GHN21}. To our knowledge, TBA equations for these scenarios are not yet fully established. These cases often involve an infinite number of BPS states. Another related area of interest is the study of 5D theories, including their BPS spectra \cite{BLR18, Longhi21} and TBA equations, which may yield valuable insights into 5D theories. We intend to explore these topics in our future work.  

The TBA equations presented here could provide solutions for spectral problems in quantum mechanics and quasi-normal modes (QNM). In particular, the TBA equations for $(A_1, D_3)$
encapsulate the quantum periods (or Voros multipliers) related to the Stark effect in the case $m=0$. In \cite{IY23}, TBA equations in the minimal chamber are explicitly written down. The TBA equations derived in this paper generalize to include any parameters of the reduced Schrödinger equation discussed therein.

\section*{Acknowledgements}

We would like to thank Kazunobu Maruyoshi for his kind comments, as well as Pietro Longhi, Hongfei Shu and Mingshuo Zhu for their valuable discussions. J.Y. is grateful to the workshop {\it New Aspects in Topological Recursion, Resurgence and Related Topics} at Kyoto University for its hospitality during the completion of this work. The work of J.Y is partially supported by the Tsubame Scholarship at the Tokyo Institute of Technology and Tokyo Tech Program for Development of Next-Generation Front-Runners with Comprehensive Knowledge and Humanity (Tokyo Tech SPRING). The work of K.I. is supported in part by Grant-in-Aid for Scientific Research 21K03570 from Japan Society for the Promotion of Science (JSPS).

\appendix
\section{Coefficients in the differential operator method}
\label{sc: pf}

In this appendix, we outline the first several coefficients $c_k^{(n)}$ in the Picard-Fuchs equation \eqref{eq:pfeq}.
For the $(A_1, D_3)$ theory, we consider $\phi_2(z)=z+u_1+\frac{u_2}{z}$ and $\ell=-\frac{1}{2}$ for simplicity. The coefficients of the first three orders in the Picard-Fuchs equation \eqref{eq:pfeq} are listed as follows:
\begin{equation}
\begin{aligned}
c_1^{(0)}&=\frac{2u_1}{3}, \quad c_2^{(0)}=\frac{4u_2}{3},\\
c_1^{(1)}&=\frac{1}{6 \Delta}-\frac{1}{12 u_2},\quad  c_2^{(1)}=\frac{u_1}{12 \Delta }, \\
c_1^{(2)}&=\frac{u_1 \left(56 u_1^6-663 u_2 u_1^4+2685 u_2^2 u_1^2-2820 u_2^3\right)}{2880 \Delta ^3 u_2^3} ,\quad
c_2^{(2)}= \frac{14 u_1^6-150 u_2 u_1^4+765 u_2^2 u_1^2-660 u_2^3}{1440 \Delta
   ^3 u_2^2}, \\
c_1^{(3)}&=\frac{-3968 u_1^{12}+82648 u_2 u_1^{10}-702093 u_2^2 u_1^8+3086713 u_2^3 u_1^6-7276850 u_2^4 u_1^4+8856288 u_2^5 u_1^2-3156384 u_2^6}{161280 \Delta ^5 u_2^5},\\
c_2^{(3)}&=
   \frac{-1984 u_1^{11}+39092 u_2 u_1^9-309393 u_2^2 u_1^7+1236305 u_2^3 u_1^5-2332120 u_2^4 u_1^3+2388624 u_2^5 u_1}{161280 \Delta ^5 u_2^4},
\end{aligned}
\end{equation}
where the discriminant of $z\phi_2(z)$ is denoted as
\begin{equation}
    \Delta=u_1^2-4u_2.
\end{equation}
We set $\phi_2(z)=z^2+u_1 z+u_2+\frac{u_3}{z}$ and $\ell=-\frac{1}{2}$ for $(A_1,D_4)$ theory, and 
denote the discriminant of $z\phi_2(z)$ as
\begin{equation}
    \Delta=\left(u_1^2-4 u_2\right) u_2^2-27 u_3^2+2 u_1 \left(9 u_2-2 u_1^2\right) u_3,
\end{equation}
the coefficients read:
\begin{equation}
    \begin{aligned}
 c_1^{(0)}&= \frac{u_1}{2},\quad c_2^{(0)}=u_2, \quad c_3^{(0)}=\frac{3 u_3}{2}, \\
 c_1^{(1)}&= \frac{1}{6} \left(\frac{2 u_1^3-9 u_2 u_1+27 u_3}{\Delta }-\frac{1}{u_3}\right),\quad   c_2^{(1)}=\frac{1}{12} \left(-\frac{\left(-2 u_1^3+9 u_2 u_1-27 u_3\right) u_1}{\Delta
   }-\frac{u_1}{u_3}\right),\\
   c_3^{(1)}&=-\frac{3 \left(u_1^2-6 u_2\right) u_3-u_1 \left(u_1^2-4 u_2\right) u_2}{12 \Delta}.
   \end{aligned}
\end{equation}

\begin{equation}
    \begin{aligned}
 c_1^{(2)}&=
 \frac{1}{1440 \Delta^3 u_3^3}
\left(
-3584 u_2^{10} + 2688 u_1^2 u_2^9 - 672 u_1^4 u_2^8 + 47808 u_1 u_3 u_2^8 
+ 56 u_1^6 u_2^7 - 65856 u_3^2 u_2^7 \right.\\
&- 34512 u_1^3 u_3 u_2^7 
- 185424 u_1^2 u_3^2 u_2^6 + 8292 u_1^5 u_3 u_2^6 + 739440 u_1 u_3^3 u_2^5 
+ 147348 u_1^4 u_3^2 u_2^5 - 663 u_1^7 u_3 u_2^5\\
&- 934416 u_3^4 u_2^4 
- 97464 u_1^3 u_3^3 u_2^4 - 34959 u_1^6 u_3^2 u_2^4 - 821448 u_1^2 u_3^4 u_2^3 
- 127557 u_1^5 u_3^3 u_2^3 + 2685 u_1^8 u_3^2 u_2^3\\
&+ 1792368 u_1 u_3^5 u_2^2 
+ 457695 u_1^4 u_3^4 u_2^2 + 37749 u_1^7 u_3^3 u_2^2 - 327564 u_1^3 u_3^5 u_2 
- 46845 u_1^6 u_3^4 u_2 - 2820 u_1^9 u_3^3 u_2\\
&\left. - 379080 u_1^2 u_3^6
- 7317 u_1^5 u_3^5 - 828 u_1^8 u_3^4\right).
\end{aligned}
\end{equation}

\begin{equation}
    \begin{aligned}
 c_2^{(2)}&=\frac{1}{2880 \Delta ^3 u_3^3}\left(-2820 u_2 u_3^3 u_1^{10}-828 u_3^4 u_1^9+2685 u_2^3 u_3^2 u_1^9+37749 u_2^2 u_3^3 u_1^8-663 u_2^5 u_3 u_1^8+56 u_2^7 u_1^7\right.\\
 &-46845 u_2 u_3^4 u_1^7-34959 u_2^4 u_3^2 u_1^7-7317 u_3^5 u_1^6-127557 u_2^3 u_3^3 u_1^6+8292 u_2^6 u_3 u_1^6-672 u_2^8 u_1^5\\
 &+457695 u_2^2 u_3^4 u_1^5+147348 u_2^5 u_3^2 u_1^5-327564 u_2 u_3^5 u_1^4-97464
   u_2^4 u_3^3 u_1^4-34512 u_2^7 u_3 u_1^4+2688 u_2^9 u_1^3\\
   &-379080 u_3^6 u_1^3-821448 u_2^3 u_3^4 u_1^3-185424 u_2^6 u_3^2 u_1^3+1792368 u_2^2 u_3^5 u_1^2+739440 u_2^5
   u_3^3 u_1^2+47808 u_2^8 u_3 u_1^2\\
   &\left.-3584 u_2^{10} u_1-934416 u_2^4 u_3^4 u_1-65856 u_2^7 u_3^2 u_1
 \right).
 \end{aligned}
\end{equation}

\begin{equation}
    \begin{aligned}
 c_3^{(2)}&=\frac{1}{1440 \Delta ^3 u_3^2}\left(-660 u_3^3 u_1^{10}+765 u_2^2 u_3^2 u_1^9+8454 u_2 u_3^3 u_1^8-150 u_2^4 u_3 u_1^8+14 u_2^6 u_1^7-13491 u_3^4 u_1^7\right.\\
 &-10119 u_2^3 u_3^2 u_1^7-21969 u_2^2 u_3^3
   u_1^6+1863 u_2^5 u_3 u_1^6-168 u_2^7 u_1^5+114111 u_2 u_3^4 u_1^5+44640 u_2^4 u_3^2 u_1^5\\
&-72009 u_3^5 u_1^4-75297 u_2^3 u_3^3 u_1^4-7620 u_2^6 u_3 u_1^4+672 u_2^8
   u_1^3-165456 u_2^2 u_3^4 u_1^3-66576 u_2^5 u_3^2 u_1^3\\
&+234252 u_2 u_3^5 u_1^2+312264 u_2^4 u_3^3 u_1^2+9936 u_2^7 u_3 u_1^2-896 u_2^9 u_1-349920 u_3^6 u_1-325296
   u_2^3 u_3^4 u_1\\
&\left.+3840 u_2^6 u_3^2 u_1+571536 u_2^2 u_3^5-123984 u_2^5 u_3^3+1344 u_2^8 u_3\right).
\end{aligned}
\end{equation}

For $(A_1,D_5)$ theory, we set $\phi_2(z)=z^3+u_1 z^2+u_2 z+u_3+\frac{u_4}{z}$ and $\ell=-\frac{1}{2}$, and denote the discriminant of $z\phi_2(z)$ as
\begin{equation}
\begin{aligned}
      \Delta&=256 u_4^3-\left(27 u_1^4-144 u_2 u_1^2+192 u_3 u_1+128 u_2^2\right) u_4^2\\
      &+2 \left(-2 \left(u_1^2-4 u_2\right) u_2^3+u_1 \left(9 u_1^2-40 u_2\right) u_3 u_2-3
   \left(u_1^2-24 u_2\right) u_3^2\right) u_4\\
   &+u_3^2 \left(\left(u_1^2-4 u_2\right) u_2^2-27 u_3^2+2 u_1 \left(9 u_2-2 u_1^2\right) u_3\right),
\end{aligned}
\end{equation}
the coefficients read:
\begin{equation}
    \begin{aligned}
c_1^{(0)}&=\frac{2 u_1}{5},\quad c_2^{(0)}=\frac{4 u_2}{5}, \quad c_3^{(0)}=\frac{6 u_3}{5}, \quad c_4^{(0)}=\frac{8 u_4}{5}.
 \end{aligned}
\end{equation}

\begin{equation}
    \begin{aligned}
c_1^{(1)}&=-\frac{1}{4 u_4}+\frac{1}{4\Delta}\left(-27 u_4 u_1^4+9 u_2 u_3 u_1^3-2 u_2^3 u_1^2-3 u_3^2 u_1^2+144 u_2 u_4 u_1^2-40 u_2^2 u_3 u_1-192 u_3 u_4 u_1\right.\\
&\left.+8 u_2^4+72 u_2 u_3^2+384 u_4^2-128 u_2^2 u_4\right),\\
c_2^{(1)}&=-\frac{u_1}{6 u_4}+\frac{1}{12\Delta}\left(-54 u_4 u_1^5+18 u_2 u_3 u_1^4-4 u_2^3 u_1^3-12 u_3^2 u_1^3+297 u_2 u_4 u_1^3-79 u_2^2 u_3 u_1^2-390 u_3 u_4 u_1^2\right.\\
&\left.+16 u_2^4 u_1+171 u_2 u_3^2 u_1+672 u_4^2 u_1-296
   u_2^2 u_4 u_1-54 u_3^3-4 u_2^3 u_3+144 u_2 u_3 u_4\right),\\
c_3^{(1)}&=-\frac{u_2}{12 u_4}+\frac{1}{12\Delta}\left(8 u_2^5-2 u_1^2 u_2^4-40 u_1 u_3 u_2^3-160 u_4 u_2^3+78 u_3^2 u_2^2+9 u_1^3 u_3 u_2^2+150 u_1^2 u_4 u_2^2-4 u_1^2 u_3^2 u_2\right.\\
&\left.+512 u_4^2 u_2-27 u_1^4 u_4 u_2-112 u_1 u_3
   u_4 u_2-9 u_1 u_3^3-72 u_1^2 u_4^2-72 u_3^2 u_4-9 u_1^3 u_3 u_4\right),\\
c_4^{(1)}&=\frac{1}{12\Delta}\left(-18 u_3 u_4 u_1^4+4 u_2 u_3^2 u_1^3+18 u_4^2 u_1^3+3 u_2^2 u_4 u_1^3-u_2^3 u_3 u_1^2+85 u_2 u_3 u_4 u_1^2-18 u_2^2 u_3^2 u_1\right.\\
&\left.-80 u_2 u_4^2 u_1-12 u_2^3 u_4 u_1-126 u_3^2
   u_4 u_1+27 u_2 u_3^3+288 u_3 u_4^2+4 u_2^4 u_3-24 u_2^2 u_3 u_4\right),\\
    \end{aligned}
\end{equation}

In the case of $\mathbb{Z}_4$ maximally symmetric point, where $u_1=u_2=u_3=0$ and $u_4=1$. the only non-zero coefficients up to $\mathcal{O}(\hbar^{24})$ are
\begin{equation}
    \begin{aligned}
        c_4^{(0)}&=\frac{8}{5}, \quad 
      c_1^{(1)}=-\frac{5}{8},\quad 
      c_2^{(2)}=\frac{133}{9216},\quad 
      c_3^{(3)}=\frac{2159}{65536},\quad 
      c_4^{(4)}=-\frac{37832553}{587202560},\\
      c_1^{(5)}&=-\frac{263717771395}{106300440576},\quad 
      c_2^{(6)}=\frac{173114893083769}{136064563937
   28},\quad 
      c_3^{(7)}=\frac{15405368950904625}{114349209288704},\\
      c_4^{(8)}&=-\frac{963145355778301522161}{292733975779082240},\quad 
      c_1^{(9)}=-\frac{345000540510358460821043845}{837885703473026039808},\\ 
      c_2^{(10)}&=\frac{5816
   28545915944940523377885317}{115661085342158888632320},\quad   c_3^{(11)}=\frac{394696439067518703928216748385}{2361183241434822606848},\\
c_4^{(12)}&=-\frac{131419681684386258272929265572848953}{1390
   2646925568235509121024}.
    \end{aligned}
\end{equation}

\bibliographystyle{JHEP}
\linespread{0.6}

\begin{thebibliography}{99}

\bibitem{SW941}
N.~Seiberg and E.~Witten, \emph{{Electric - magnetic duality, monopole
  condensation, and confinement in N=2 supersymmetric Yang-Mills theory}},
  \href{http://dx.doi.org/10.1016/0550-3213(94)90124-4}{\emph{Nucl. Phys. B}
  {\bf 426} (1994) 19--52}, [\href{http://arxiv.org/abs/hep-th/9407087}{{\tt
  hep-th/9407087}}].

\bibitem{SW942}
N.~Seiberg and E.~Witten, \emph{{Monopoles, duality and chiral symmetry
  breaking in N=2 supersymmetric QCD}},
  \href{http://dx.doi.org/10.1016/0550-3213(94)90214-3}{\emph{Nucl. Phys. B}
  {\bf 431} (1994) 484--550}, [\href{http://arxiv.org/abs/hep-th/9408099}{{\tt
  hep-th/9408099}}].

\bibitem{AD95}
P.~C. Argyres and M.~R. Douglas, \emph{{New phenomena in SU(3) supersymmetric
  gauge theory}},
  \href{http://dx.doi.org/10.1016/0550-3213(95)00281-V}{\emph{Nucl. Phys. B}
  {\bf 448} (1995) 93--126}, [\href{http://arxiv.org/abs/hep-th/9505062}{{\tt
  hep-th/9505062}}].

\bibitem{ADSW95}
P.~C. Argyres, M.~R. Plesser, N.~Seiberg and E.~Witten, \emph{{New N=2
  superconformal field theories in four-dimensions}},
  \href{http://dx.doi.org/10.1016/0550-3213(95)00671-0}{\emph{Nucl. Phys. B}
  {\bf 461} (1996) 71--84}, [\href{http://arxiv.org/abs/hep-th/9511154}{{\tt
  hep-th/9511154}}].

\bibitem{EHIY96}
T.~Eguchi, K.~Hori, K.~Ito and S.-K. Yang, \emph{{Study of N=2 superconformal
  field theories in four-dimensions}},
  \href{http://dx.doi.org/10.1016/0550-3213(96)00188-5}{\emph{Nucl. Phys. B}
  {\bf 471} (1996) 430--444}, [\href{http://arxiv.org/abs/hep-th/9603002}{{\tt
  hep-th/9603002}}].

\bibitem{Gaiotto09}
D.~Gaiotto, \emph{{N=2 dualities}},
  \href{http://dx.doi.org/10.1007/JHEP08(2012)034}{\emph{JHEP} {\bf 08} (2012)
  034}, [\href{http://arxiv.org/abs/0904.2715}{{\tt 0904.2715}}].

\bibitem{GMN09}
D.~Gaiotto, G.~W. Moore and A.~Neitzke, \emph{{Wall-crossing, Hitchin systems,
  and the WKB approximation}},
  \href{http://dx.doi.org/10.1016/j.aim.2012.09.027}{\emph{Adv. Math.} {\bf
  234} (2013) 239--403}, [\href{http://arxiv.org/abs/0907.3987}{{\tt
  0907.3987}}].

\bibitem{Bonelli:2011aa}
G.~Bonelli, K.~Maruyoshi and A.~Tanzini,
\emph{{Wild Quiver Gauge Theories}},
\href{https://doi.org/10.1007/JHEP02(2012)031}{\emph{JHEP} \textbf{02} (2012) 031},
[\href{https://arxiv.org/abs/1112.1691}{1112.1691}].

\bibitem{Xie12}
D.~Xie, \emph{{General Argyres-Douglas Theory}},
  \href{http://dx.doi.org/10.1007/JHEP01(2013)100}{\emph{JHEP} {\bf 01} (2013)
  100}, [\href{http://arxiv.org/abs/1204.2270}{{\tt 1204.2270}}].

\bibitem{WX15}
Y.~Wang and D.~Xie, \emph{{Classification of Argyres-Douglas theories from M5
  branes}}, \href{http://dx.doi.org/10.1103/PhysRevD.94.065012}{\emph{Phys.
  Rev. D} {\bf 94} (2016) 065012}, [\href{http://arxiv.org/abs/1509.00847}{{\tt
  1509.00847}}].

\bibitem{Alim11}
M.~Alim, S.~Cecotti, C.~Cordova, S.~Espahbodi, A.~Rastogi and C.~Vafa,
  \emph{{BPS Quivers and Spectra of Complete N=2 Quantum Field Theories}},
  \href{http://dx.doi.org/10.1007/s00220-013-1789-8}{\emph{Commun. Math. Phys.}
  {\bf 323} (2013) 1185--1227}, [\href{http://arxiv.org/abs/1109.4941}{{\tt
  1109.4941}}].

\bibitem{KLPV96}
A.~Klemm, W.~Lerche, P.~Mayr, C.~Vafa and N.~P. Warner, \emph{{Selfdual strings
  and N=2 supersymmetric field theory}},
  \href{http://dx.doi.org/10.1016/0550-3213(96)00353-7}{\emph{Nucl. Phys. B}
  {\bf 477} (1996) 746--766}, [\href{http://arxiv.org/abs/hep-th/9604034}{{\tt
  hep-th/9604034}}].

\bibitem{KKV96}
S.~H. Katz, A.~Klemm and C.~Vafa, \emph{{Geometric engineering of quantum field
  theories}},
  \href{http://dx.doi.org/10.1016/S0550-3213(97)00282-4}{\emph{Nucl. Phys. B}
  {\bf 497} (1997) 173--195}, [\href{http://arxiv.org/abs/hep-th/9609239}{{\tt
  hep-th/9609239}}].

\bibitem{Nekrasov02}
N.~A. Nekrasov, \emph{{Seiberg-Witten prepotential from instanton counting}},
  \href{http://dx.doi.org/10.4310/ATMP.2003.v7.n5.a4}{\emph{Adv. Theor. Math.
  Phys.} {\bf 7} (2003) 831--864},
  [\href{http://arxiv.org/abs/hep-th/0206161}{{\tt hep-th/0206161}}].

\bibitem{NS09}
N.~A. Nekrasov and S.~L. Shatashvili, \emph{{Quantization of Integrable Systems
  and Four Dimensional Gauge Theories}},  in \emph{{16th International Congress
  on Mathematical Physics}}, pp.~265--289, 2010.
\newblock \href{http://arxiv.org/abs/0908.4052}{{\tt 0908.4052}}.
\newblock \href{http://dx.doi.org/10.1142/9789814304634_0015}{DOI}.

\bibitem{WKB}
J.~M.~L. Dunham, \emph{The Wentzel-Brillouin-Kramers method of solving the wave
  equation}, {\emph{Physical Review} {\bf 41} (1932) 713--720}.

\bibitem{BW69}
C.~M. Bender and T.~T. Wu, \emph{{Anharmonic oscillator}},
  \href{http://dx.doi.org/10.1103/PhysRev.184.1231}{\emph{Phys. Rev.} {\bf 184}
  (1969) 1231--1260}.

\bibitem{BPV79}
R.~Balian, G.~Parisi and A.~Voros, \emph{Quartic oscillator},  in \emph{Feynman
  Path Integrals} (S.~Albeverio, P.~Combe, R.~H{\o}egh-Krohn, G.~Rideau,
  M.~Sirugue-Collin, M.~Sirugue et~al., eds.), (Berlin, Heidelberg),
  pp.~337--360, Springer Berlin Heidelberg, 1979.

\bibitem{Voros81}
A.~Voros, \emph{Spectre de l'{\'e}quation de Schr{\"o}dinger et m{\'e}thode
  BKW},  Publications Math{\'e}matiques d'Orsay, 1981.

\bibitem{Voros83}
A.~Voros, \emph{The return of the quartic oscillator. the complex WKB method},
  {\emph{Annales de l'I.H.P. Physique th\'eorique} {\bf 39} (1983) 211--338}.

\bibitem{DDP93}
H.~Dillinger, E.~Delabaere and F.~Pham, \emph{R\'esurgence de {V}oros et
  p\'eriodes des courbes hyperelliptiques}, {\emph{Annales de l'Institut
  Fourier} {\bf 43} (1993) 163}.

\bibitem{DDP97}
E.~Delabaere, H.~Dillinger and F.~Pham, \emph{{Exact semiclassical expansions
  for one-dimensional quantum oscillators}},
  \href{http://dx.doi.org/10.1063/1.532206}{\emph{J. Math. Phys.} {\bf 38}
  (1997) 6126}.

\bibitem{DP99}
E.~Delabaere and F.~Pham, \emph{Resurgent methods in semi-classical
  asymptotics}, {\emph{Annales de l'I.H.P. Physique th\'eorique} {\bf 71}
  (1999) 1--94}.

\bibitem{KT05}
T.~Kawai and Y.~Takei, \emph{Algebraic analysis of singular perturbation
  theory}, vol.~227.
\newblock American Mathematical Soc., 2005.

\bibitem{IN14}
K.~Iwaki and T.~Nakanishi, \emph{{Exact WKB analysis and cluster algebras}},
  {\emph{J. Phys. A} {\bf 47} (2014) 474009}.

\bibitem{IN15}
K.~Iwaki and T.~Nakanishi, \emph{Exact {WKB} analysis and cluster algebras
  {II}: Simple poles, orbifold points, and generalized cluster algebras},
  \href{http://dx.doi.org/10.1093/imrn/rnv270}{\emph{International Mathematics
  Research Notices} {\bf 2016} (oct, 2015) 4375--4417}.

\bibitem{SKMU20}
N.~Sueishi, S.~Kamata, T.~Misumi and M.~Ünsal, \emph{On exact-{WKB} analysis,
  resurgent structure, and quantization conditions},
  \href{http://dx.doi.org/10.1007/jhep12(2020)114}{\emph{JHEP} {\bf 12} (2020) 114 }.

\bibitem{SKMU21}
N.~Sueishi, S.~Kamata, T.~Misumi and M.~Ünsal, \emph{Exact-{WKB}, complete
  resurgent structure, and mixed anomaly in quantum mechanics on $S^1$},
  \href{http://dx.doi.org/10.1007/jhep07(2021)096}{\emph{JHEP} {\bf 07} (2021) 096 }.

\bibitem{GGM19}
A.~Grassi, J.~Gu and M.~Mari\~no, \emph{{Non-perturbative approaches to the
  quantum Seiberg-Witten curve}},
  \href{http://dx.doi.org/10.1007/JHEP07(2020)106}{\emph{JHEP} {\bf 07} (2020)
  106}, [\href{http://arxiv.org/abs/1908.07065}{{\tt 1908.07065}}].

\bibitem{GHN21}
A.~Grassi, Q.~Hao and A.~Neitzke, \emph{{Exact WKB methods in SU(2) N$_{f}$ =
  1}}, \href{http://dx.doi.org/10.1007/JHEP01(2022)046}{\emph{JHEP} {\bf 01}
  (2022) 046}, [\href{http://arxiv.org/abs/2105.03777}{{\tt 2105.03777}}].

\bibitem{Yan20}
F.~Yan, \emph{{Exact WKB and the quantum Seiberg-Witten curve for 4d N = 2 pure
  SU(3) Yang-Mills. Abelianization}},
  \href{http://dx.doi.org/10.1007/JHEP03(2022)164}{\emph{JHEP} {\bf 03} (2022)
  164}, [\href{http://arxiv.org/abs/2012.15658}{{\tt 2012.15658}}].

\bibitem{ML22}
F.~Del~Monte and P.~Longhi, \emph{{The threefold way to quantum periods: WKB,
  TBA equations and q-Painlev\'e}},
  \href{http://dx.doi.org/10.21468/SciPostPhys.15.3.112}{\emph{SciPost Phys.}
  {\bf 15} (2023) 112}, [\href{http://arxiv.org/abs/2207.07135}{{\tt
  2207.07135}}].

\bibitem{GMN08}
D.~Gaiotto, G.~W. Moore and A.~Neitzke, \emph{{Four-dimensional wall-crossing
  via three-dimensional field theory}},
  \href{http://dx.doi.org/10.1007/s00220-010-1071-2}{\emph{Commun. Math. Phys.}
  {\bf 299} (2010) 163--224}, [\href{http://arxiv.org/abs/0807.4723}{{\tt
  0807.4723}}].

\bibitem{Gaiotto14}
D.~Gaiotto, \emph{{Opers and TBA}},  \href{http://arxiv.org/abs/1403.6137}{{\tt
  1403.6137}}.

\bibitem{Fioravanti:2019vxi}
D.~Fioravanti and D.~Gregori, \emph{{Integrability and cycles of deformed
  ${\cal N}=2$ gauge theory}},
  \href{http://dx.doi.org/10.1016/j.physletb.2020.135376}{\emph{Phys. Lett. B}
  {\bf 804} (2020) 135376}, [\href{http://arxiv.org/abs/1908.08030}{{\tt
  1908.08030}}].

\bibitem{Imaizumi:2020fxf}
K.~Imaizumi, \emph{{Exact WKB analysis and TBA equations for the Mathieu
  equation}},
  \href{http://dx.doi.org/10.1016/j.physletb.2020.135500}{\emph{Phys. Lett. B}
  {\bf 806} (2020) 135500}, [\href{http://arxiv.org/abs/2002.06829}{{\tt
  2002.06829}}].

\bibitem{OS22}
H.~Ouyang and H.~Shu, \emph{{TBA-like equations for non-planar scattering
  amplitude/Wilson lines duality at strong coupling}},
  \href{http://dx.doi.org/10.1007/JHEP05(2022)099}{\emph{JHEP} {\bf 05} (2022)
  099}, [\href{http://arxiv.org/abs/2202.10700}{{\tt 2202.10700}}].

\bibitem{GMN12}
D.~Gaiotto, G.~W. Moore and A.~Neitzke, \emph{{Spectral networks}},
  \href{http://dx.doi.org/10.1007/s00023-013-0239-7}{\emph{Annales Henri
  Poincare} {\bf 14} (2013) 1643--1731},
  [\href{http://arxiv.org/abs/1204.4824}{{\tt 1204.4824}}].

\bibitem{LP16}
P.~Longhi and C.~Y. Park, \emph{{ADE Spectral Networks}},
  \href{http://dx.doi.org/10.1007/JHEP08(2016)087}{\emph{JHEP} {\bf 08} (2016)
  087}, [\href{http://arxiv.org/abs/1601.02633}{{\tt 1601.02633}}].

\bibitem{IK20}
K.~Iwaki and O.~Kidwai, \emph{{Topological recursion and uncoupled BPS
  structures I: BPS spectrum and free energies}},
  \href{http://dx.doi.org/10.1016/j.aim.2022.108191}{\emph{Adv. Math.} {\bf
  398} (2022) 108191}, [\href{http://arxiv.org/abs/2010.05596}{{\tt
  2010.05596}}].

\bibitem{IK21}
K.~Iwaki and O.~Kidwai, \emph{{Topological Recursion and Uncoupled BPS
  Structures II: Voros Symbols and the $\tau $-Function}},
  \href{http://dx.doi.org/10.1007/s00220-022-04563-y}{\emph{Commun. Math.
  Phys.} {\bf 399} (2023) 519--572},
  [\href{http://arxiv.org/abs/2108.06995}{{\tt 2108.06995}}].

\bibitem{MPY13}
K.~Maruyoshi, C.~Y. Park and W.~Yan, \emph{{BPS spectrum of Argyres-Douglas
  theory via spectral network}},
  \href{http://dx.doi.org/10.1007/JHEP12(2013)092}{\emph{JHEP} {\bf 12} (2013)
  092}, [\href{http://arxiv.org/abs/1309.3050}{{\tt 1309.3050}}].

\bibitem{DT96}
P.~Dorey and R.~Tateo, \emph{{Excited states by analytic continuation of TBA
  equations}},
  \href{http://dx.doi.org/10.1016/S0550-3213(96)00516-0}{\emph{Nucl. Phys. B}
  {\bf 482} (1996) 639--659}, [\href{http://arxiv.org/abs/hep-th/9607167}{{\tt
  hep-th/9607167}}].

\bibitem{DT98}
P.~Dorey and R.~Tateo, \emph{{Anharmonic oscillators, the thermodynamic Bethe
  ansatz, and nonlinear integral equations}},
  \href{http://dx.doi.org/10.1088/0305-4470/32/38/102}{\emph{J. Phys. A} {\bf
  32} (1999) L419--L425}, [\href{http://arxiv.org/abs/hep-th/9812211}{{\tt
  hep-th/9812211}}].

\bibitem{DT07}
P.~Dorey, C.~Dunning and R.~Tateo, \emph{{The ODE/IM Correspondence}},
  \href{http://dx.doi.org/10.1088/1751-8113/40/32/R01}{\emph{J. Phys. A} {\bf
  40} (2007) R205}, [\href{http://arxiv.org/abs/hep-th/0703066}{{\tt
  hep-th/0703066}}].

\bibitem{IMS18}
K.~Ito, M.~Mariño and H.~Shu, \emph{{TBA equations and resurgent Quantum
  Mechanics}}, \href{http://dx.doi.org/10.1007/JHEP01(2019)228}{\emph{JHEP}
  {\bf 01} (2019) 228}, [\href{http://arxiv.org/abs/1811.04812}{{\tt
  1811.04812}}].

\bibitem{Emery20}
Y.~Emery, \emph{{TBA equations and quantization conditions}},
  \href{http://dx.doi.org/10.1007/JHEP07(2021)171}{\emph{JHEP} {\bf 07} (2021)
  171}, [\href{http://arxiv.org/abs/2008.13680}{{\tt 2008.13680}}].

\bibitem{IS19}
K.~Ito and H.~Shu, \emph{{TBA equations for the Schr\"odinger equation with a
  regular singularity}},
  \href{http://dx.doi.org/10.1088/1751-8121/ab96ee}{\emph{J. Phys. A} {\bf 53}
  (2020) 335201}, [\href{http://arxiv.org/abs/1910.09406}{{\tt 1910.09406}}].

\bibitem{Klassen:1989ui}
T.~R. Klassen and E.~Melzer, \emph{{Purely Elastic Scattering Theories and
  their Ultraviolet Limits}},
  \href{http://dx.doi.org/10.1016/0550-3213(90)90643-R}{\emph{Nucl. Phys. B}
  {\bf 338} (1990) 485--528}.

\bibitem{Zam91}
{Al. B. Zamolodchikov}, \emph{{On the thermodynamic Bethe ansatz equations for
  reflectionless ADE scattering theories}},
  \href{http://dx.doi.org/10.1016/0370-2693(91)91737-G}{\emph{Phys. Lett. B}
  {\bf 253} (1991) 391--394}.

\bibitem{IKO19}
K.~Ito, S.~Koizumi and T.~Okubo, \emph{{Quantum Seiberg-Witten curve and
  Universality in Argyres-Douglas theories}},
  \href{http://dx.doi.org/10.1016/j.physletb.2019.03.024}{\emph{Phys. Lett. B}
  {\bf 792} (2019) 29--34}, [\href{http://arxiv.org/abs/1903.00168}{{\tt
  1903.00168}}].

\bibitem{Dorigoni14}
D.~Dorigoni, \emph{{An Introduction to Resurgence, Trans-Series and Alien
  Calculus}}, \href{http://dx.doi.org/10.1016/j.aop.2019.167914}{\emph{Annals
  Phys.} {\bf 409} (2019) 167914}, [\href{http://arxiv.org/abs/1411.3585}{{\tt
  1411.3585}}].

\bibitem{KS08}
M.~Kontsevich and Y.~Soibelman, \emph{{Stability structures, motivic
  Donaldson-Thomas invariants and cluster transformations}},
  \href{http://arxiv.org/abs/0811.2435}{{\tt 0811.2435}}.

\bibitem{Dorey:1999uk}
P.~Dorey and R.~Tateo, \emph{{On the relation between Stokes multipliers and
  the T-Q systems of conformal field theory}},
  \href{http://dx.doi.org/10.1016/S0550-3213(99)00609-4}{\emph{Nucl. Phys. B}
  {\bf 563} (1999) 573--602}, [\href{http://arxiv.org/abs/hep-th/9906219}{{\tt
  hep-th/9906219}}].

\bibitem{Zam90}
{Al. B. Zamolodchikov}, \emph{{Thermodynamic Bethe Ansatz in Relativistic
  Models. Scaling Three State Potts and Lee-yang Models}},
  \href{http://dx.doi.org/10.1016/0550-3213(90)90333-9}{\emph{Nucl. Phys. B}
  {\bf 342} (1990) 695--720}.

\bibitem{BD15}
G.~Ba\c{s}ar and G.~V. Dunne, \emph{{Resurgence and the Nekrasov-Shatashvili
  limit: connecting weak and strong coupling in the Mathieu and Lam\'e
  systems}}, \href{http://dx.doi.org/10.1007/JHEP02(2015)160}{\emph{JHEP} {\bf
  02} (2015) 160}, [\href{http://arxiv.org/abs/1501.05671}{{\tt 1501.05671}}].

\bibitem{CM16}
S.~Codesido and M.~Mari\~no, \emph{{Holomorphic Anomaly and Quantum Mechanics}},
  \href{http://dx.doi.org/10.1088/1751-8121/aa9e77}{\emph{J. Phys. A} {\bf 51}
  (2018) 055402}, [\href{http://arxiv.org/abs/1612.07687}{{\tt 1612.07687}}].

\bibitem{BD17}
G.~Ba\c{s}ar, G.~V. Dunne and M.~\"Unsal, \emph{{Quantum Geometry of Resurgent
  Perturbative/Nonperturbative Relations}},
  \href{http://dx.doi.org/10.1007/JHEP05(2017)087}{\emph{JHEP} {\bf 05} (2017)
  087}, [\href{http://arxiv.org/abs/1701.06572}{{\tt 1701.06572}}].

\bibitem{GY21}
B.~Gabai and X.~Yin, \emph{{Exact quantization and analytic continuation}},
  \href{http://dx.doi.org/10.1007/JHEP03(2023)082}{\emph{JHEP} {\bf 03} (2023)
  082}, [\href{http://arxiv.org/abs/2109.07516}{{\tt 2109.07516}}].

\bibitem{Fendley:1997ys}
P.~Fendley, \emph{{Excited state energies and supersymmetric indices}},
  \href{http://dx.doi.org/10.4310/ATMP.1997.v1.n2.a2}{\emph{Adv. Theor. Math.
  Phys.} {\bf 1} (1998) 210--236},
  [\href{http://arxiv.org/abs/hep-th/9706161}{{\tt hep-th/9706161}}].

\bibitem{IY23}
K.~Ito and J.~Yang, \emph{{Exact WKB Analysis and TBA Equations for the Stark
  Effect}}, \href{http://dx.doi.org/10.1093/ptep/ptad154}{\emph{PTEP} {\bf
  2024} (2024) 013A02}, [\href{http://arxiv.org/abs/2307.03504}{{\tt
  2307.03504}}].

\bibitem{IS24}
K.~Ito and H.~Shu, \emph{{TBA equations and exact WKB analysis in deformed
  supersymmetric quantum mechanics}},
  \href{http://dx.doi.org/10.1007/JHEP03(2024)122}{\emph{JHEP} {\bf 03} (2024)
  122}, [\href{http://arxiv.org/abs/2401.03766}{{\tt 2401.03766}}].

\bibitem{IKKS21}
K.~Ito, T.~Kondo, K.~Kuroda and H.~Shu, \emph{{WKB periods for higher order ODE
  and TBA equations}},
  \href{http://dx.doi.org/10.1007/JHEP10(2021)167}{\emph{JHEP} {\bf 10} (2021)
  167}, [\href{http://arxiv.org/abs/2104.13680}{{\tt 2104.13680}}].

\bibitem{IKS21}
K.~Ito, T.~Kondo and H.~Shu, \emph{{Wall-crossing of TBA equations and WKB
  periods for the third order ODE}},
  \href{http://dx.doi.org/10.1016/j.nuclphysb.2022.115788}{\emph{Nucl. Phys. B}
  {\bf 979} (2022) 115788}, [\href{http://arxiv.org/abs/2111.11047}{{\tt
  2111.11047}}].

\bibitem{Cecotti:2014zga}
S.~Cecotti and M.~Del~Zotto, \emph{{$Y$ systems, $Q$ systems, and 4D
  $\mathcal{N}=2$ supersymmetric QFT}},
  \href{http://dx.doi.org/10.1088/1751-8113/47/47/474001}{\emph{J. Phys. A}
  {\bf 47} (2014) 474001}, [\href{http://arxiv.org/abs/1403.7613}{{\tt
  1403.7613}}].

\bibitem{Ito:2017ypt}
K.~Ito and H.~Shu, \emph{{ODE/IM correspondence and the Argyres-Douglas
  theory}}, \href{http://dx.doi.org/10.1007/JHEP08(2017)071}{\emph{JHEP} {\bf
  08} (2017) 071}, [\href{http://arxiv.org/abs/1707.03596}{{\tt 1707.03596}}].

\bibitem{BLR18}
S.~Banerjee, P.~Longhi and M.~Romo, \emph{{Exploring 5d BPS Spectra with
  Exponential Networks}},
  \href{http://dx.doi.org/10.1007/s00023-019-00851-x}{\emph{Annales Henri
  Poincare} {\bf 20} (2019) 4055--4162},
  [\href{http://arxiv.org/abs/1811.02875}{{\tt 1811.02875}}].

\bibitem{Longhi21}
P.~Longhi, \emph{{Instanton Particles and Monopole Strings in 5D SU(2)
  Supersymmetric Yang-Mills Theory}},
  \href{http://dx.doi.org/10.1103/PhysRevLett.126.211601}{\emph{Phys. Rev.
  Lett.} {\bf 126} (2021) 211601}, [\href{http://arxiv.org/abs/2101.01681}{{\tt
  2101.01681}}].

\end{thebibliography}

\end{document}